\documentclass[12pt,preprint]{aastex}

\slugcomment{Submitted to ApJ 06/11/2004}

\usepackage{times}
\usepackage{mathptm}
\usepackage{bm}

\newcommand{\bolE}{{\bm  E}}
\newcommand{\bolD}{{\bm  D}}
\newcommand{\bolB}{{\bm  B}}
\newcommand{\bolJ}{{\bm  J}}
\newcommand{\bolV}{{\bm  V}}

\newcommand{\beq}{\begin{equation}}
\newcommand{\eeq}{\end{equation}}
\newcommand{\beqn}{\begin{eqnarray}}
\newcommand{\eeqn}{\end{eqnarray}}
\newcommand{\beqno}{\begin{equation*}}
\newcommand{\eeqno}{\end{equation*}}
\newcommand{\beqnno}{\begin{eqnarray*}}
\newcommand{\eeqnno}{\end{eqnarray*}}

\makeatletter
\def\ddigit#1{\setbox0=\hbox{9}\setbox1=\hbox{#1}%
  \ifdim \wd0<\wd1 #1\else 0#1\fi}
\newcounter{hour}
\newcounter{HOUR}
\newcounter{minuite}
 \divide \c@hour by 60
 \multiply \c@HOUR by 60
 \advance \c@minuite by -\c@HOUR

\def\filename{{\tt \jobname.tex}}
\makeatother

\shorttitle{Nonlinear Poynting Flux Dominated Jets}
\shortauthors{Nakamura \& Meier}

\begin{document}

\title{Poynting Flux Dominated Jets in Decreasing Density Atmospheres.\\
       I. The Non-relativistic Current-driven Kink Instability\\
       and the Formation of ``Wiggled'' Structures}

\author{Masanori Nakamura\altaffilmark{1} and David L. Meier}
\affil{Jet Propulsion Laboratory\\ California Institute of Technology\\
       Pasadena, CA 91109}
\email{Masanori.Nakamura@jpl.nasa.gov}
\altaffiltext{1}{NRC Research Fellow}

\begin{abstract}
Non-relativistic three-dimensional magnetohydrodynamical (MHD) simulations of 
Poynting flux dominated (PFD) jets are presented. 
Our study focuses on the propagation of strongly magnetized hypersonic, 
but sub-Alfv\'enic ($C^{2}_{\rm s} \ll V^{2}_{\rm jet} < V^{2}_{\rm A}$) flow  
and on the subsequent development of a current-driven (CD) kink instability.  
This instability may be responsible for the ``wiggled'' structures seen 
in sub-parsec scale (VLBI) jets. 
In the present paper, we investigate the nonlinear behavior of PFD jets 
in a variety of external ambient magnetized gas distributions,  
including those with density, pressure, and temperature gradients.
Our numerical results show that the jets can develop CD distortions 
in the trans-Alfv\'enic flow case, even when the flow itself is still 
strongly magnetically dominated.
An internal non-axisymmetric body mode grows on time scales of order of 
the Alfv\'en crossing time and distorts the structure and magnetic 
configuration of the jet. 
The kink ($m=1$) mode of the CD instability, driven  by the radial component 
of the Lorentz force, grows faster than other higher order modes ($m>1$).  
In the jet frame the mode grows locally and expands radially at each axial 
position where the jet is unstable: the instability, therefore, 
does not propagate as a wave along the jet length. 
CD instabilities have a number of features that make them an attractive 
explanation for the helical jet structure observed in AGN and pulsars:  
1) because the magnetic field 
remains strong, CD instabilities do not develop into full MHD turbulence; 
2) the helical structures saturate and advect with the bulk flow;  
3) they distort the body of the jet, not merely its interface with the ambient medium; 
4) local plasma flow, then, follows a helical path along the kinked 
magnetic field backbone.
A naturally-occurring, external helically magnetized wind, which is (quasi-) axially
current-free, surrounds the well-collimated current-carrying jet and
reduces velocity shear between the jet and external medium.  This stabilizes
the growth of MHD Kelvin-Helmholtz surface modes in the inner jet flow.
\end{abstract}

\keywords{Instabilities---galaxies: active---galaxies: jets---
methods: numerical---magnetohydrodynamics (MHD)}

\section{INTRODUCTION}

Magnetohydrodynamic (MHD) acceleration mechanisms often are invoked 
as a model for the launching and initial acceleration and collimation 
of winds and jet outflows from Young Stellar Objects (YSOs), 
X-ray binaries (XRBs), Active Galactic Nuclei (AGNs), Microquasars, and 
Quasars (QSOs) \citep[see, {\it e.g.},][and references therein]{MEI01}. 
There has been a growing recognition in recent years, however, that 
the influence of strong magnetic fields within the jet may extend well 
beyond the central engine into the region where the jet freely propagates, 
influenced only by internal and ambient hydro- and magnetohydrodynamic 
forces.  This is particularly evident in observations of jets in AGNs, 
QSOs, winds from pulsars, and $\gamma$-ray burst sources (GRBs)
\citep[{\it e.g.},][]{PER84, CON93, HES02, COB03}.

Strongly magnetized jets, particularly those with a strong 
toroidal field encircling the collimated flow, are often referred to as 
``current-carrying'' or ``Poynting flux dominated'' (PFD) jets.  
A large current flowing parallel to the jet flow is responsible for 
generating a strong, tightly-wound helical magnetic field.  
Continued rotation of the entire magnetized plasma about the jet axis also plays 
an important role in jet dynamics.  
Rotation of the helical field drives a ``barber pole''-like, torsional 
Alfv\'en wave (TAW) forward in the direction of the jet flow, 
carrying electromagnetic energy and further accelerating the plasma.  
In a PFD jet, the Poynting flux energy carried by this TAW can greatly exceed 
the kinetic energy flux in the hydrodynamic (HD) part of the flow.  

It is well-known that a cylindrical plasma column with a helical magnetic
configuration is subject to MHD instabilities.  These are usually divided
into pressure-driven (PD), Kelvin-Helmholtz (KH), and current-driven
(CD) instabilities \citep[see {\it e.g.},][for details]{KAD66, BAT80,
FRE82}.  PD instabilities are related to the interplay between gas
pressure and curvature of magnetic field lines \citep[]{KER00, LON03},
and have not been considered to be very important for a supersonic jet.
KH instabilities occur because of the presence of velocity gradients in
the flow \citep[see {\it e.g.},][for details]{LAN59, CHA61}.  They may
play an important role at the shearing boundary between the flowing jet
and external medium, particularly in kinetic flux dominated (KFD) jets
where the hydrodynamical forces dominate over magnetic.  PFD jets, on the
other hand, should be especially susceptible to CD instabilities because
of the presence of the strong axial electric current.  The investigation
into the destructive influence of CD instabilities on jet flow, therefore,
recently has become an important avenue of research in the study of
astrophysical jets.

The purpose of the present paper is an in-depth, 3-D MHD, non-relativistic
numerical investigation of the nonlinear development of CD instabilities
in PFD jets --- particularly the CD kink ($m=1$) mode.  Our jets are
endowed with most of the properties now thought to characterize those
in AGN, QSOs, pulsars, and GRBs:  strongly magnetized, hyper-sonic (but
sub-Alfv\'enic; $C_{\rm s}^2 \ll V_{\rm j}^2 < V_{\rm A}^2$) flow, driven
by a Poynting flux dominated, torsional Alfv\'en wave that carries most
of the energy.  In a previous paper \citep[]{NAK01}, the basic behavior
of TAWs and PFD jets was studied under simplified atmospheric conditions.
We now assume more realistic atmospheric situations, including density,
pressure, magnetic field, and temperature gradients in the ambient medium. 
While these jets are strictly
``{\em propagating}'' (we do not consider the central engine itself), it is
important to realize that the inclusion of toroidal magnetic fields and
rotation can create a local acceleration and collimation process within
the jet that can increase the speed of the jet downstream up to and beyond
the Alfv\'en speed.  {\em Unlike previous studies of propagating jets,
therefore, the downstream properties of the jet will be determined not
only by the flux of kinetic energy injected at the jet throat ({\it
i.e.}, the initial jet speed and Mach number), but also by the amount
of Poynting flux injected in the TAW itself.}

In section 2, we give a review of observational and theoretical work on
the subject, relating previous results to the current work.  Section 3
outlines our basic numerical methods and model.  Section 4 gives a
comprehensive report and discussion on our results. 
Our conclusions are summarized in section 5.

\section{MAGNETOHYDRODYNAMIC {\it vs.} HYDRODYNAMIC JETS}

\subsection{Observational Evidence for Strongly Magnetized Jets}

There are several observational results which indicate the presence
of strong magnetic fields in astrophysical jets. The first class of
observations concerns the structure of the magnetic field.
In the case of AGN jets, this is determined from radio polarization of
the synchrotron emission, especially the Rotation Measure (RM) and the
projected magnetic field vectors \citep[{\it e.g.},][]{PER84, OWE89,
PERLM99, FERE99, EIL02, KRA04}.  \citet[]{ASA02} showed that the RM
distribution for the quasar 3C 273 jet on parsec scales has a systematic
gradient across the jet, and the projected magnetic field vector is
systematically tilted from the direction of the central axis of jet.
They conclude that the sign reversal of the RM across jet indicates the
presence of a toroidal (azimuthal) component of the field inside jet.
In the case of $\gamma$-ray burst (GRB) jets, \citet[]{COB03} observed
linear polarization in the prompt $\gamma$-ray emission from GRB021206
and found a value of $80 \pm 20\%$.  This is the theoretical maximum
possible polarization for a magnetized plasma, indicating that the field
is so strong that the entire $\gamma$-ray-emitting region is organized
around the field structure. The flow clearly is magnetically dominated,
and the authors further suggest that the GRB explosion itself is powered
by magnetic fields.

A second class of observations concerns the morphological structure of
the jets themselves.  High resolution VLBI observations show that many AGN
and quasar jets display wiggles or kinks on sub-parsec to parsec
scales \citep[{\it e.g.},][]{KRI90, KRI92, HUM92, CON93, ROO93, JON96,
KRI98, MAN99, MUR99, HUT01, LOB01, STI03}.  Such a helical distortion
might be caused either by MHD instabilities, or by precession
of the jet ejection axis due to the existence of a binary Black Hole
(BBH) \citep[]{BEG80}, or by an encounter with another galactic core.
On the other hand, the precession and galaxy interaction models tend
to operate on rather long ($10^{8}$ yr) time scales, while the kinks
appear to occur on significantly shorter ($\ll 10^{7}$ yr) time scales
\citep[{\it e.g.},][]{POT80}.  Based on both theoretical considerations
and observational results, therefore, MHD instabilities appear to
be the most plausible model for the observed wiggled or kinked
structures.

The third indicator of possible dynamical importance of magnetic fields
is the presence of thermal overpressures in jets.  There are several
such cases in which the jet thermal pressure significantly exceeds
the surrounding X-ray gas pressure \citep[see {\it e.g.},][]{POT80,
OWE89, BIR93}.  The required self-confinement of this pressure is easily
understood in terms of the magnetically-dominated jet model.  The toroidal
field component $B_{\phi}$ provides collimation of the
MHD jet and confinement of high pressure gas via the ``hoop-stress''
$(-B_{\phi}^{2}/4\pi r)$, which is a part of the magnetic tension force
[$({\bolB} \cdot \nabla) {\bolB}$].

\subsection{Basic Structure of a Current-carrying Jet}

The global picture of a current-carrying jet with a closed current system 
linking magnetosphere and hot spots, was introduced by \citet[]{BEN78} and applied 
to AGN double radio sources.
A closed current system includes a pair of current circuits, each containing both a forward 
electric current path (the jet flow itself,  with its toroidal magnetic field, toward the lobe), 
and a return electric current path (along some path back to the AGN core). 
The stability of current-carrying jets for two types of return current distributions 
have been discussed:  
(1) the returning currents are assumed to flow within the jet itself 
(the jets are thermally confined by external medium) \citep[]{CHI89}, 
and, 
(2) the returning currents are assumed to flow around the jet, such as in a magnetized 
cocoon \citep[]{BEN78}.

\subsection{Stability of HD Jets}

Most of the early work on jet stability concentrated on the purely hydrodynamical (HD) 
KH stability for simple configurations such as ``top-hat'' 
velocity profiles \citep[]{GIL65, HAR79, HAR83, PAY85, HAR88, ZHA92a, HAR97b} and other flows 
\citep[see {\it e.g.},][and references therein]{BIR91, FERR98}.
Two types of KH waves are considered disruptive in jets: the fundamental surface wave 
and the reflected body wave \citep[]{BOD82, ZHA92a}.
Surface mode waves are excited in the presence of non-zero velocity gradients and/or 
discontinuities at the boundaries between the body of jets and external medium.
Body waves propagate through the body of a fluid or only exist in the interior 
of the medium, such as acoustic/magnetosonic waves.
The existence of body waves does not depend on the presence of boundaries, but 
the reflection and/or refraction of body waves by such boundaries leads to the 
growth of modes that could eventually cause the disruption of jets 
\citep[]{ZHA92a, HAR97a}.

Beginning with the pioneering work by M. L. Norman and his co-investigators \citep[]{NOR82}, 
numerical simulations have been performed to investigate the nonlinear development of 
KH instabilities for propagating jets 
\citep[]{KOS88, CIO92, MAS96, CAR02a, CAR02b, KRA03}. 
\citet[]{NOR82} performed the axisymmetric 2-D HD simulations and found that the jet is 
decelerated by a Mach-disc shock wave front that is, in general, much stronger than the bow shock 
of the jet. 
Backflow from the working surface builds an extensive cocoon or lobe surrounding the jet.
Hypersonic jet flow itself is largely stable to KH instabilities, which grow slowly 
compared to the jet propagation speed.  The transonic 
lobes and cocoons, however, are generally KH {\em un}stable and mix with the external gas 
at a ragged boundary. 

Another type of numerical approach investigates the nonlinear development of KH instabilities 
in pre-formed ``{\em equilibrium}'' jets which begin in radial force equilibrium.  
A thermally confined top-hat velocity profiled jet, 
surrounded by a uniform unmagnetized external medium, was investigated with 
both 2-D \citep[]{NOR88, ZHA92b, BOD94, BOD95, STO97} 
and 3-D simulations \citep[]{HAR92a, BOD98, XU00, MIC00}.
These jets are perturbed at the surface layer between jet flow and external medium, 
and axisymmetric and non-axisymmetric surface and body modes of KH instabilities 
sometimes grow under a variety of conditions.

\subsection{Stability of MHD Jets}

\subsubsection{Kelvin-Helmholtz instabilities}

The MHD KH instability of jets has been examined using linear stability analysis 
\citep[]{RAY81, FERR81, COH83, FIE84, BOD89, HAR92b, BOD96}.
In general, surface non-axisymmetric modes ($m > 0$) are stable
against MHD KH instability during sub-Alfv\'enic flow.  However, in 
super-Alfv\'enic but trans-fast magnetosonic flow, they can be unstable.
The surface symmetric mode ($m = 0$) is predicted to be MHD KH unstable in 
sub-Alfv\'enic and super-Alfv\'enic flow, although with a relatively small growth rate 
\citep[]{BOD89, HAR92b, HAR97a}.
However, body mode waves can become important and affect the jet interior in 
the following situations: (1) if jet velocity exceeds the super-fast magnetosonic 
speed ($V_{\rm FM} < V_{\rm j}$) or (2) if the flow velocity is slightly below the 
slow magnetosonic speed [$C_{\rm s} V_{\rm A}/(C^{2}_{\rm s}+V^{2}_{\rm A})^{1/2} 
< V_{\rm j} < V_{\rm SM}$] \citep[]{HAR99}.

The growth of KH instabilities of super-fast magnetosonic jets containing 
force-free helical magnetic field has been shown to be reduced by the presence
of a toroidal magnetic field \citep[]{APP92}.
Similar investigations extended to the sub-fast magnetosonic regime \citep[]{APP96} showed
that the toroidal field exhibits a {\em de}stabilizing behavior at small velocities.
At least for force-free helical magnetic configurations in linear regime, it appears that 
the KH mode exhibits faster growth than CD mode \citep[]{APP96}.  
\citet[]{BOD96} investigated the axially magnetized rotating super-fast magnetosonic jets, 
and they found that these jets could be stabilized against the non-axisymmetric surface mode 
by jet rotation.  If there exists an external wind between the jet and the ambient medium, 
these results are modified considerably.  

\citet[]{CLA86}, \citet[]{LIN89}, and \citet[]{KRA01} performed axisymmetric 2-D simulations of 
propagating MHD jets  
with a purely toroidal magnetic field  and an unmagnetized external medium. 
\citet[]{KOS90} investigated the MHD jets with a helical (poloidal + toroidal) 
magnetic field (with the external medium poloidally magnetized) 
using axisymmetric 2-D simulations.
The results of these investigations showed that a strong toroidal magnetic field 
makes the backflow weak in the cocoon and, consequently, reduces the growth rate of the 
KH instabilities. 
\citet[]{TOD92} also performed axisymmetric 2-D simulations of MHD jets assuming a 
force-free helical magnetized external medium with the injection of super-Alfv\'enic flows into 
the computational domain.  
In their investigation, the KH instability is entirely suppressed if the the magnetic field 
is parallel to the velocity shear between cocoon and shroud and if the following criterion 
is satisfied \citep[{\it e.g.},][]{CHA61}: 
$\rho_{\rm 1} \rho_{\rm 2} (V_{\rm 1}-V_{\rm 2})^{2} 
\leq (\rho_{\rm 1}-\rho_{\rm 2})B^{2}/8\pi$. 
Here $\rho$, $V$, and $B$ are the density, velocity, and magnetic field strength 
parallel to the velocity, respectively, with subscripts denoting the two different fluids. 
For the present case, subscript 1 corresponds to the cocoon and 2 represents the shroud.
\citet[]{TOD92} derived a qualitative tendency in which the dense external medium 
($\rho_{1}/\rho_{2} \la 0.1$) promotes the KH instability, whereas the large scale magnetic field 
suppresses it.

Recently, using 3-D simulations, \citet[]{OUY03} investigated the ejection and 
propagation of a MHD jet from a pseudo-Keplerian disk 
({\it i.e.}, the disk is treated as a fixed boundary condition which has a Keplerian-like azimuthal 
velocity profile, but does not allow accretion) 
\citep[see {\it e.g.},][]{BEL95, UST95, MEI97, OUY97a, OUY97b, OUY99}.
Their results showed that the MHD jet beyond the Alfv\'en surface temporarily becomes 
unstable to the non-axisymmetric ($m > 0$) KH instability. 
However, the jet maintains long-term stability via a self-limiting process 
that keeps the average Alfv\'en Mach number within the jet to of order unity.
This occurs because the poloidal magnetic field becomes concentrated along the central axis 
of the jet, forming a ``backbone'' in which the Alfv\'en speed is high enough 
to reduce the average jet Alfv\'enic Mach number to unity.

Of particular interest are the axially magnetized 2-D slab equilibrium jets 
\citep[]{HAR92b, HAR95} and the helically magnetized 3-D equilibrium cylindrical jets 
investigated by P. E. Hardee and his collaborators \citep[]{HAR97a}.  
In the super-Alfv\'enic regime, if the jet is {\em also} super-fast magnetosonic, then it 
becomes more stable with increasing fast magnetosonic Mach number ($M_{\rm FM}$), and 
the destabilization length ($L$) varies approximately proportional to $M_{\rm FM}$ 
($L \propto M_{\rm FM} R_{\rm j}$, where $R_{\rm j}$ is the jet radius).
\citet[]{HAR99} investigated helically magnetized 3-D trans-Alfv\'enic ``light'' jets 
(density ratio $\rho_{\rm j}/\rho_{\rm e} \sim 0.03$, 
where subscript ``j'' corresponds to the jet itself and ``e'' corresponds to the external medium). 
These experienced considerable slowing as denser material was entrained following 
destabilization; provided the jet is super-Alfv\'enic, their KH growth rates also 
increase as $M_{\rm FM}$ decreases.  
However, the jets are nearly completely stabilized to these instabilities 
when the jet is sub-Alfv\'enic.

These numerical results show that the super-Alfv\'enic but trans-fast magnetosonic flow region 
is a potential zone of enhanced KH instabilities just downstream of the Alfv\'en point.  
They also show that magnetic tension can significantly modify the development of 
the KH instability in nonlinear regime.
An increase of density ratio would stabilized the MHD jets beyond the Alfv\'en point and, in general, 
denser jets have been found to be more robust than their less dense counterparts 
\citep[]{ROS99, ROS00}.
\citet[]{HAR02} confirmed the stabilizing influence of a surrounding magnetized wind against 
the non-axisymmetric KH surface modes of the helically magnetized 3-D trans-Alfv\'enic jets.  
They concluded that the jets could be stabilized entirely to the non-axisymmetric 
KH surface modes if the velocity shear, $\Delta V \equiv V_{\rm j} -V_{e}$ is 
less than a ``surface'' Alfv\'en speed, $V_{\rm As} \equiv [(\rho_{\rm j}+\rho_{\rm e})
(B^{2}_{\rm j}B^{2}_{\rm e})/(4 \pi \rho_{\rm j} \rho_{\rm e})]^{1/2}$. 

\subsubsection{Current-driven instabilities}

Prior to this point in time, much less consideration has been given to CD instabilities 
than to KH instabilities in the investigation of the disruption of the interior of 
astrophysical jets.  
This has been because, until recently, jets were believed to be in super-Alfv\'enic or 
super-fast magnetosonic flow ({\it i.e.}, kinetic energy was expected to exceed magnetic energy).  
Some analytic studies on CD instabilities as a possible explanation for jet disruption had been done 
\citep[]{EIC93, SPR97, BEG98, LYU99}.
\citet[]{APP00} also analyzed, for a large range of magnetic pitch ($r B_{z}/B_{\phi}$), the 
CD linear growth rate for configurations with a force-free helical magnetic field and a constant 
distribution of density and velocity.  (These assumptions exclude PD and KH instabilities.) 
They concluded that the properties of the fastest growing CD kink ($m=1$) mode are nearly 
independent of the details of the pitch profile.  However, they also concluded that this was
an internal mode that does not cause a significant distortion of jet.

\citet[]{LER00a} studied the nonlinear development of CD instability for 
cold super-fast magnetosonic equilibrium jets based on their linear analysis \citep[]{APP00}.
It was found that the current density is redistributed within the inner part of the jet radius 
on a characteristic time scale of order the Alfv\'en crossing time in the jet frame. 
Nothing in their numerical results indicated a possible disruption of jet 
by the CD sausage ($m=0$) or kink ($m=1$) mode.
However, for numerical reasons, their simulations were limited to early nonlinear phases; 
a full investigation of the nonlinear development of the instability was not carried out.

\citet[]{TOD93} performed 3-D MHD simulations of a model for HH objects in which 
super-Alfv\'enic propagating jets are injected into a pre-formed force-free 
helically magnetized ambient medium.  
They found that YSO jets can be disrupted into a large scale wiggled structure 
by the CD kink instability.
In a similar study, \citet[]{NAK01} performed 3-D simulations of propagating MHD jets 
to investigate the formation of wiggled structures in AGN jets. 
They found that the propagation of nonlinear torsional Alfv\'en waves (TAWs) can produce 
a slender jet shape by the ``sweeping magnetic twist'' mechanism of Uchida \& Shibata 
\citep[]{SHI85, SHI86, UCH85, UCH86}.
In addition, wiggles in the jet can be produced by the CD kink instability though 
the interaction between the TAW and the ambient medium.
Recently, \citet[]{BAT02} confirmed the interaction of KH and CD instabilities 
in helically magnetized super-fast magnetosonic equilibrium jets 
by performing 3-D MHD simulations.
This nonlinear interaction can contribute to jet survival, and the large-scale 
magnetic deformations associated with CD mode development can effectively saturate 
KH surface vortices and prevent jet disruption.

\subsection{Theoretical Arguments for Poynting Flux Dominated Jets}

There has been a growing recognition in recent years that Poynting flux dominated (PFD) flow 
plays an important role in the jets in AGNs, QSOs, winds from pulsars, and 
possibly $\gamma$-ray burst sources.  
This is in strong contrast to the kinetic energy flux dominated (KFD) jets that 
had been expected in these sources, and still are believed to be relevant for YSO jets.
The energy carrier of PFD jets is primarily the electromagnetic field, while in KFD jets it is 
the kinetic flux.

The concept of PFD jets is based on the theory of magnetically driven outflow, 
proposed (in the electromagnetic regime) by \citet[]{BLA76} and \citet[]{LOV76} and subsequently 
applied to rotating black holes \citep[]{BLA77} and to magnetized accretion disks \citep[]{BLA82}.
By definition, these outflows initially are dominated by electromagnetic forces close to the central engine.  
In these and subsequent models of magnetically driven outflows (jets/winds),
the plasma velocity passes successively through the hydrodynamic (HD) sonic, slow-magnetosonic,
Alfv\'enic, and fast-magnetosonic critical points (surfaces).
In very strongly magnetized flows ($C_{\rm s} \ll V_{\rm A}$), the hydrodynamic and slow-magnetosonic
points almost coincide, and the Alfv\'en and fast points also occur close to each other.  The
distance between these two clusters of critical points (HD/slow and Alfv\'en/fast),
and therefore the nature of much of the outflow, depends on the relative dominance of the
advected magnetic field, {\it i.e.}, on how long the high level of Poynting flux domination 
can be maintained as the flow propagates.  

In many early models of steady, axisymmetric MHD outflow,
it was expected that the flow would smoothly pass the Alfv\'enic point (surface) at a distance
very close to the central object.  After that the flow would become super-Alfv\'enic and,
therefore, kinetic energy flux dominated.
However, according to recent theoretical studies, this picture is probably not correct.
Of particular interest here is the current-carrying core of the outflow near the central axis (the jet).
In steady non-relativistic solutions \citep[]{FEN96, KRA99} the innermost part of the outflow remains sub- to
trans-Alfv\'enic ($M_{\rm A} \lesssim 1$), {\it i.e.}, Poynting flux dominated.
Furthermore, in relativistic steady models of MHD outflows from disks, the flow can remain PFD for
many hundreds to thousands of outer disk radii --- tens of parsecs in AGN \citep[]{VK01, VK03a, VK03b, VK04}.
Even in dynamical simulations, a self-organized process occurs whereby the Alfv\'en speed increases
due to the concentration of poloidal magnetic flux ${\bolB}_{\rm p}$ along the central axis of the outflow
\citep[]{OUY03}.  This reduces $M_{\rm A}$ near the axis and forestalls the passage of the flow
through the Alfv\'en point.

So, it appears likely that the axial part of the outflow (the current-carrying jet) remains in
sub- to trans-Alfv\'enically PFD flow for a long distance. In the case of AGN jets this translates
into a few to several tens of parsecs.  {\em Jets observed by VLBI, therefore, may be Poynting flux dominated,
making our simulations herein directly applicable.}

Further theoretical development of PFD jets from magnetized accretion disks has been performed by 
\citet[]{BEG84}; \citet[]{LOV87}; \citet[]{LI92}; \citet[]{LYN96}; \citet[]{ROM97}; \citet[]{LEV98};
\citet[]{COL99}; \citet[]{LI01}; \citet[]{LOV02}; \citet[]{HEY03a, HEY03b, HEY03c}; 
\citet[]{LOV03}; and \citet[]{VLA03}.
Two-dimensional axisymmetric (``2.5-D'') non-relativistic MHD simulations of PFD jets have been 
performed for opening magnetic loops threading a Keplerian disk \citep{ROM98, UST00}.
And relativistic 2.5-D simulations of propagating PFD jets carrying a toroidal field component only 
were performed by \citet{KOM99} in order to compare their simulations with the 
non-relativistic simulations of \citet{LIN89}.
Finally, \citet[]{LI00}, \citet[]{TOM01}, and \citet[]{WAN04} have considered analytically the 
CD instability (the so-called ``screw instability'') of black hole magnetospheres. 

To our knowledge, however, no study has been made of the nonlinear behavior of PFD jets and 
the related CD instabilities in full 3-D dynamics.

\section{NUMERICAL METHODS}

\subsection{Basic Astrophysical Model}

In this paper we study the structure, dynamics, and stability of propagating MHD jets.  
Our simulations concentrate on the region in an AGN in which a collimated 
jet-like flow (at much greater than the escape velocity) has been established, but that 
flow still is dominated by magnetic forces and has not yet achieved a 
super-(fast magneto) sonic velocity. 
The energy carried by the jet, therefore, is dominated by Poynting flux rather than 
kinetic energy flux, and 
the Alfv\'en speed in the flow is much higher than the local sound speed, $C_{\rm s} \ll V_{\rm A}$. 

Numerous theoretical investigations of the central engine itself have shown that 
a rotating, magnetic structure can be created a few Schwarzschild radii ($10^{14-15}$ cm) 
from the black hole \citep[]{BLA77, BLA82, KOI02}.
While the physical connection between this central region and the sub-parsec region 
is still poorly understood, it is reasonable to suppose that both strong magnetic and 
rotational fields will be influential at the base of the sub-parsec flow.  
Our basic jet model, therefore, contains a plasma with a strong, rotating poloidal 
magnetic field at its base.  
Nonlinear TAWs propagate out along this field, creating and propagating a magnetically 
driven collimated outflow via the sweeping magnetic twist mechanism of \citet[]{UCH86}.

In contrast with the many numerical stabilities studies, in which the magnetized jets
are assumed to be confined thermally by a non-magnetized external ambient medium, 
we assume a large scale poloidal magnetic field in the ambient medium surrounding the jet.
The origin of such a galactic magnetic field is not yet is fully understood, 
but its existence is suggested by both synchrotron emission and Faraday rotation observations.
The magnetic field assumed here might either be a part of the primordial inter-stellar 
field \citep[]{KUL92} brought into the central part of the protogalaxy 
in the process, or the central part of the amplified field by a galactic turbulent dynamo 
process, which are argued by many authors 
\citep[for reviews, see][and references therein]{KRO94, HAN02}, or a field structure 
carried out from the central engine by a lower-velocity magnetized disk wind.

Models of the central parsec and sub-parsec regions of galaxies indicate that the ambient 
medium is characterized by strongly decreasing gradients in density and pressure and, 
presumably, magnetic field. 
We therefore are especially interested in the behavior of PFD jets in decreasing density atmospheres.  

\subsection{MHD Equations and Simulation Code}

In the present study we assume non-relativistic ideal MHD and neglect the effect of gravity.  
We solve the nonlinear system of time-dependent MHD equations numerically 
in 3-D Cartesian coordinate system ($x,\,y,\,z$):
\beqn
\label{eq:mass}
\frac{\partial \rho}{\partial t}+\nabla \cdot (\rho {\bolV})=0,\\
\label{eq:momentum}
\rho \left[\frac{\partial {\bolV}}{\partial t}+({\bolV} \cdot \nabla) 
{\bolV} \right]=-\nabla p + {\bolJ} \times {\bolB},\\
\label{eq:induction}
\frac{\partial {\bolB}}{\partial t}=\nabla \times ({\bolV}\times{\bolB}),\\
\label{eq:energy}
\frac{\partial p}{\partial t}+\nabla \cdot (p {\bolV})= -(\Gamma-1)p \nabla \cdot {\bolV}.
\eeqn
Here, $\rho$ is the mass density, $p$ the gas pressure, ${\bolV}$ the fluid velocity.
${\bolB}$ is the magnetic field and ${\bolJ}=
(\nabla \times {\bolB})/4\pi$ is the corresponding current density. 
$\Gamma$ is the ratio of specific heats (a value of $5/3$ is used).

We normalize all physical quantities with the unit length scale $L_{0}$, 
typical density $\rho_{0}$, typical velocity $V_{0}$ in the system, and their
combinations, {\it e.g.}, $\rho^{\prime}=\rho/\rho_{0}$, etc. The normalizing factors 
are shown in Table \ref{tbl:unit}.
A factor of $4 \pi$ has been absorbed into the scaling for both the magnetic field ${\bolB}$ and the current density ${\bolJ}$.
This normalization does not change the form of the basic equations (\ref{eq:mass}) -- (\ref{eq:energy}), and 
hereafter, we will use the normalized variables and will omit the primes on
physical quantities. 
(We will write the dimensional variables with subscript "0":  $\rho_{\rm 0}$, for example.)
The system of dimensionless equations is integrated in time by using a two-step 
Lax-Wendroff scheme \citep[]{RUB67} with artificial viscosity term \citep[]{LAP67}.
The original MHD code was developed by K. Shibata and extended by his co-workers 
\citep[{\it e.g.},][]{SHI83, SHI85, SHI86, MAT96}.
Parallelization of the code was done using MPI. 

The total computational domain is taken to be $|x| \leq x_{\rm max}$, $|y| \leq y_{\rm max}$, 
and $z_{\rm min} \leq z \leq z_{\rm max}$, where $x_{\rm max}, \ y_{\rm max} 
\simeq 16.0$, $z_{\rm min} \simeq -1.0$, and $z_{\rm max} \simeq 20.0$.
The numbers of grid points in the simulations reported here are
$N_{x}\times N_{y}\times N_{z}=261 \times 261 \times 729$, where the grid
points are distributed non-uniformly in the $x$, $y$, and $z$ directions.
The grid spacing in the initial propagation region near the jet axis is uniform, 
$\Delta x=\Delta y=\Delta z=0.025$ for 
$|x|, \ |y| \leq 1.5$ and $|z| \leq 16.0$ ($121 \times 121 \times 681$ cells are assigned here), 
and then stretched by 5\% per each grid step for the regions $|x|, \ |y| > 1.5$ and $|z| > 16.0$.
Forty uniform cells are distributed across the initial jet diameter in the transverse direction, 
$L \simeq 1$ at $z=0$.
Low-resolution exploratory 
computations were performed on the Jet Propulsion Laboratory Origin 2000 machines.  
High-resolution 3-D computations were performed on the FUJITSU VPP 5000/32R at the National 
Astronomical Observatory in Japan 
(9.6 GFLOP/s peak speed per node) requiring about 6 CPU hours each.  

\subsection{Initial Conditions: Non-uniform magnetized atmospheres}

In the non-uniform, magnetized atmosphere we adopt a current- (and therefore force-) free 
magnetic configuration (${\bolJ}=0$), 
by placing one pair of current loops on both the upper and lower $z$-boundaries 
of the computational domain. 
This constrains the magnetic field to have only a field $z$-component 
at $z=z_{\rm max}$, thereby avoiding numerical errors when physical MHD waves pass though the
upper ($z=z_{\rm max}$) boundary. 
Such a field configuration does not disturb the force equilibrium of the hydrostatically 
stable atmosphere because $\bolJ \times \bolB = 0$.
The field configuration chosen is explicitly given in the Appendix.

For our initial density distribution, we assume that $\rho$ varies as a power of 
the magnetic field strength 
\beqn
\label{eq:power_rho_B}
\rho \propto |{\bolB}|^{\alpha},
\eeqn
where $\alpha$ is a free parameter in this paper.
If $\alpha=2$, the Alfv\'en speed $V_{\rm A} (\equiv |{\bolB}|/\sqrt{\rho})$ will be 
constant throughout the computational domain. If $\alpha \neq 2$, the Alfv\'en speed
will decrease ($\alpha < 2$) or increase ($\alpha > 2$) with distance from origin (0, 0, 0).  
This power law model embraces several accretion and collapse models for the formation 
of the interstellar medium (ISM). 
\begin{itemize}
\item{For example, if the ISM had formed by the conservative spherical 
contraction of a protogalactic gas cloud, the magnetic field strength $|\bolB|$ 
would be amplified as $|\bolB| \propto \rho^{2/3}$ or $\alpha = 3/2$.  
Combining this value for $\alpha$ with our initial magnetic field distribution, 
the polar (axial) distribution of Alfv\'en velocity $V_{{\rm A}z} (z) (\equiv B_{z}/\sqrt{\rho})$ 
would gradually decrease as $V_{{\rm A}z}(z) \propto z^{-1/2}$ 
$[B_{z}(z) \propto z^{-2},~\rho (z) \propto z^{-3}]$.}
\item{In the case of interstellar magnetized clouds (isothermal gravitational contractions), 
$2 \leq \alpha \leq 3$ is theoretically \citep[]{MOU76}, numerically \citep[]{SCO80}, 
and observationally inferred \citep[]{CRU99} rather than $\alpha = 3/2$.
However, $\alpha$ varies widely with position $(r, z)$ in the contracting gas, and 
is smaller than 2 near the central ($z$) axis, because the field there is being 
``dragged along'' by the collapse in the $r$-direction.
This leads to a large amplification of the field with no corresponding large increase in
$\rho$; thus $\alpha$ becomes small \citep[]{MOU76, SCO80}.
$\alpha \sim 2$ $[\rho (r) \propto B_{z}(r)^{2}]$ on the equatorial plane and 
$\alpha \sim 1$ $[\rho (z) \propto B_{z}(z)]$ on the central axis of collapsing gas, also 
have been found with numerical simulations \citep[]{TOM96}.}
\item{A third possibility is that the ISM may have formed in an advection-dominated accretion 
flow \citep[]{NAR98}, 
where $\rho \propto r^{-3/2}$ and $|\bolB| \propto r^{-5/4}$, or $\alpha = 6/5$.} 
\end{itemize}

We therefore choose the following two representative cases:  
``{\em decreasing} $V_{\rm A}$'' ($\alpha = 1$) and 
``{\em constant}   $V_{\rm A}$'' ($\alpha = 2$) as the initial ambient medium. 

For the initial gas pressure distribution we make the polytropic assumption 
\beqn
\label{eq:polytropic}
p \propto \rho^{\Gamma},
\eeqn
where $\Gamma$ is the polytropic index (we use $\Gamma=5/3$ throughout this paper).
The sound speed $C_{s}$ ($\equiv \sqrt{\Gamma p/\rho} \propto T^{1/2}$, with T 
being the temperature) will decrease with distance from origin.
The atmosphere is artificially confined to prevent it from expanding under its own 
pressure gradient by imposing a pseudo-gravitational potential ($\tilde{\phi}$) 
designed to `hold on to' the atmosphere without significantly impeding the advancing jet 
\citep[]{CLA97}:
\beqn
\label{eq:pseudo_gravity}
\tilde{\phi} \equiv \frac{\tilde{p}}{\tilde{\rho}}.
\eeqn
Here, $\tilde{p}$ and $\tilde{\rho}$ are the initial gas pressure and density distributions.
By design, the quantity $\rho \tilde{\phi}$ exactly cancels the initial gas pressure gradients 
in the stratified atmosphere, $\nabla (p-\rho \tilde{\phi})=0$. Assuming a current-free 
(${\bolJ}=0$) field, the right-hand side of equation (\ref{eq:momentum}) exactly equals zero.

We now consider the ``plasma-$\beta$'' parameter:
\beqn
\label{eq:beta}
\beta \equiv \frac{2 p}{|{\bolB}|^{2}} =\frac{2 C_{\rm s}^{2}}{\Gamma V_{\rm A}^{2}}
\eeqn
as the ratio of the gas to the magnetic pressure.
In all the situations we investigate in this paper, 
the thermal energy density is much less than the magnetic energy density, so 
$\beta \sim C^{2}_{\rm s}/V^{2}_{\rm A}\ll 1$ in the whole computational domain.

\subsection{Boundary Conditions}

In the lower ``boundary zone'' [$z_{\rm min} \leq  z<0$, where $r=(x^2+y^2)^{1/2}$] 
we set the velocity field to be
\beqn
\label{eq:vel_jet}
{\bolV}_{\rm j}=V_{\phi} (r,\,z) \hat{\phi} +V_{z} (r,\,z) \hat{z}.
\eeqn
This zone is the injection region, in which the hypersonic but sub-Alfv\'enic flow
({\it i.e.}, the PFD jet) is formed. 
Equation (\ref{eq:vel_jet}) represents a cylindrical MHD jet, powered by a nonlinear 
TAW, entering the upper region ($z \geq 0$) of the computational domain.
In all our simulations the injection speed of the PFD jet in the boundary zone is 
sub-Alfv\'enic, but super-slow magnetosonic. 

To stabilize the numerical calculations in the injection region, we employed a
``damping zone'' to prevent unnecessary reflections and interactions from below.
The disturbances in all physical quantities except for the magnetic field are damped 
at each time steps as follows:
\beqn
\label{eq:damping}
Q(x,\,y,\,z,\,t)=\left\{1-f_{Q}(z)\right\}Q(x,\,y,\,z,\,t) \nonumber \\
+f_{Q}(z)Q(x,\,y,\,z,\,0), \\
f_{Q}(z) \equiv 
\frac 12\left\{\cos \left(\frac{z-z_{1}}
{z_{2}-z_{1}} \pi \right)+1\right\}.
\eeqn
The damping factor $f_{Q}$ for quantity $Q$ is defined such that 
it is equal to zero at the upper end of the damping zone, 
$z \geq z_{2}$, and to unity at the lower end, $z \leq z_{1}$. 
We set the values to $z_{1}=-0.975$ and $z_{2}=-0.025$, respectively.
We use about 40 axial cells in this boundary zone.
Note that we leave the magnetic field ${\bolB}(x,\,y,\,z,\,t)$ unmodified so that 
the induction equation (\ref{eq:induction}) still holds.

At the bottom of this zone ($z=z_{\rm min}$), we impose symmetry 
for $\rho$, $V_{x}$, $V_{y}$, $B_z$, and $p$, and antisymmetry 
for $V_{z}$, $B_{x}$, $B_{y}$.
The free time-dependent evolution of the flow physical variables occurs only in 
the upper part of the computational domain ($z \geq 0$).  

At the outer boundaries ($x=x_{\rm min}$, $x=x_{\rm max}$, $y=y_{\rm min}$, 
$y=y_{\rm max}$, and $z=z_{\rm max}$) 
we set free conditions for each quantity $Q$ as follows \citet[]{SHI83}: 
\beqn
\label{eq:free_boundary}
\frac{\partial \Delta Q}{\partial x}=
\frac{\partial \Delta Q}{\partial y}=
\frac{\partial \Delta Q}{\partial z}=0, \nonumber \\
\quad \Delta Q \equiv Q(x,\,y,\,z,\,t + \Delta t) - Q(x,\,y,\,z,\,t).
\eeqn

\section{RESULTS AND DISCUSSION}

In this section we provide a detailed description of the evolution of four different 
simulations of PFD jets.  
We will concentrate on the solutions in the ``jet-propagation'' region, 
$-2.25 \leq x,~ y \leq 2.25$, and $0.0 \leq z \leq 18.0$. 
An important unit time scale in these dimensionless systems is the Alfv\'en crossing time 
$\tau_{\rm A}$ ($\equiv L/V_{\rm A}$), normalized with $\tau_{\rm A 0} \equiv L_{0} / V_{\rm A0}$.

\subsection{The Four Initial Models}

We have carried out a number of simulations with varying physical conditions, evolving 
four distinct models:  shallow-atmosphere models (A) and steep-atmosphere
models (B), each with a highly and a mildly Poynting flux dominated flow case.
The values of their physical parameters listed in Table \ref{tbl:model} and discussed more 
fully below.

\begin{figure}
\begin{center}
\includegraphics[scale=1.0, angle=-90]{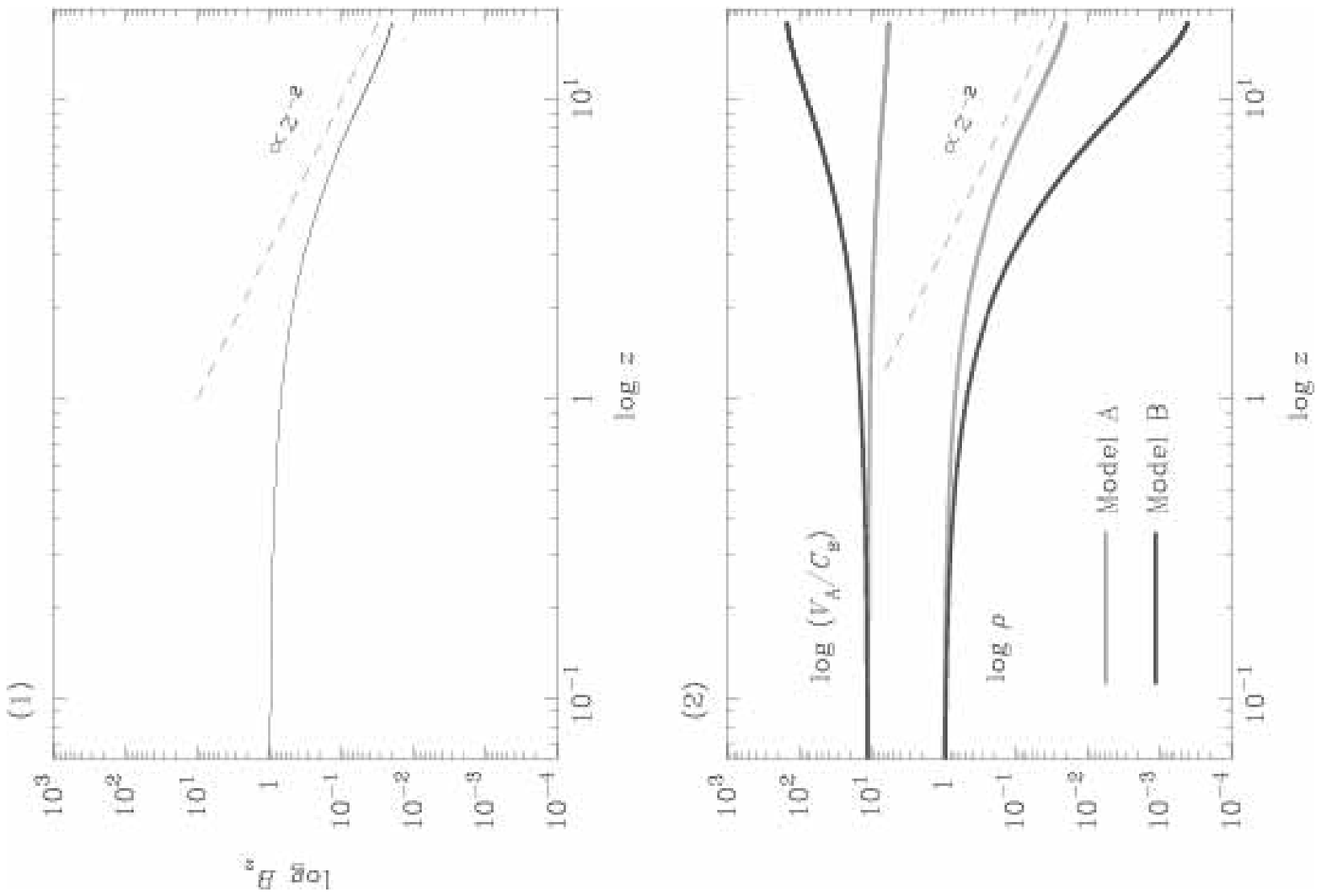}
\caption{\label{fig:init_zaxis}
Initial profiles along the $z$-axis of (1) the axial component of the magnetic field $B_{z}$ 
for all models and (2) density $\rho$ and ratio of Alfv\'en velocity to 
sound velocity $V_{\rm A}/C_{\rm s}$ in logarithmic scale for models A ({\em light gray solid
 line}) and B ({\em dark gray solid line}). 
The broken line indicates a $z^{-2}$ power law variation for comparison.
}
\end{center}
\end{figure}

The main purpose of this investigation is to study the effects of a 
decreasing density, magnetized ambient medium on the growth of current-driven 
instabilities in dynamically propagating PFD jets. 
Figure \ref{fig:init_zaxis} shows the initial ($t=0.0$) distribution of the physical quantities 
along the $z$-axis.
For the A models ($\alpha = 1$; {\em decreasing} $V_{\rm A}$), $\rho$ and $B_{z}$
decrease gradually along the $z$-axis and asymptote to $\sim z^{-2}$ for $z > 1$.
On the other hand, for the B models, $B_{z}$ decreases in the same manner as model A, 
but $\rho$ decreases faster than $\sim z^{-2}$ for large $z$ ($\alpha = 2$; {\em constant} $V_{\rm A}$).
We set the plasma beta ($\beta = 10^{-2}$) 
at the origin for all of the models.
The ratio $V_{\rm A} / C_{\rm s}$ decreases slightly along $z$-axis for model A 
but gradually increases for model B.

\begin{figure}
\begin{center}
\includegraphics[scale=1.0, angle=-90]{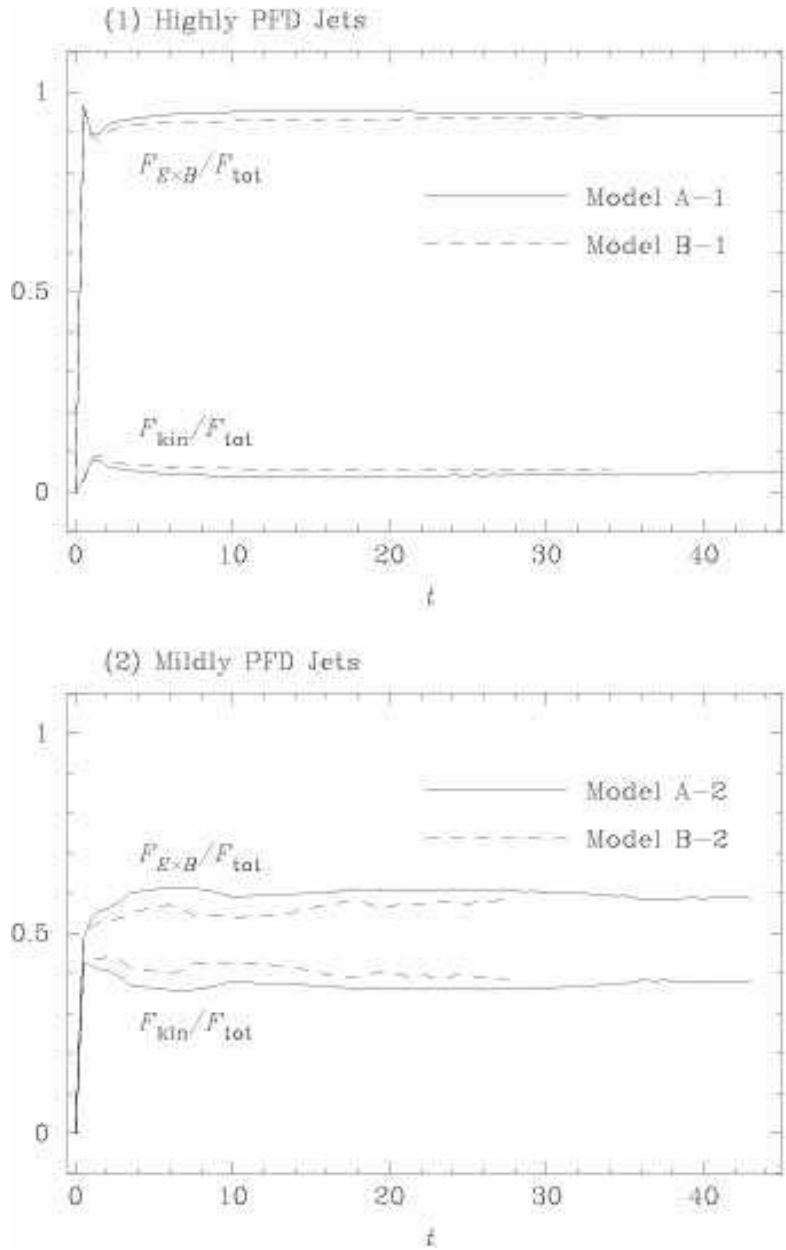}
\caption{\label{fig:inj_eflux}
Time variation of the energy fluxes injected though the $z=0$ plane.
Kinetic energy flux $F_{\rm kin}$ and Poynting flux $F_{E \times B}$ are 
normalized by the total energy flux $F_{\rm tot}= F_{\rm kin}+F_{E \times B}+F_{\rm th}$ 
(thermal energy flux).  Both Case 1 ({\em Highly PFD Jets}) and Case 2 ({\em Mildly PFD Jets})
are shown.
The solid line shows Model A ({\em slowly decreasing ambient density}) and 
the broken line Model B ({\em steeply decreasing ambient density}).
}
\end{center}
\end{figure}

We also consider two different types of flow in each atmosphere model, 
{\it i.e.} the ratio of 
Poynting flux $F_{E \times B}$ to the total energy flux $F_{\rm tot}$ in MHD jets:  
\begin{enumerate}
\item Highly PFD jets ($F_{E \times B}/F_{\rm tot} \sim 0.9$) (Case 1)
\item Mildly PFD jets ($F_{E \times B}/F_{\rm tot} \sim 0.6$) (Case 2)
\end{enumerate}
Figure \ref{fig:inj_eflux} shows the time variation of the ratio of the injected energy fluxes into 
the upper ``evolved region'' from the lower boundary zone.
Note that the quasi-stationary PFD flow is injected throughout the time 
evolution of the system.
The net amount of injected {\em Poynting flux} for the mildly PFD jets is 
almost equal to the case of the highly PFD jets.  However, the amount of {\em kinetic 
energy flux} injected in Case 2 is considerably larger than in Case 1. 

\subsection{Comparative Overview of Simulation Evolution}

Before considering our numerical results in detail, it is 
instructive to give a brief comparative overview of the time development of
the jet systems in the four models.
This can be done by inspecting the logarithmic density $\rho$ and the velocity field
($V_{x}, V_{z}$) normalized by $V_{\rm A0}$. 
Figures \ref{fig:A1_dens} -- \ref{fig:B2_dens} display these quantities at various times in 
the simulation evolution using two-dimensional $x-z$ slices at $y=0$.  
Note that these slices contain the jet ($z$) axis.  

\subsubsection{Early jet evolution: propagation of the nonlinear MHD waves}

\begin{figure}
\begin{center}
\includegraphics[scale=0.89, angle=0]{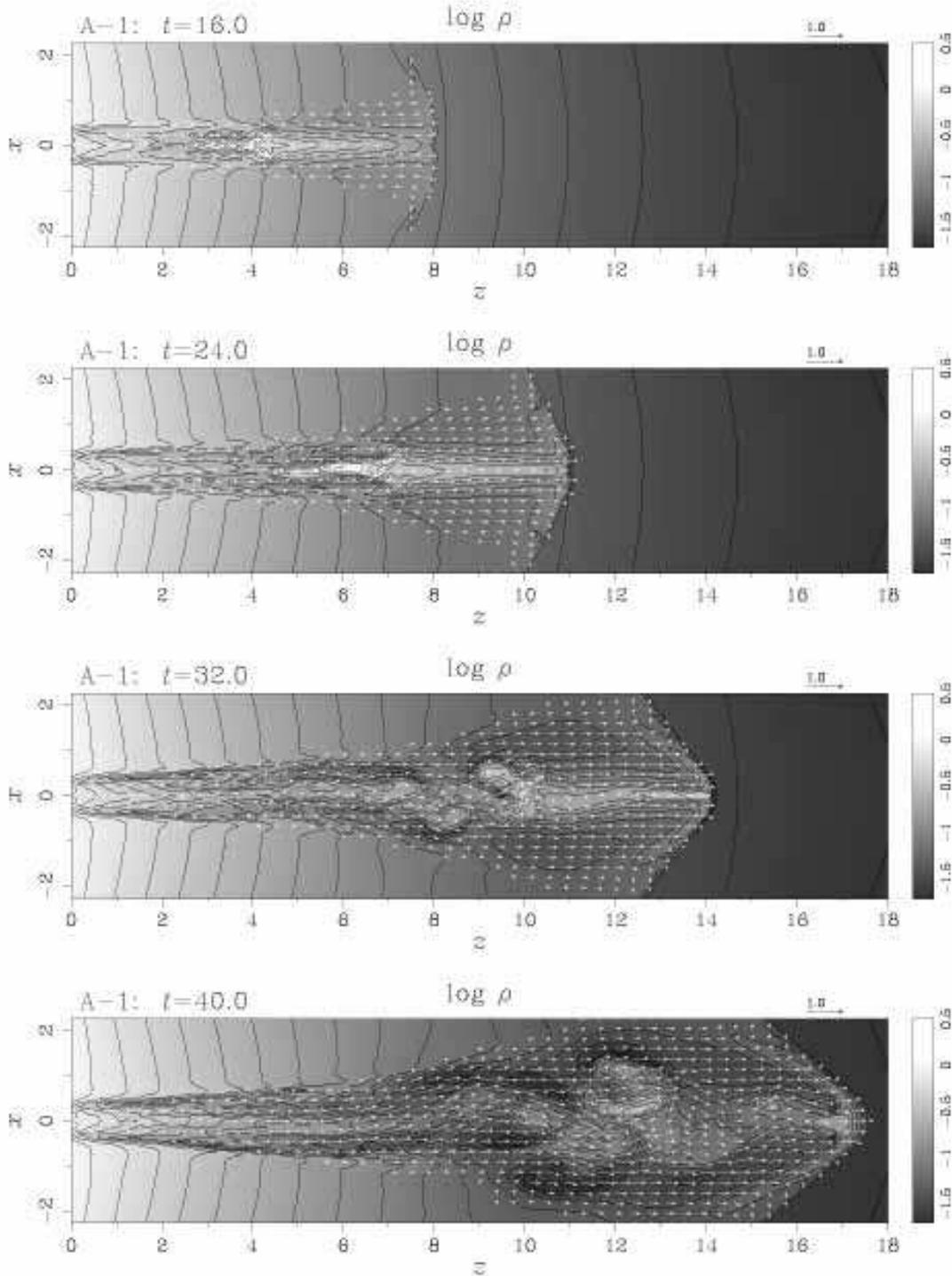}
\caption{\label{fig:A1_dens}
Time evolution of the region ($|x| \leq 2.25,~0.0 \leq z \leq 18.0$) for Model A -- 1.
Contour images of the density distribution (logarithmic scale) are shown along with the 
poloidal velocity field in the $x-z$ plane.  Model times are given at upper left in each 
of the four panels.  The length of the arrow at upper right in each panel shows the unit 
velocity.  
}
\end{center}
\end{figure}

\begin{figure}
\begin{center}
\includegraphics[scale=0.89, angle=0]{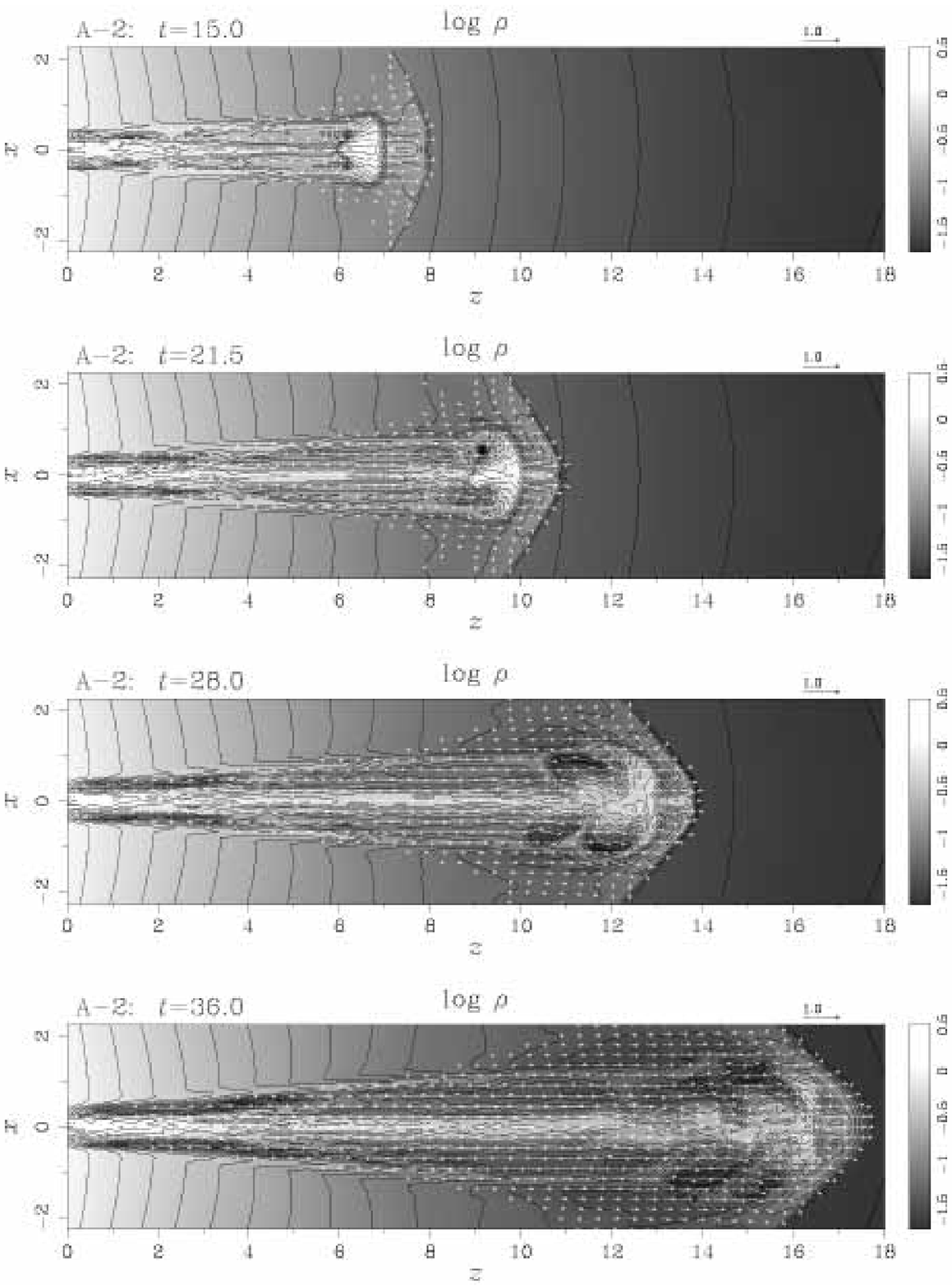}
\caption{ \label{fig:A2_dens}
Similar to Fig. \ref{fig:A1_dens}, but for Model A -- 2.
}
\end{center}
\end{figure}

At early times in the evolution ({\em Top} part of each Figs. \ref{fig:A1_dens}--\ref{fig:B2_dens}), 
the well-collimated 
helically PFD jets powered by the TAWs advance into the poloidally magnetized ambient medium.
Figure \ref{fig:V_axial_before} displays for all models the axial distributions of the Alfv\'en and
sound speeds, as well as each component of the jet velocity, in the $x-z$ plane close to the jet axis. 
The propagation of three distinct fronts of compressive MHD waves 
is clearly visible for all models: a fast-mode MHD wave (at the head of the PFD jet or TAW)
and a pair of slow-mode MHD waves further upstream.  
We characterize these waves as propagating in both the forward ({\sf F}) and reverse ({\sf R}) 
directions relative to a reference frame that co-moves with the jet, and give them the names 
forward fast-mode ({\sf F--F}), forward slow-mode ({\sf F--S}), and reverse slow-mode ({\sf R--S}) waves.
In addition, a contact discontinuity travels in this co-moving frame between the two slow-mode waves.  
Because the jet is injected at a super-slow magnetosonic velocity 
in the boundary zone, the forward and backward (reverse) propagating slow-mode compressive waves 
seen in each of the models quickly become slow-mode shock waves early in the simulation.  

In models A -- 1 and A -- 2, due to a gradual decreasing ambient $V_{\rm A}$, the {\sf F--F} 
compressive wave front decelerates and its wave amplitude gradually increases as it propagates forward.  
Through this nonlinear process, this front steepens into a {\em fast-mode MHD bow shock} 
when the phase velocity becomes super-fast magnetosonic (see also the 
{\em 2nd} panels in Fig. \ref{fig:A1_dens} and \ref{fig:A2_dens}).

On the other hand, only a very low amplitude {\sf F--F} compressive wave front can be (barely) seen in models 
B -- 1 and B -- 2 in Figure \ref{fig:V_axial_before}.
This wave front never reaches a super-fast magnetosonic velocity during our simulations of models B.
The velocity amplitude associated with the {\sf F--F} compressive wave becomes much weaker
as the ratio $V_{\rm A}/C_{s}$ in the ambient medium [$\equiv (2/\Gamma \beta)^{1/2}$] 
along $z$-axis increases.  
The phase velocity of a fast-mode compressive MHD wave is $V_{\rm ph} = V_{\rm FM} \sim V_{\rm A}$ 
when $C_{\rm s}^{2} \ll V_{\rm A}^{2}$. 
Because the initial distribution of the Alfv\'en velocity in Models B -- 1 and B -- 2 is 
constant with $z$ ($V_{\rm A} = 1.0$), the {\sf F--F} compressive wave will propagate quickly compared 
with models A -- 1 and A -- 2.  In the B models the wave reaches $z=18.0$ at a time 
$t \sim 18.0$.

Several features of PFD MHD jet flow have counterparts in the well-known supersonic hydrodynamic 
jet flow. The {\sf F--F} compressive/shock wave, of course, propagates through the external medium at a 
speed faster than the other two shock waves, playing the same role as the bow shock does in HD 
jets.  The {\sf F--S} shock wave propagates at a slower speed, further compressing and heating the 
ambient medium.  This second shock wave has no direct counterpart in HD jets, although it acts 
similarly to the bow shock, especially when the {\sf F--F} front is a compressive wave.  
The reverse ({\sf R--S}) shock wave plays the same role as the Mach disk does in HD jets, decelerating 
and heating the magnetized jet flow itself.  From the point of view of the jet material, the 
flow can reach the {\sf R--S} shock wave and enter the compressed region, but it can never 
reach the {\sf F--S} shock wave.  Similarly, the external ambient material crossing the 
{\sf F--S} shock wave also reaches this heated region, but can never reach the {\sf R--S} 
shock wave.  Instead, ambient material is entrained between the two slow-mode shock waves, along with 
the accumulated jet material.  The contact discontinuity between the two accumulations, 
therefore, plays the same role as the contact discontinuity in HD jets.  We often 
identify this interface as defining the rest frame of the jet flow.  

\subsubsection{Intermediate jet evolution: energy conversion though the MHD shocks}
\label{sect:energy_conv}

\begin{figure}
\begin{center}
\includegraphics[scale=0.89, angle=0]{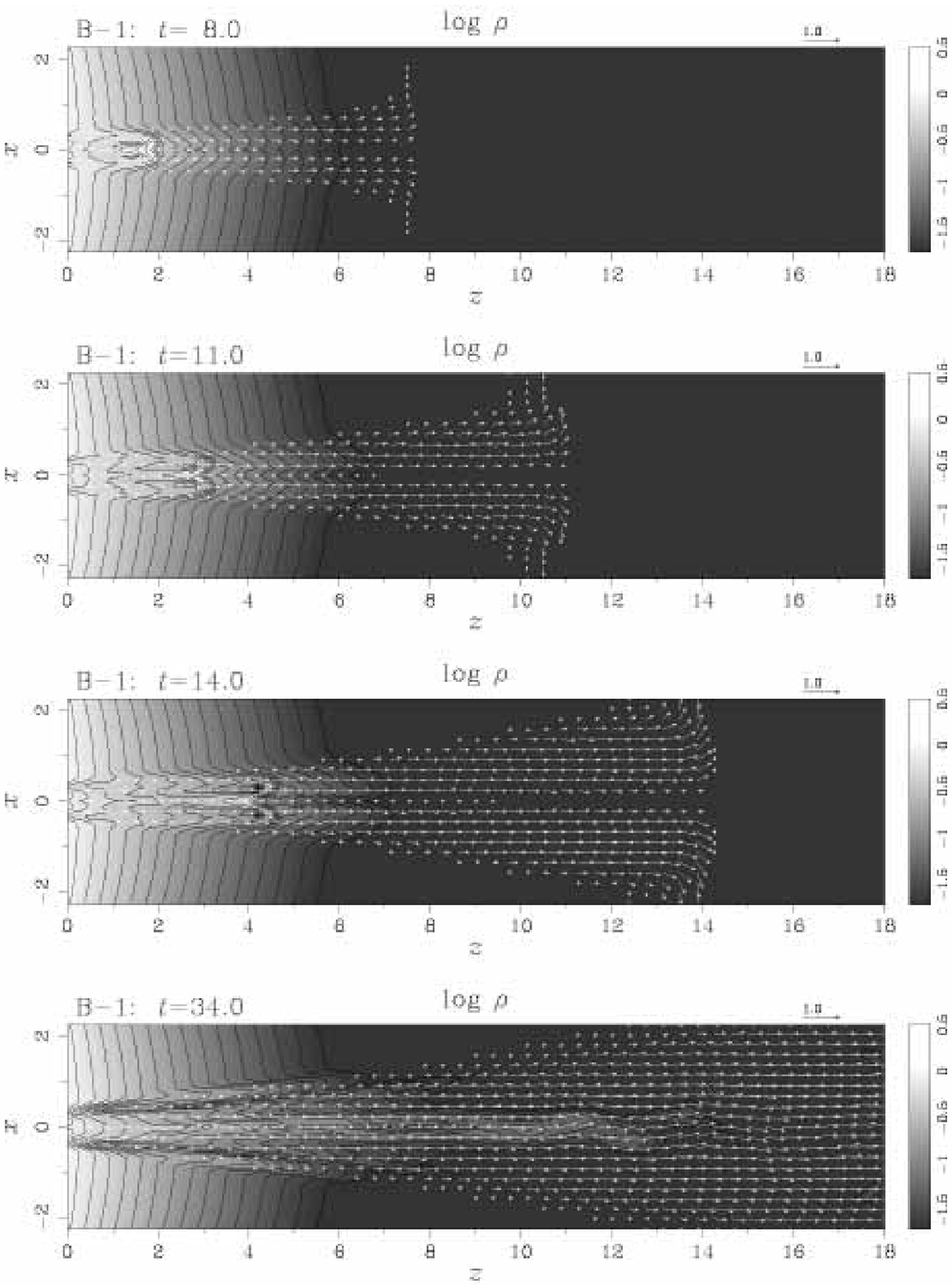}
\caption{ \label{fig:B1_dens}
Similar to Fig. \ref{fig:A1_dens}, but for Model B -- 1.
}
\end{center}
\end{figure}

\begin{figure}
\begin{center}
\includegraphics[scale=0.89, angle=0]{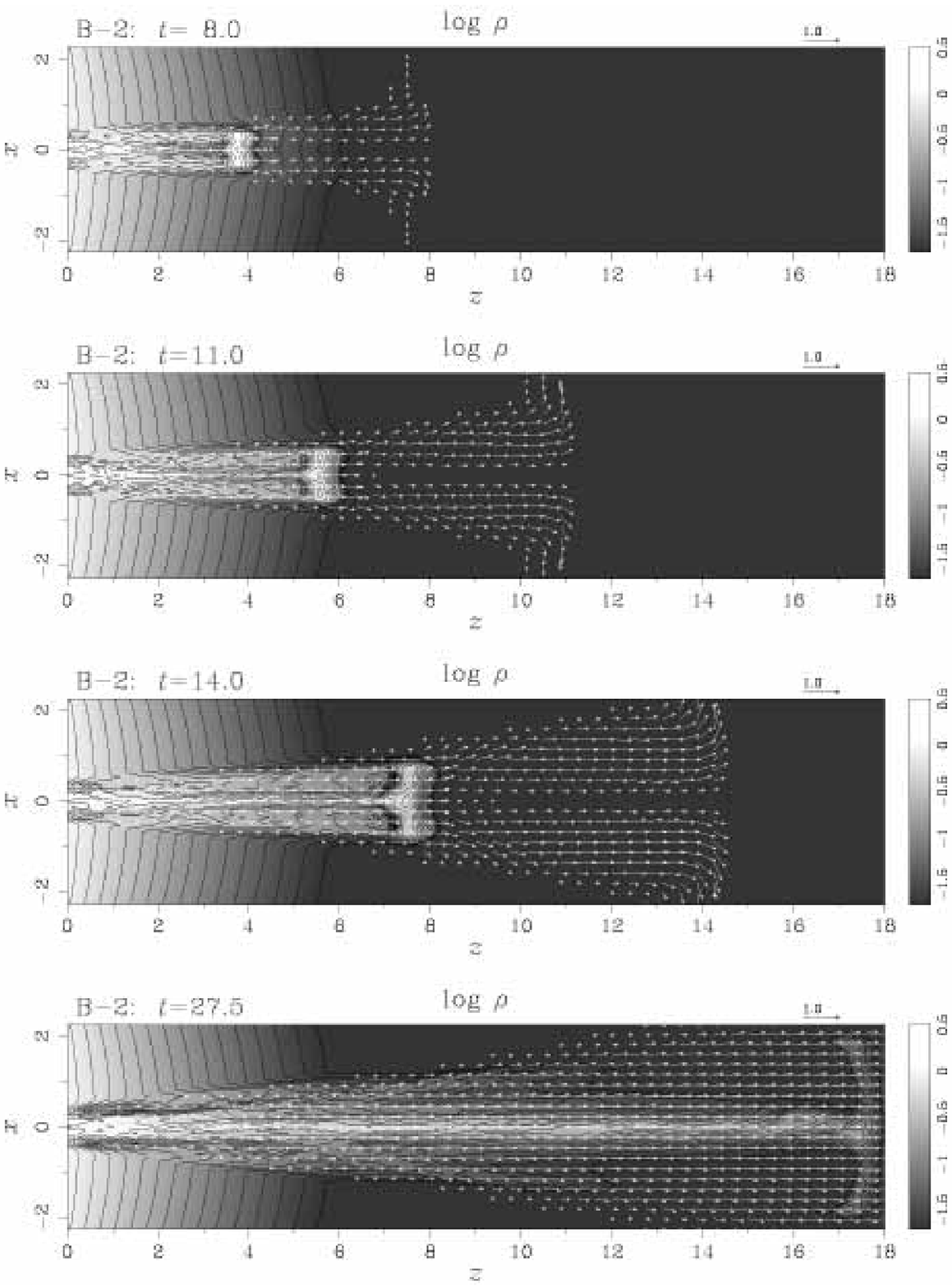}
\caption{ \label{fig:B2_dens}
Similar to Fig. \ref{fig:A1_dens}, but for Model B -- 2.
}
\end{center}
\end{figure}

\begin{figure}
\begin{center}
\includegraphics[scale=0.7, angle=-90]{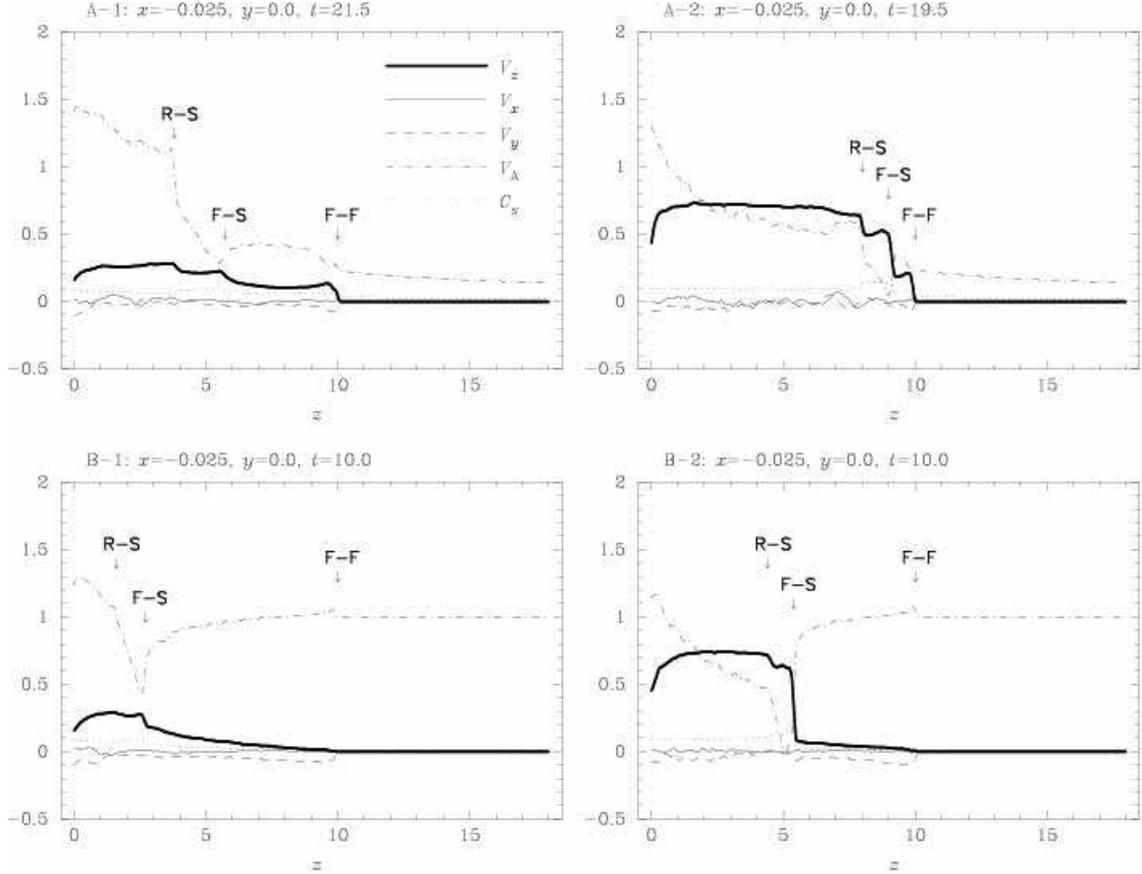}
\caption{ \label{fig:V_axial_before}
Offset axial velocity profiles parallel the $z$-axis at $(x,\,y)=(-0.025,\,0.0)$
in the intermediate stage for each of the models.  Shown are the components 
of velocity $(V_{x},\,V_{y},\,V_{z})$ and the local $V_{\rm A}$ and $C_{\rm s}$ at model times 
$t=21.5$ ({\em Upper-Left}: A -- 1), 
$t=19.5$ ({\em Upper-Right}: A -- 2), 
$t=10.0$ ({\em Lower-Left}: B -- 1), and 
$t=10.0$ ({\em Lower-Right}: B -- 2). 
The labels {\sf F--F}, {\sf F--S}, and {\sf R--S} indicate the positions of 
the forward fast-mode, forward slow-mode , and reverse slow-mode compressive MHD 
(magnetosonic) waves, respectively.
Some of these will steepen into MHD shock waves during the dynamical evolution. 
Each time selected corresponds to when the {\sf F--F} compressive/shock wave fronts reach approximately 
$z=10.0$.
}
\end{center}
\end{figure}

In the "intermediate stages" of the flow ({\em 2nd} panel each in Figs. \ref{fig:A1_dens}--\ref{fig:B2_dens}), 
the gas is weakly compressed in the $z$ direction across the first 
({\sf F--F}) compressive/shock wave, and then is strongly compressed further across the 
second {\rm (}{\sf F--S}{\rm )} shock wave.  
In addition, the gas crossing the third ({\sf R--S}) shock wave also 
is expanded by a large ratio, but in the opposite direction.
Because the greatest compression occurs when the gas crosses the slow shock waves, 
the material undergoes extensive heating and accumulates in the region 
between these two wave fronts. 
In the shallow-atmosphere models A the structure of the pair of slow
shock waves does not change during the simulation. However, in models B
this structure undergoes an expansion in the transverse direction 
because of the steeply decreasing gradient in the external ambient 
pressure.

\begin{figure}
\begin{center}
\includegraphics[scale=0.89, angle=0]{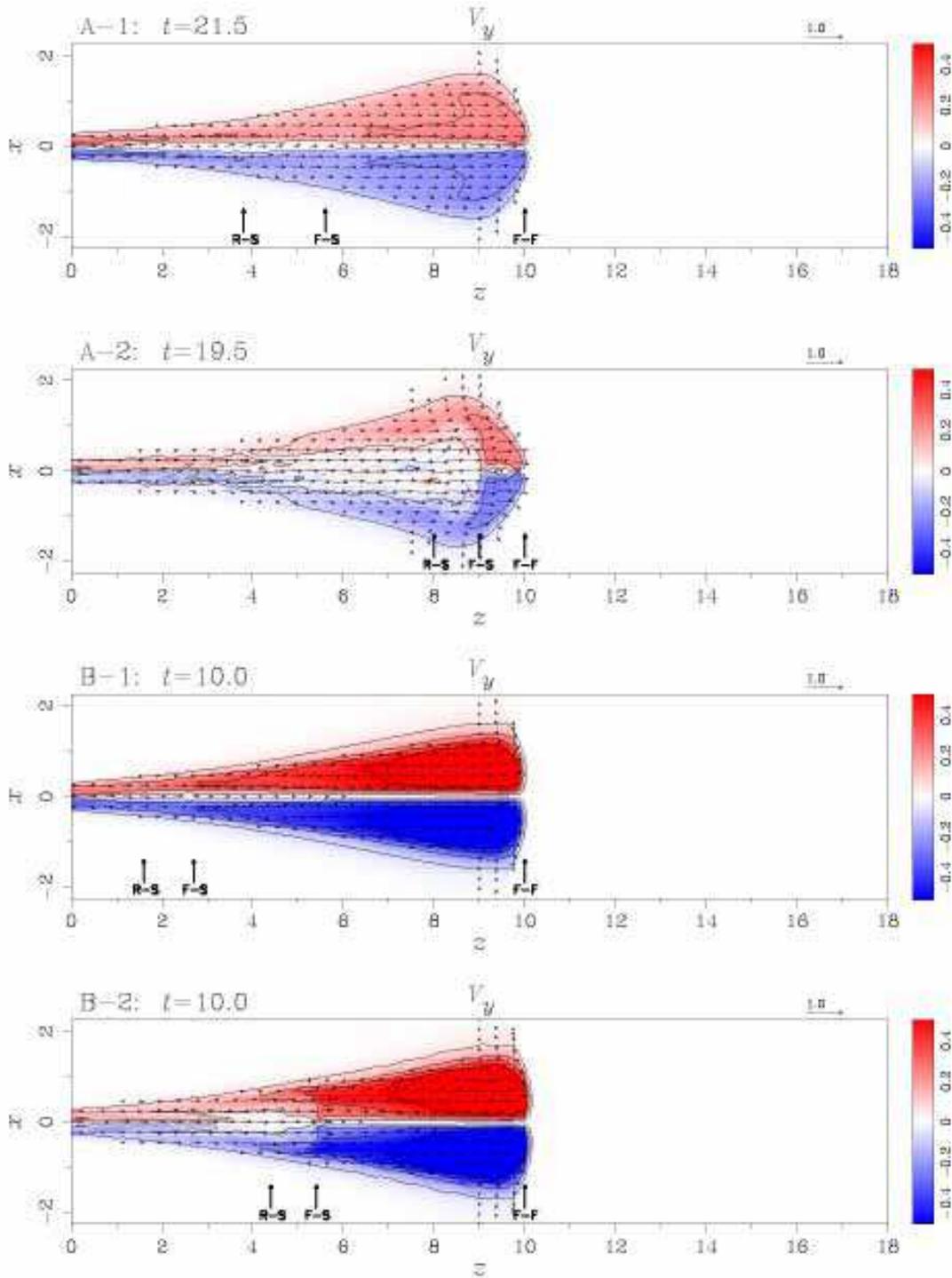}
\caption{ \label{fig:Vy_before}
Snapshots of the azimuthal velocity $V_{y}$ in the intermediate stage for each of the models. 
The labels {\sf F--F}, {\sf F--S}, and {\sf R--S} are the same as those in figure \ref{fig:V_axial_before}.
}
\end{center}
\end{figure}

\begin{figure}
\begin{center}
\includegraphics[scale=0.89, angle=0]{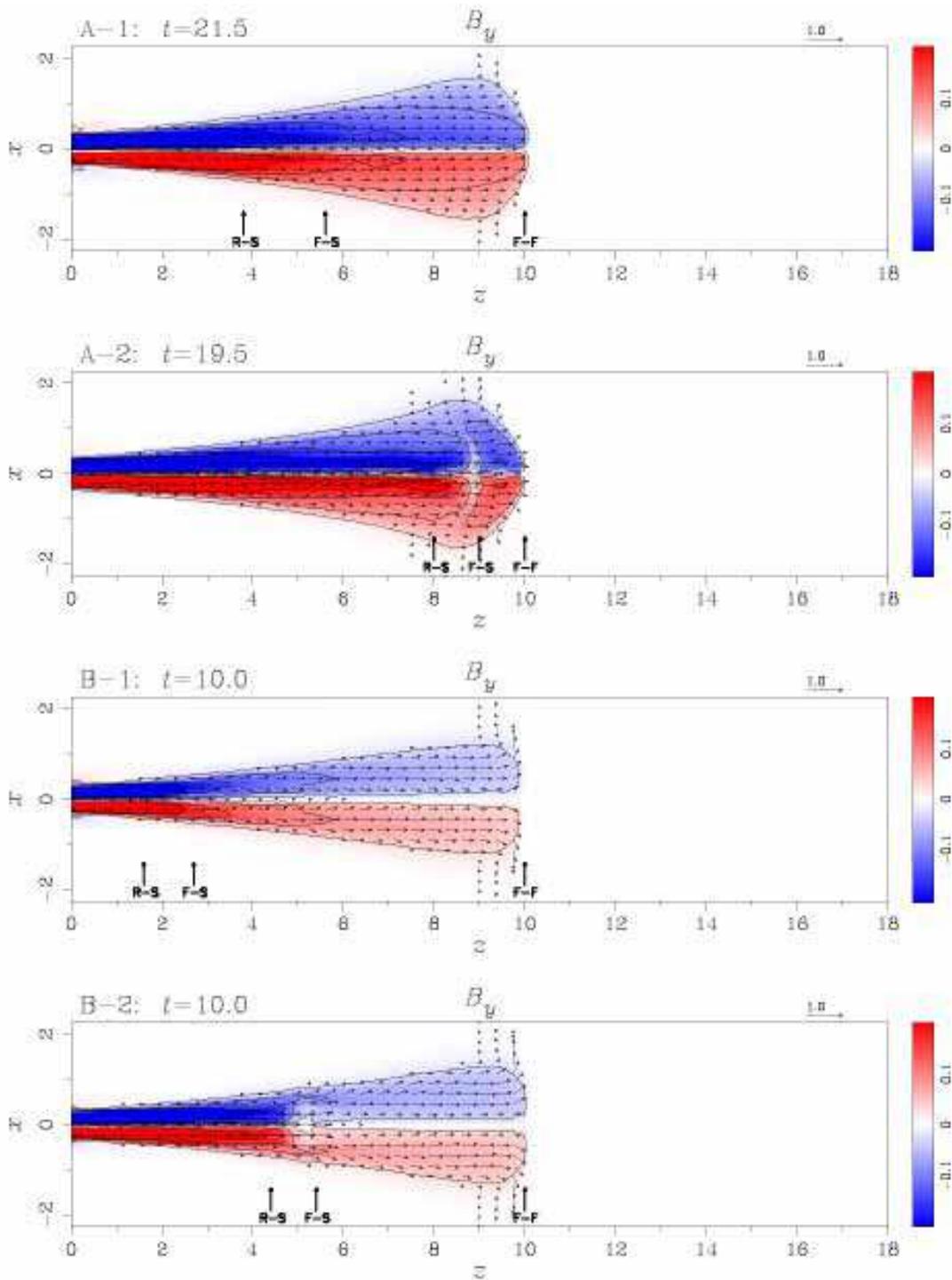}
\caption{ \label{fig:By_before}
Similar to figure \ref{fig:Vy_before}, but for $B_{y}$.  
}
\end{center}
\end{figure}

Figures \ref{fig:Vy_before} and \ref{fig:By_before} show snapshots of $V_{y}$ and $B_{y}$ 
(similar to Fig. \ref{fig:A1_dens} -- \ref{fig:B2_dens}) at 
the time when the head of the PFD jet reaches $z=10.0$ in each of the models. 
($V_{y}$ and $B_{y}$ correspond to $V_{\phi}$ and $B_{\phi}$ in cylindrical coordinates.)
Below the flow through each compressive/shock wave is described in the shock frame (in the negative $z$ 
direction for the {\sf F--F} and {\sf F--S} shock waves and in the positive $z$ direction for 
the {\sf R--S} shock wave).
We now can compare the energy conversion in each of Figs. \ref{fig:Vy_before} and \ref{fig:By_before}: 
\begin{enumerate}
\item {The toroidal component of the magnetic field ($B_{y}$) increases strongly as 
the flow crosses the {\sf F--F} {\em shock} wave front (in the -$z$ direction; models 
A -- 1 and A -- 2).  However, that component increases only weakly across 
the {\sf F--F} {\em compressive} wave (models B -- 1 and B -- 2). Therefore, 
{\em the toroidal (rotational) kinetic energy of a PFD jet is strongly converted into 
toroidal magnetic energy across the {\sf F--F} shock wave in the shallow-atmosphere models A, but 
not so much in the steep-atmosphere models B}.}
\item {The toroidal component of the magnetic field {\em decreases} across the {\sf F--S} shock wave, 
and the toroidal component of the velocity field also decreases across it.  Therefore, 
{\em the toroidal kinetic and magnetic energy of a PFD jet are converted into thermal 
and axial kinetic energy across the {\sf F--S} shock wave}.}

\item {The toroidal component of the magnetic field also decreases as the flow crosses 
the {\sf R--S} shock wave (in the +$z$ direction).   However, unlike the first slow-mode shock wave, 
the toroidal component of the velocity field does {\em not} decrease across the second.  
It actually {\em increases} as material flows into the region between the two slow-mode 
shock waves.  
(Note the relative strength of the toroidal component of the velocity beyond the {\sf F--S} shock 
wave and behind the {\sf R--S} shock wave in the frame co-moving with 
the contact discontinuity for Case A -- 2 in Fig. \ref{fig:Vy_before}.)  Therefore, 
{\em axial kinetic and toroidal magnetic energy are converted into thermal 
and toroidal kinetic energy across the {\sf R--S} shock wave}.
\footnote{At first it would appear that the behavior of two slow-mode shock waves is strangely 
asymmetric.  However, if one transforms to the translating {\em and rotating} frame of the 
contact discontinuity between these two shock waves, one finds that the behavior of all energy 
components (axial and toroidal kinetic, toroidal magnetic, and thermal) are all precisely 
symmetric.  In particular, at both slow-mode shock waves toroidal kinetic energy {\em decreases} 
in this frame, and is converted into thermal energy, as the flow crosses into the region 
between them.  There is only one real asymmetry:  the change in the sense of rotation 
across the two slow-mode shock waves is {\em anti}-symmetric.  This antisymmetry is cause by 
the axial speeds of the two shock waves being opposite in this frame while the handedness of the 
magnetic field pitch, and therefore $V_{\phi}$, is the same.}}
\end{enumerate}

So, we expect that the magnetic field between the second and the third shock waves will be 
less twisted, whereas that between the first and second, and behind the third, shock waves 
will be more twisted.
This physical picture is similar to the results of 1.5-dimensional MHD simulations 
performed by \citet[]{UCH92}.

\subsubsection{Nature of PFD jets as current-carrying jets}

\begin{figure}
\begin{center}
\includegraphics[scale=0.89, angle=0]{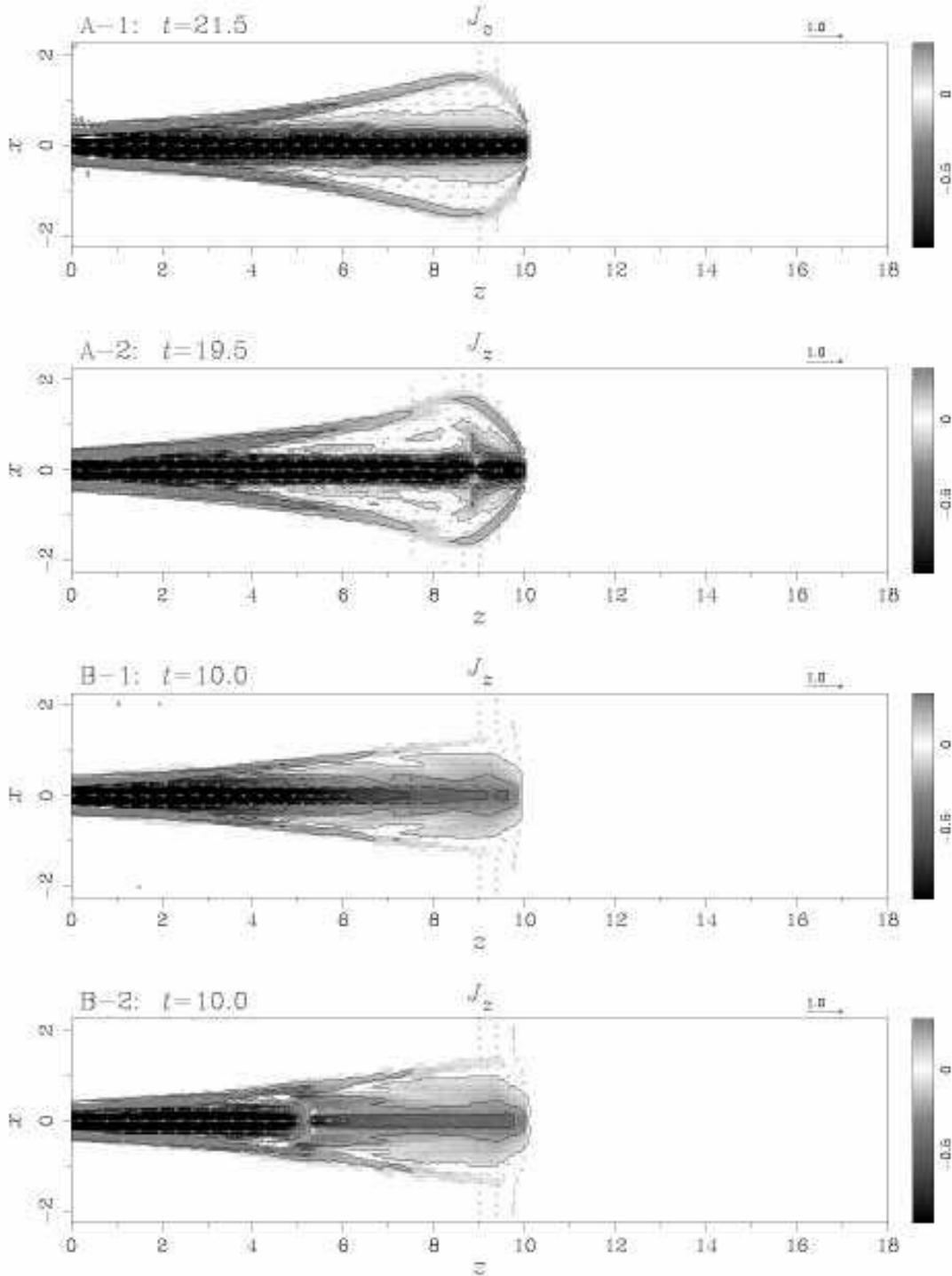}
\caption{ \label{fig:Jz_before}
Snapshots of the axial current density $J_{z}$ in the intermediate stage for each of the models. 
The figures show a closed circulating current system, in which one current path occurs near the 
central ($z$) axis and co-moves with the PFD jet (${\bolJ}^{\rm jc}$), while the other 
conically-shaped current returns outside (${\bolJ}^{\rm rc}$). 
}
\end{center}
\end{figure}

We now consider the behavior of PFD jets as current-carrying systems.
Figure \ref{fig:Jz_before} shows snapshots of the axial current density $J_{z}$ 
(similar to Fig. \ref{fig:Vy_before} and \ref{fig:By_before})
at the time when the head of the PFD jet reaches $z=10.0$ in each models.
Clearly, the flow displays a closed circulating current system, in which one path occurs close to the 
central axis, and co-moves with the PFD jet (the ``{\em forward jet current density}'' 
${\bolJ}^{\rm jc}$) and another conically-shaped path that flows outside 
(the ``{\em return current density}'' ${\bolJ}^{\rm rc}$).

\begin{figure}
\begin{center}
\includegraphics[scale=0.7, angle=-90]{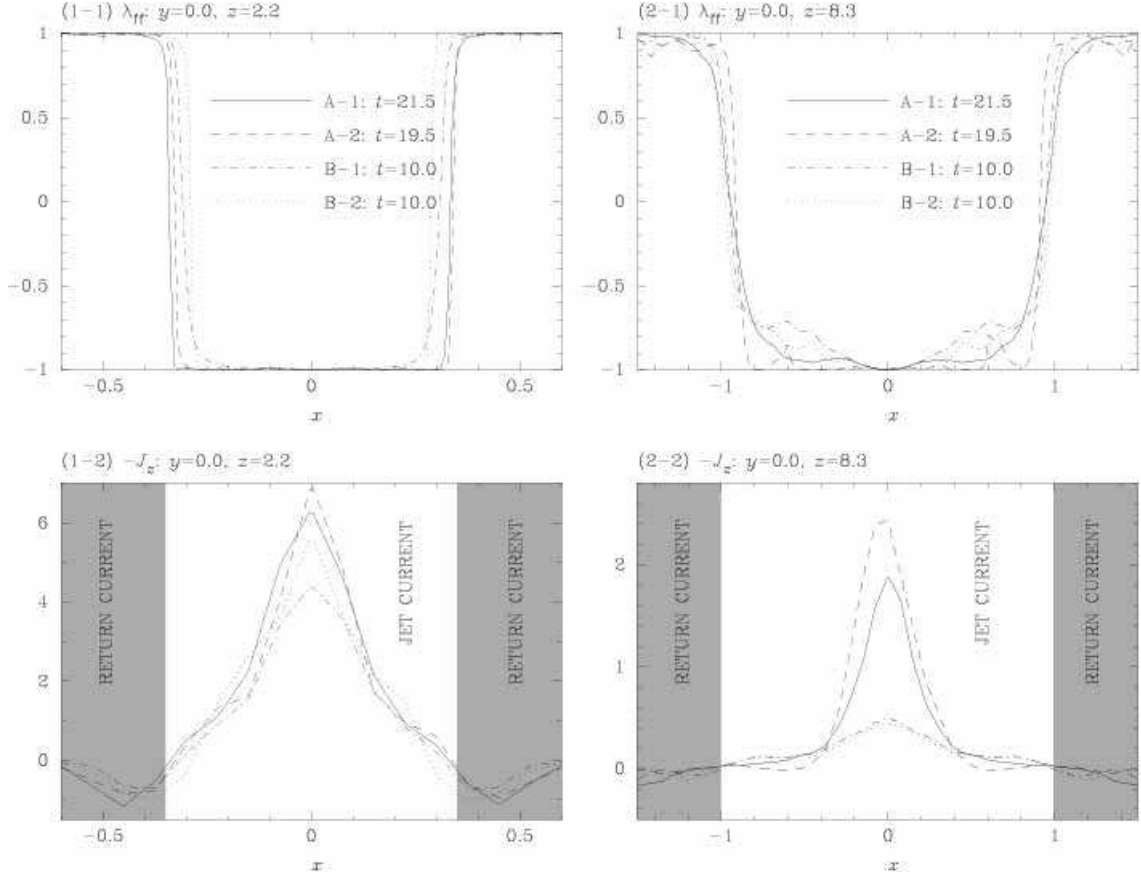}
\caption{ \label{fig:Jz_transverse_before}
Transverse profiles parallel to the $x$-axis of the force-free parameter 
$\lambda_{\rm ff}$ ({\em Upper}) and of the negative of the axial current 
density $-J_{z}$ ({\em Lower}) in the intermediate stage.  Two axial positions are shown:  
$z=2.2$ ({\em Left}) and $z=8.3$ ({\em Right}).
$|\lambda_{\rm ff}|=1$ indicates that ${\bolJ}$ is parallel or anti-parallel 
to the local ${\bolB}$ direction ({\it i.e.}, force-free).  
The reversal of the sign of $\lambda_{\rm ff}$ from -1 to 1 (seen in the {\em Upper} 
panels) indicates a change in sign of the current flow 
from ${\bolJ}^{\rm jc}$ to ${\bolJ}^{\rm rc}$ 
({\em light gray colored regions} seen in both {\em Lower} panels).
}
\end{center}
\end{figure}

Figure \ref{fig:Jz_transverse_before} shows (at the same times as in Fig. \ref{fig:Jz_before}) 
the transverse profile of the force-free parameter $\lambda_{\rm ff}$ ($\equiv {\bolJ} \cdot 
{\bolB}/|{\bolJ}| |{\bolB}| $)
and the axial current density $J_{z}$ near the lower ($z=2.2$) and upper ($z=8.3$) regions. 
The diagram shows that the jets relax into radial force balance 
with the ambient medium:  there is a current-carrying core ${\bolJ}^{\rm jc}$
in the jet itself and a return-current skin ${\bolJ}^{\rm rc}$ outside, both of 
which are in approximate force-free equilibrium (${\bolJ} \times {\bolB} \simeq 0$).
(The reversal of sign from $\lambda_{\rm ff} = -1$ to 1 indicates a change in sign 
in the direction of axial current flow.)
Moreover, an almost (axially) current-free ($J_{z} \simeq 0$) sheath
lies between them, as can be seen in panel (2-2) of
Fig. \ref{fig:Jz_transverse_before} ($0.4 \lesssim |x| \lesssim 1.0$).  
The net axial current $I_{z}$ must be zero to avoid the accumulation of
electrical charge at the end point: 
\beqn
I_{z}=\int_{0}^{r_{\rm max}}\,2\pi\,(J_{z}^{\rm jc}+J_{z}^{\rm rc})\,r\,dr=0.
\eeqn
$r_{\rm max}$ is located at the radius where $B_{\phi}(r_{\rm max})=0$.
This is similar to the transverse analytical structure used by \citet[]{LIN89} as 
inflow conditions for their axisymmetric, toroidal field-only simulations.  
At intermediate times, therefore, 
our three-dimensional numerical simulations have the same dynamical structure as a 
current-carrying equilibrium jet in force balance.

A remarkable feature of the $J_{z}^{\rm jc}$ distribution is the high 
axial current density in the region $5 \lesssim z \lesssim 10$ 
in models A -- 1 and A -- 2 ({\em Top} and {\em 2nd} panels in Fig. \ref{fig:Jz_before}), 
and low axial current density there in models B -- 1 and B -- 2 ({\em 3rd} and {\em Bottom} 
panels in Fig. \ref{fig:Jz_before}) [see also (2-2) of Fig. \ref{fig:Jz_transverse_before}].
This is directly related to how the toroidal component of the field $B_{\phi}$
increases across the first ({\sf F--F}) shock/compressive and third ({\sf R--S}) shock waves 
[{\em or} decreases across the second ({\sf F--S}) shock wave].
A strong accumulation of $B_{\phi}$ produces a large hoop-stress ($-B_{\phi}^{2}/r$), 
pinching the jet towards the central axis.
As a result, the gradient of the $B_{\phi}$ in the transverse direction
becomes larger, and $J_{z}^{\rm jc}$ along the central axis is therefore enhanced. 

As discussed above, we identify an almost (axially) current-free region between the
forward jet current ${\bolJ}^{\rm jc}$ and the return current
${\bolJ}^{\rm rc}$ densities, where $J_{z}$ is close to 0 [see
Fig. \ref{fig:Jz_before} and (2-2) of Fig. \ref{fig:Jz_transverse_before}].
In this region the transverse component of the field $B_{y}$ and velocity $V_{y}$ have a maximum value
and the magnetic lines of force are highly twisted (with the ratio $B_{y}/B_{z}$ much 
larger than that in the central jet).
We will refer to this region as the ``external magnetized wind''. 
(The total outflow consists of both the jet {\em and} the wind.
Here we distinguish the wind from the jet by the fact that it is 
not current-carrying.)  
In a later section we will discuss the wind's effects on the stabilization of
the jet against disruption by KH instabilities. 

\subsubsection{Final evolutionary stages: destabilization of PFD jets}

In the late stages of jet evolution 
({\em 3rd} and {\em Bottom} panels of each of Figs. \ref{fig:A1_dens}--\ref{fig:B2_dens}),
the jets are deformed into wiggled structures in both the density and velocity fields.
For the shallow-atmosphere, highly PFD model A -- 1, the jet is distorted both beyond the second shock wave and 
behind the third shock wave.  In addition, as the jet instabilities become fully developed, 
the high temperature region containing the pair of two slow-mode shock waves and the contact 
discontinuity between them, is fully disrupted also. 

On the other hand, for models A -- 2 (the shallow-atmosphere, mildly PFD case) and 
B -- 1 (the steep-atmosphere, highly PFD case) the jets are distorted behind the second 
shock wave only.  The high temperature region between the two slow-mode shock waves appears unaffected.  

The instability appears to have the slowest growth rate in the steep-atmosphere, 
mildly PFD case (model B -- 2), with only a slight distortion appearing behind 
the {\sf R--S} shock wave.  

\begin{figure}
\begin{center}
\includegraphics[scale=0.89, angle=0]{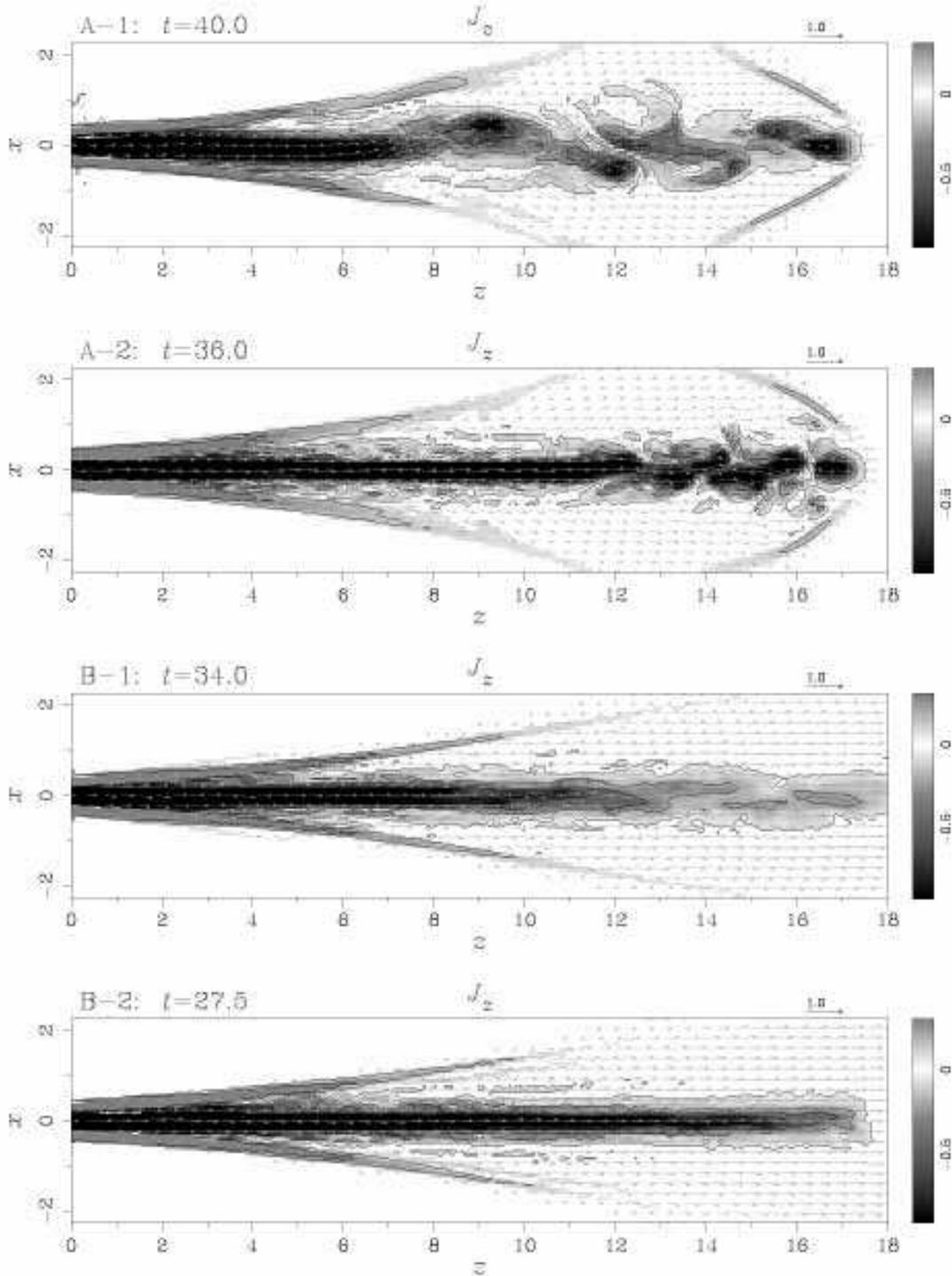}
\caption{ \label{fig:Jz_after}
Snapshots of the axial current density $J_{z}$ in the final stage for each of the models. 
Each time selected corresponds to when the {\sf F--F} (A -- 1 and A -- 2) 
or the {\sf F--S} (B -- 1 and B -- 2) shock wave fronts approach or exceed approximately $z=18.0$. 
Note that these figures are 2-D meridional slices only through the 3-D jet structure. 
While the distributions of ${\bolJ}^{\rm jc}$ appear asymmetric in models A -- 1, A -- 2, and B -- 1, the actual jet 
structures in these models are a three-dimensional spiral helix. 
}
\end{center}
\end{figure}

\begin{figure}
\begin{center}
\includegraphics[scale=0.7, angle=-90]{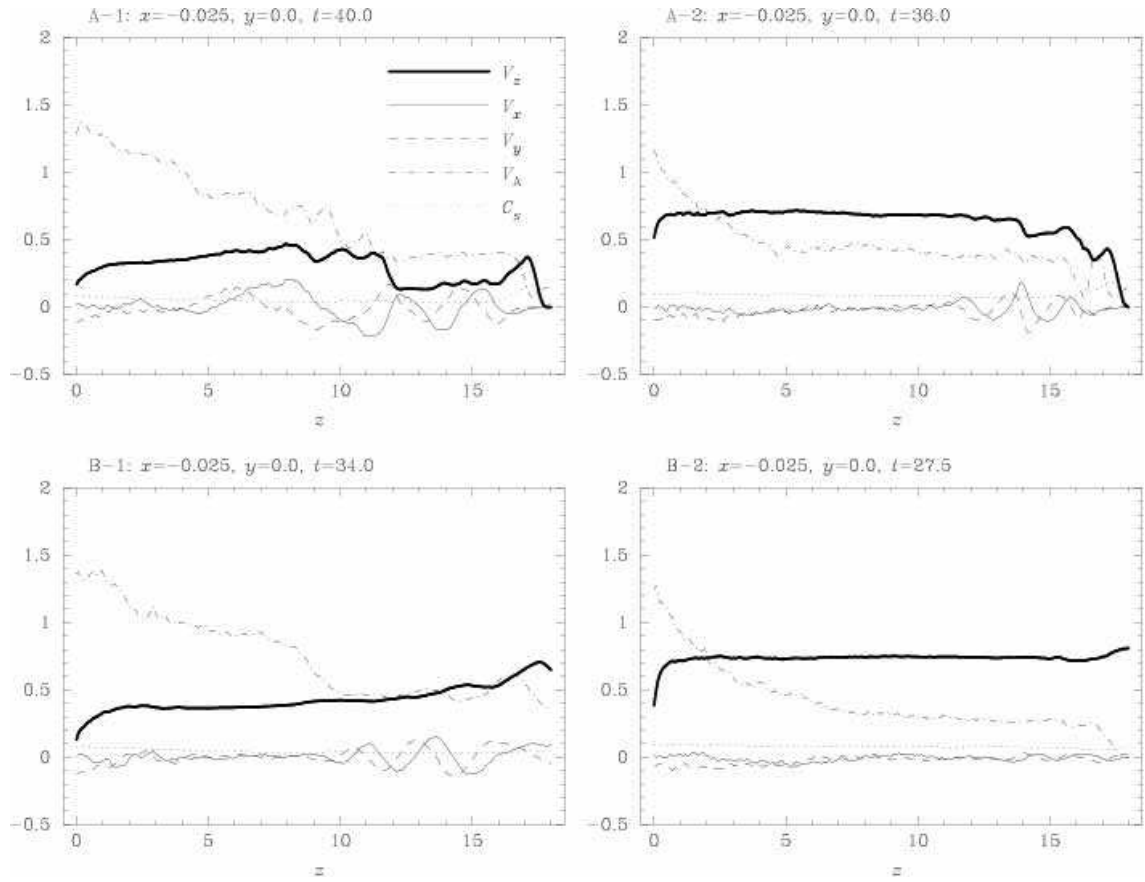}
\caption{ \label{fig:V_axial_after}
Similar to figure \ref{fig:V_axial_before}, but for the final stage in each of the models. 
}
\end{center}
\end{figure}

Figure \ref{fig:Jz_after} shows snapshots of $J_{z}$ in the "final stages" of all the models.
Here we also see that the jet current density $J_{z}^{\rm jc}$ is distorted into the 
wiggled structure seen in the density and velocity fields of models A -- 1, A -- 2, and B -- 1.
In Fig. \ref{fig:V_axial_after} we also show (in the final stages in all models and in the 
$x-z$ plane close to the jet axis) the distributions of the Alfv\'en and sound speeds, 
as well as each component of the jet velocity.   We identify the following features in 
each of the models: 
\begin{itemize}
\item {The axial speed of the jet $V_{z}$ in model A -- 1 is still sub-Alfv\'enic except at
the jet head itself (the head is a {\sf F--F} shock wave, and therefore,
super-Alfv\'enic and super-fast magnetosonic); that is, the highly PFD jet remains 
Poynting flux dominated in this shallow atmosphere.}
\item {$V_{z}$ in model A -- 2 is super-Alfv\'enic beyond $z > 2$ (the jet head is also a
{\sf F--F} shock wave); that is, the mildly PFD jet has switched to a mildly kinetic energy
flux dominated (KFD) jet.}
\item {$V_{z}$ in model B -- 1 is trans-Alfv\'enic beyond $z > 10$, where the jet
is distorted; that is, the highly PFD jet in this steep atmosphere has switched to an 
equipartition state between the kinetic and Poynting fluxes.}
\item {$V_{z}$ in model B -- 2 is super-Alfv\'enic beyond $z > 2$; that is, 
the mildly PFD jet in the steep atmosphere has switched to a KFD jet.}
\end{itemize}

We conclude, therefore, that jets propagating in the trans-Alfv\'enic region, 
before they become kinetic energy flux dominated 
($V_{\rm A} \sim V_{\rm FM} \lesssim V_{\rm j}$), can be deformed into wiggled structures.  
We consider the physical mechanism for these destabilization in the following sections.

\subsection{Nonlinear Growth of CD Instabilities}

\subsubsection{Computation of the power spectrum}

To better understand the nonlinear behavior of CD instabilities in our jet 
simulations, we begin by defining the forward jet current density 
${\bolJ}^{\rm jc}$ to be equal to the total current density ${\bolJ}$ in those 
places where $J_{z}$ has a negative sign and zero otherwise:
\beqn
&&{\bolJ}^{\rm jc} \equiv \cases{{\bolJ} & ($J_{z} < 0$) \cr
	   0 & ($J_{z} \geq 0$) \cr}.
\eeqn
We distinguish ${\bolJ}^{\rm jc}$ from the jet return current density
\beqn
&&{\bolJ}^{\rm rc} \equiv \cases{{\bolJ} & ($J_{z} > 0$) \cr
	   0 & ($J_{z} \leq 0$) \cr},
\eeqn
which differs in sign from ${\bolJ}^{\rm jc}$.  Note that ${\bolJ}^{\rm jc}$ (and 
${\bolJ}^{\rm rc}$) is a vector and can have additional radial and azimuthal components 
$J^{\rm jc}_r$ and $J^{\rm jc}_\phi$ with {\em any} sign. 
(These components are computed by transforming to a cylindrical coordinate system from 
the simulation Cartesian coordinates.) 

We now proceed to analyze the modal structure of ${\bolJ}^{\rm jc}$ by forming the 
volume-averaged Fourier transform of its magnitude 
\beqn
\label{eq:fourier_1}
\tilde{{\bolJ}^{\rm jc}}(m,\,k) = \frac{1}{V_{\rm cl}} \int\!\!\!\!\!\!\int\!\!\!\!\!\!\int_{V_{\rm cl}}
|{\bolJ}^{\rm jc}| e^{i(m\phi+kz)}\,r\,dr\,d\phi\,dz,
\eeqn
where 
$|{\bolJ}^{\rm jc}| \equiv \bigl[(J^{\rm jc}_r)^{2}+(J^{\rm jc}_{\phi})^{2}+(J^{\rm jc}_z)^{2}\bigr]^{1/2}$ 
and $V_{\rm cl}$ is the annular volume that encloses ${\bolJ}^{\rm jc}$ 
\beqn
\label{eq:fourier_2}
V_{\rm cl}=
\int_{r_{\rm a}}^{r_{\rm b}}\!\!\!
\int_{0}^{2 \pi}\!\!\!
\int_{z_{\rm a}}^{z_{\rm b}}\,r\,dr\,d\phi\,dz.
\eeqn
We use the values $r_{\rm a}=0.075$, $r_{\rm b}=2.25$, $z_{\rm a}=0.0$, 
and $z_{\rm b}=18.0$. 

$\tilde{{\bolJ}^{\rm jc}}(m,\,k)$ is now a function of the azimuthal mode $m$ and axial 
wave number $k=2 \pi/ \lambda$ (corresponding to a characteristic wave length $\lambda$).
Like ${\bolJ}^{\rm jc}$, $\tilde{{\bolJ}^{\rm jc}}$ is a function of time. 

Finally, we identify the power spectrum as the squared amplitude of $\tilde{{\bolJ}^{\rm jc}}$ 
\beqn
\label{eq:fourier_3}
&&|\tilde{{\bolJ}^{\rm jc}}(m,\,k)|^{2}=
\left\{{\rm Re} \bigl[\tilde{{\bolJ}^{\rm jc}} (m,\,k) \bigr] \right\}^{2} +
\left\{{\rm Im} \bigl[\tilde{{\bolJ}^{\rm jc}} (m,\,k) \bigr] \right\}^{2},
\eeqn
\beqn
\label{eq:fourier_4}
&&{\rm Re} \bigl[\tilde{{\bolJ}^{\rm jc}} (m,\,k) \bigr] \nonumber \\
&&\quad =\frac{1}{V_{\rm cl}} \int\!\!\!\!\!\!\int\!\!\!\!\!\!\int_{V_{\rm cl}} 
|{\bolJ}^{\rm jc}| \cos\,(m\phi+kz)\,r\,dr\,d\phi\,dz,\\
\label{eq:fourier_5}
&&{\rm Im} \bigl[\tilde{{\bolJ}^{\rm jc}} (m,\,k) \bigr] \nonumber \\
&&\quad =\frac{1}{V_{\rm cl}} \int\!\!\!\!\!\!\int\!\!\!\!\!\!\int_{V_{\rm cl}} 
|{\bolJ}^{\rm jc}| \sin\,(m\phi+kz)\,r\,dr\,d\phi\,dz.
\eeqn
For each model we now will examine the time-dependent behavior of the power 
spectrum and investigate the nonlinear growth of each mode. 

\subsubsection{Growth of azimuthal modes}

\begin{figure}
\begin{center}
\includegraphics[scale=0.7, angle=-90]{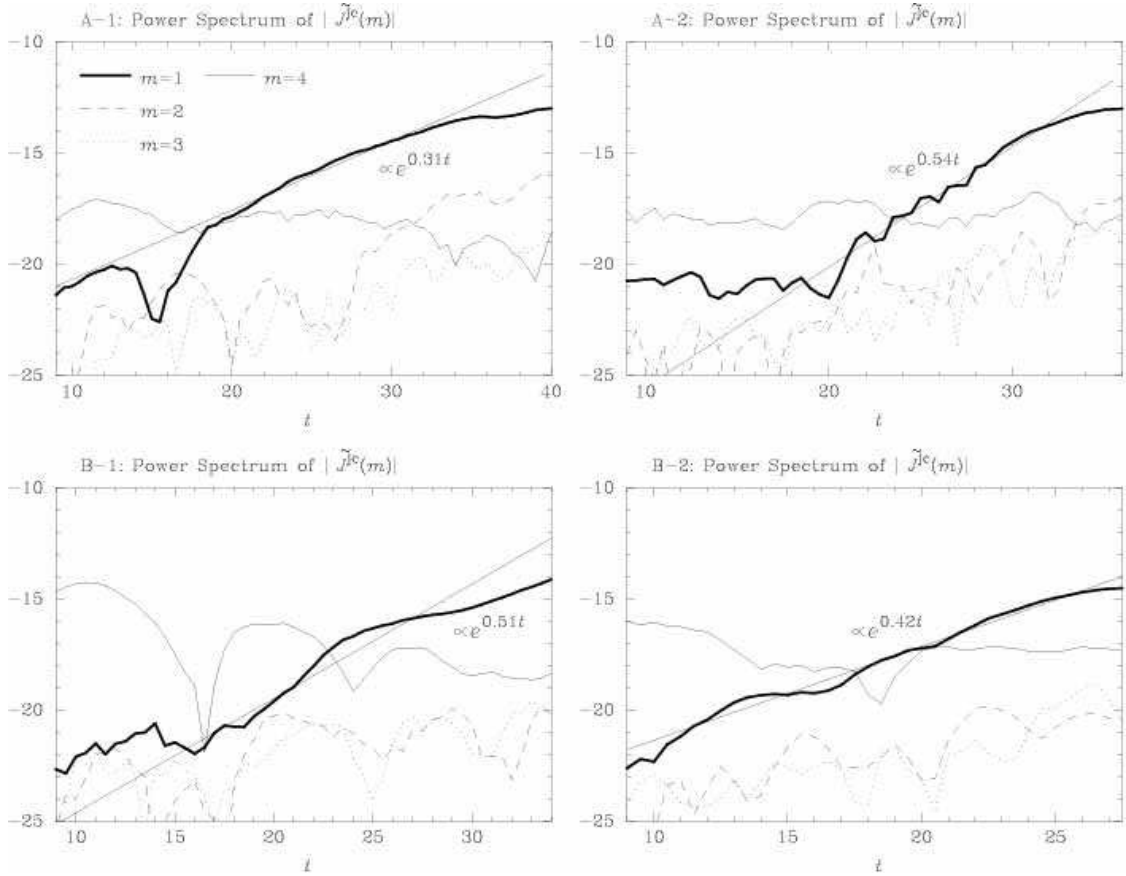}
\caption{ \label{fig:FPS}
Growth of unstable CD modes for each model sequence, showing the time variation of 
the azimuthal Fourier power in modes $m=1-4$, on a natural log scale. 
Solid lines are derived by fitting the slopes of the $m=1$ mode growth to a linear function.
}
\end{center}
\end{figure}

We first will consider only azimuthal modes ($m$) --- {\it i.e.}, the power spectrum 
$|\tilde{{\bolJ}^{\rm jc}}(m)|$ for infinitesimal wave numbers ($k \rightarrow 0$). 
Several modes ($m=1 - 4$) are plotted as functions of time in Fig. \ref{fig:FPS}.
Note that, while we have not imposed any explicit perturbations to our jets, they 
are subject to the implicit perturbations created by the 3--D Cartesian grid.  
As a result of our resolving an initially, rotating cylindrically symmetric object 
on a Cartesian coordinate, 
the $m=4$ mode appears before the physical growth of the other asymmetric modes.
We find, however, that the power in the $m=4$ mode remains constant or decreases with 
time on all models.  
We therefore identify the level of power in the $m=4$ mode as our (conservative) noise 
level and identify all power greater than that (and in any other mode) as spectral power 
generated by the physical evolution of the jet. 

In all models the kink ($m=1$) mode grows faster than the other higher order modes ($m=2,\,3$).
In Fig. \ref{fig:FPS} the time at which the kink mode appears above the ``noise'' 
level in each of the models is as follows: 
$t \sim 18$ (A -- 1), $t \sim 23$ (A -- 2), $t \sim 23$ (B -- 1), and $t \sim 17$ (B -- 2). 
In addition, in A -- 1 both the $m=1$ and $m=2$ modes 
appear around $t \sim 30$, and
the $m=3$ mode saturates at or below the noise level. 
In A -- 2, the $m=2$ and 3 modes track one another and saturate at our chosen noise level. 
In B -- 1 and B -- 2, the $m=2$ and 3 modes also track one another 
and saturate at a level {\em lower} than that of the $m=4$ mode.

The $m=1$ mode for each model exhibits approximately exponential growth 
on a dynamical time scale. 
A linear fit using a least-squares method has been performed on the time-sequenced data to 
extract a linear growth rate 
\beqn
{\rm Im}\,(\omega) \sim \frac{d\,{\rm ln}\,|\tilde{{\bolJ}^{\rm jc}}(m)|^{2}}{dt}.
\eeqn
The estimated growth rates are given in Table \ref{tbl:growth} and shown on Fig. \ref{fig:FPS}. 
The growth rate appears to be strongest for A -- 2 [${\rm Im}\,(\omega) \sim 0.54$], 
and weakest for A -- 1 [${\rm Im}\,(\omega) \sim 0.31$].
As seen in Fig. \ref{fig:FPS}, 
the instability appears to have saturated in the final stages for both A -- 2 and B -- 2 
(the mildly PFD models), 
but still appears to be growing for both of the highly PFD models A -- 1 and B -- 1. 

How do these growth rates compare with the Alfv\'en crossing time scale, which is typically 
the time scale on which pure CD modes can develop \citep[]{BEG98,APP00,LER00b}? 
The spatial length scale of the CD instability in our results (in the radial direction) 
is of the order of a jet width, $L \sim 1.0$ (see Fig. \ref{fig:Jz_after}).
The local $V_{\rm A}$ in the distorted jet portion is of order $0.3 - 0.8$, 
as seen in Figs. \ref{fig:V_axial_before} and \ref{fig:V_axial_after}.
So, the inverse of the Alfv\'en crossing time $\tau_{\rm A}^{-1}$ in the distorted jets is approximately given by
\beqn
\tau_{\rm A}^{-1}=\frac{V_{\rm A}}{L} \sim 0.3 - 0.8.
\eeqn
This is, in general, consistent with the time scales of the growing $m=1$ modes derived from Fig. \ref{fig:FPS},
\beqn
{\rm Im}\,(\omega) \approx O\,(\tau_{\rm A}^{-1}).
\eeqn

\subsubsection{Development of the $m=1$ helical kink}

\begin{figure}
\begin{center}
\includegraphics[scale=0.7, angle=-90]{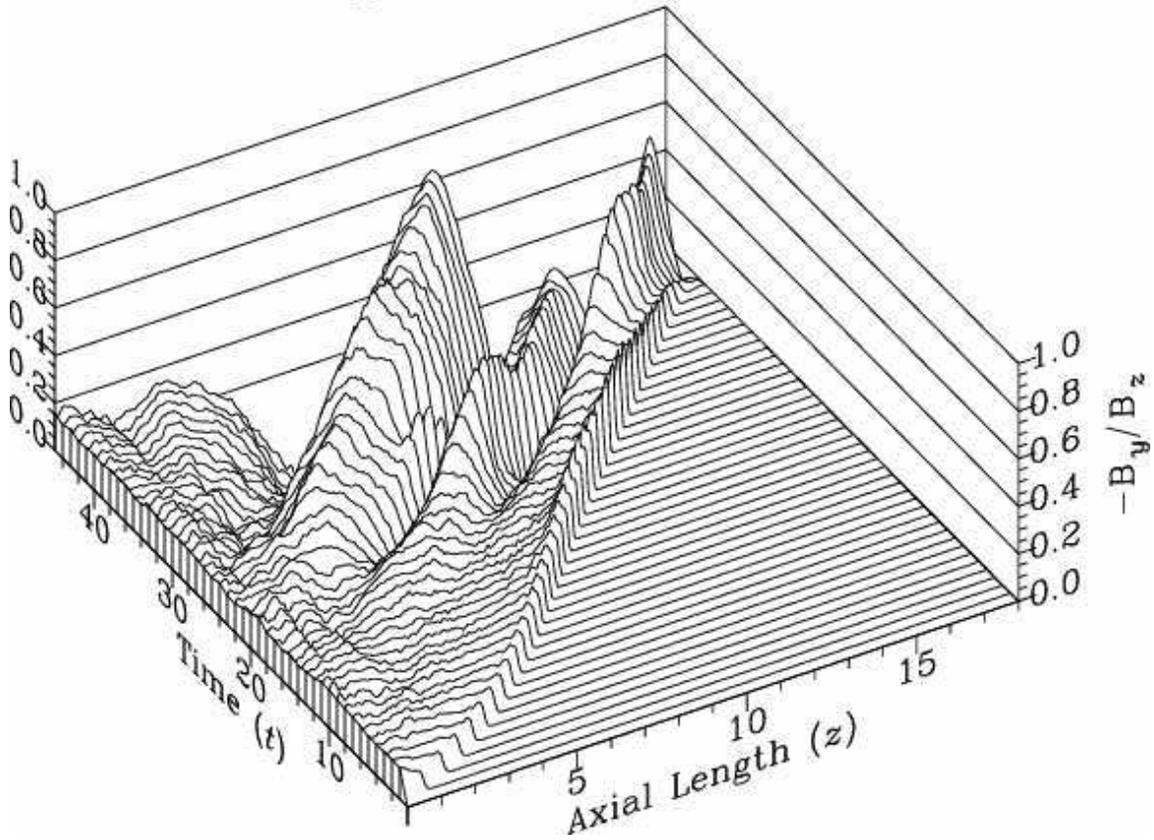}
\caption{ \label{fig:space_time_twist}
Space-time diagram showing the growth of the ratio of transverse and axial magnetic 
field $-B_{y}/B_{z}$ during the interval $t=0.0-45.0$.  Values are measured 
near, and parallel to, the $z$ axis ($x=-0.025,\,y=0.0$) for Model A -- 1.
}
\end{center}
\end{figure}

Figure \ref{fig:space_time_twist} shows a space-time ($z,t$) plot of the
ratio of the transverse and axial magnetic field components $-B_{y}/B_{z}$
for model A -- 1, illustrating the dynamical evolution of the helically kinked
magnetic field.  Propagation of the jet head, corresponding to 
a {\sf F--F} compressive wave front, can be seen.  The wave gradually
decelerates due to the decreasing ambient $V_{\rm A}$ until $t \sim 15$;
after that it propagates with an almost constant phase velocity.  The amplitude
of the {\sf F--F} compressive wave grows significantly with time
and could steepen into a {\sf F--F} shock wave (bow shock) with subsequent evolution. 
After $t \sim 20$, the kinked part of the field appears behind the {\sf
R--S} first, and later, it appears in the upstream region.  Finally, the
flow downstream (between the {\sf F--F} and the {\sf F--S}) is also kinked.
We can see that {\em the kinks in these three regions grow independently, and
the peaks of the kinked magnetic field simply co-move with the jet}.
These characteristics indicate that {\em the CD 
kink mode does not propagate in the jet rest frame} \citep[]{APP96}.  This
behavior is quite different from that of the KH instability.  In the latter case,
the disturbance propagates {\em backwards} in the jet rest frame, because
the phase velocity of the KH modes is smaller than the jet velocity,
which must be super-Alfv\'enic to stimulate the KH unstable modes.

Figure \ref{fig:JxB_after} shows snapshots of the radial specific power generated by the
Lorentz force on two-dimensional $x-z$ slices at $y=0$.
This is defined as the work done per unit mass 
in the $r$-direction in a cylindrical coordinate system ($r,\phi,z$):
\beqn
({\bolJ} \times {\bolB})_{r}\,V_{r}/\rho,
\eeqn
where 
\beqn
({\bolJ} \times {\bolB})_{r}=-J_{z}B_{\phi}+J_{\phi}B_{z}.
\nonumber
\eeqn
If we consider only radial variations in the Lorentz force, then
\beqn
&&-J_{z}\,B_{\phi} = - \frac{B_{\phi}}{r}\frac{\partial}{\partial r} (rB_{\phi}), 
\nonumber
\eeqn
and
\beqn
&&J_{\phi}\,B_{z} = - B_{z}\frac{\partial B_{z}}{\partial r}.
\nonumber
\eeqn
There is a strong correlation between this power distribution and the
wiggled structures seen in A -- 1, A -- 2, and B -- 1.
This is consistent with our early finding that the disruption occurs only 
locally where the MHD jet is CD unstable.  That is, {\em distortions do not propagate}. 
These snapshots also lead us to conclude, as we did in Fig. \ref{fig:FPS}, 
that the CD instability is still growing for the highly PFD jets A -- 1 and B -- 1, 
but is almost saturated for the mildly PFD flows A -- 2 and B -- 2.
This means that the magnetic field in the flow should be in ``radial'' force-free equilibrium 
(${\bolJ}\times{\bolB} \simeq 0$) in the saturated models.

We conclude that CD instabilities grow during the simulation evolution even without 
the addition of small perturbations to the system.
This is because we have performed computations with high enough resolution to
amplify asymmetric modes from the numerical background noise.
We close this section by noting that the instabilities reported here do not arise 
if we choose a grid spacing that is two or more times larger than the current one. 
The effects of numerical diffusion appear to damp out the growth of the CD 
instability in this case. 

\subsection{Stabilization of the CD Instability with Rapid Rotation}

\subsubsection{The Classical Kruskal--Shafranov Criterion}

According to the well-known Kruskal--Shafranov (K--S) criterion \citep[]{SHA57, KRU58}, which is  
based on a skin-current model,
a magnetized flux tube (along the $z$ axis) is stable to the CD kink ($m=1$) mode
as long as the magnetic twist angle $\Phi(r)$ is below some critical value $\Phi_{\rm crit}$,
\beqn
\label{eq:K-S_limit}
\Phi(r) \equiv \frac{L B_{\phi}}{r B_{z}} < \Phi_{\rm crit},
\eeqn
where, $L$ is the length of the current-carrying magnetic flux system, 
$r$ is the radius of the flux tube, and 
$B_{z}$ and $B_{\phi}$ are the axial and azimuthal field components in cylindrical coordinates, 
respectively.
Many theoretical and numerical investigations have been performed on the current-driven 
stability problem of solar coronal loops, and the K--S criterion plays an important 
role in those studies.  

In the original K--S criterion, $\Phi_{\rm crit}$ is set equal to $2 \pi$.
The effects of line-tying, {\it i.e.}, anchoring the axial boundary conditions on the magnetic 
field in a high-density photosphere, however, raise the stability threshold.
The modified critical values for a force-free configuration are between $2 \pi$ and $10 \pi$ 
\citep[]{HOO79, EIN83}.

For a jet configuration, which has $L \gg r$, we replace $L$ by a spectrum of 
axial wavelengths $\lambda$, and we define the critical wavelength to be the one that 
is marginally stable
\beqn
\label{eq:K-S_crit_wavelength}
\lambda_{\rm crit} & \equiv & \frac{\Phi_{\rm crit} \, r \, B_{z}}{B_{\phi}}
\\
& = & \frac{2 \pi \, r \, B_{z}}{B_{\phi}}.
\nonumber
\eeqn
The K--S stability criterion (eq. \ref{eq:K-S_limit}) then becomes
\beqn 
\lambda < \lambda_{\rm crit},
\eeqn
so that wavelengths greater than $\lambda_{\rm crit}$ are unstable to the CD instability. 
From a computational point of view, $\lambda$ must lie in the range 
\beqn
\Delta \le \lambda \le L(t),
\eeqn
where $\Delta$ is the minimum grid width and $L (t)$ is time-dependent jet length, 
with maximum value $L_{\rm max} (t)$ dependent on the computational domain. 
In practice, however, shock waves and other MHD phenomena provide effective domain 
boundaries, so that the maximum value of $\lambda$ could be only a fraction of 
the entire jet length. 

\subsubsection{Force Balance in Rotating Jet Systems}

Because real astrophysical jet systems may be rotating, the K--S criterion 
may not be an adequate stability condition for the CD kink mode. 
Rotation will modify the radial force balance inside jet.
Although rotation is often neglected in theoretical models 
of jet equilibrium \citep[]{BEG98, APP00}, it could play an important role in the stability analysis. 
Let us consider the time-independent, force-balance equation (\ref{eq:momentum}) in 
the radial direction.  In an inertial frame co-moving with the jet we have 
\beqn
\label{eq:radial_1}
- {\nabla} P + \bolJ \times \bolB + \hat{r} \rho \Omega^{2} r = 0,
\eeqn
where $\Omega$ is the equilibrium angular velocity and $\hat{r}$ is the unit vector 
in the $r$ direction.
It is clear that the axial flow velocity $V_{z}$ does not affect the radial force balance, 
but the presence of an {\em azimuthal} velocity component can have a significant influence
on the equilibrium, modifying the competition between 
the pressure gradient and the Lorentz force.
Moreover, our PFD jets are strongly magnetized ($\beta \ll 1$), so 
the pressure gradient term on the left-hand side of equation (\ref{eq:radial_1}) can be 
neglected.  This yields 
\beqn
\label{eq:radial_2}
-\frac{\partial}{\partial r} \left(\frac{B_{\phi}^{2}+B_{z}^{2}}{2}\right)
-\frac{B_{\phi}^{2}}{r}+\frac{\rho V_{\phi}^{2}}{r} = 0,
\eeqn
where $V_{\phi}$ is the equilibrium azimuthal velocity.
The first term on the right-hand side in equation (\ref{eq:radial_2}) is the magnetic pressure gradient 
in both the azimuthal and axial components.
These forces are positive or negative depending on whether $B_{z}$ and $B_{\phi}$ decrease 
or increase with $r$, respectively.
The second term is the magnetic tension force (hoop-stress), which is always directed
inward (in the $-r$ direction).
The third is the centrifugal force, which always acts in the $+r$ direction, 
opposite to the hoop-stress.

One key parameter of radial equilibrium in rotating jets is the radial index $\kappa$ of the 
magnetic pressure $p_{\rm m}^{\ast}=(B_{\phi}^{2}+B_{z}^{2})/2$ gradient 
\beqn
\label{eq:kappa} 
\kappa \equiv \frac{\partial~{\rm log}~p_{\rm m}^{\ast}}{\partial~{\rm log}~r}. 
\eeqn
Under the assumption that the magnetic field strength deceases outward near the jet surface, 
$\kappa$ should be less than zero. 
Furthermore, for specific cases of analytically asymptotic solutions for cylindrically 
symmetric axial flows with power-law radial distributions of the physical quantities, 
a lower limit on $\kappa$ must exist \citep[]{OST97}. 
The magnetic field strength $|B^{\ast}| [\equiv (B_{\phi}^{2}+B_{z}^{2})^{1/2}]$ must decline with $r$ 
more slowly than $r^{-1}$ to ensure force balance and cylindrical collimation.  (That is, the magnetic 
field cannot have a gradient so steep that the corresponding magnetic pressure force will exceed the 
hoop-stress.)  Taken together, these conditions require $-2 \leq \kappa < 0$. 

We now introduce another key parameter, the azimuthal (toroidal) Alfv\'en Mach number:
\beqn
\label{eq:M_A_phi_1} 
M_{\rm A \phi} \equiv \frac{V_{\phi}}{V_{\rm A \phi}},
\eeqn
where $V_{\rm A \phi} \equiv |B_{\phi}| / \sqrt{\rho}$.
In the present paper, we have excluded the situation that jets are confined solely by the 
external hot ambient medium. 
So, to prevent radial expansion of the self-collimated jets with $\kappa < 0$, an upper-limit 
on $M_{\rm A \phi}$ should be set:
\beqn
\label{eq:M_A_phi_2}
M_{\rm A \phi} \leq 1.
\eeqn

In the following sections, we will consider these two key parameters as diagnostics of 
radial force equilibrium in the rotating MHD jets.

\subsubsection{Non-rotating strongly magnetized jets}

If the jet has a low plasma--$\beta$ ($\ll 1$) and 
is non-rotating ($M_{\rm A \phi}=0$), the magnetic field in the jet will be 
in force--free equilibrium 
\beqn
\label{eq:radial_3}
-\frac{\partial}{\partial r} \left(\frac{B_{\phi}^{2}+B_{z}^{2}}{2}\right)
-\frac{B_{\phi}^{2}}{r}= 0,
\eeqn
That is, the gradient of the magnetic pressure will be balanced by the magnetic hoop-stress.
Various types of the force--free equilibrium in jets have been investigated 
for different radial profiles of the magnetic twist $\Phi (r)$ \citep[]{APP00, LER00a}.
For astrophysical jets of magnetic origin, the azimuthal field is likely to be dominant, and 
the properties of the fastest growing CD kink ($m=1$) mode become nearly independent of the details 
of the pitch profile \citep[]{APP00}.

\subsubsection{Rotating strongly magnetized jets}

If the jet is rotating ($M_{\rm A \phi} > 0$), an equilibrium {\em force--free} magnetic field will not 
be possible. 
The inertia of the rotating plasma will play an important role in jet stability as $M_{\rm A \phi}$ 
becomes large.  This situation is quite different from the one that occurs in solar coronal loops. 

For an azimuthally sub--Alfv\'enic flow ($0 < M_{\rm A \phi} < 1$) in equilibrium, equation 
(\ref{eq:radial_2}) will be satisfied.
The hoop-stress is the only force acting inward and this must be balanced by the outward 
centrifugal and magnetic pressure gradient forces ($\kappa < 0$). 

For an azimuthally trans--Alfv\'enic flow $M_{\rm A \phi} \sim 1$, the gradient of the magnetic pressure
becomes zero ({\it i.e.}, $\kappa \sim 0$), and force--equilibrium between the hoop-stress 
and the centrifugal force holds:
\beqn
\label{eq:radial_4}
-\frac{B_{\phi}^{2}}{r}+\frac{\rho V_{\phi}^{2}}{r} = 0.
\eeqn
If the radius $r$ of the magnetized loop increases slightly, the centrifugal force will fall off faster 
than the hoop-stress, causing an inward-directed net magnetic tension that returns the loop to its original size. 
On the other hand, if $r$ decreases slightly, the hoop-stress increase less rapidly than the centrifugal force, 
causing an outward-directed net centrifugal force that again returns the loop to its original size.  
Rapid rotation, therefore, has a potentially stabilizing effect on traditionally unstable twisted 
magnetic configurations. 

\subsubsection{Numerical results for non-rotating and rotating jets:  
growth of axial modes}

We now inspect our numerical results more closely and analyze 
the stabilizing effects of rotation against the CD kink mode.
We will continue to use the Fourier power spectrum analysis, 
applying it to the $m=1$ mode and also extending 
it to growing modes in the axial ($k$) direction. 
We also will re-examine the distribution of $J_{z}$ and study a correlation
between the maximum allowable magnetic twist $\Phi (r)$ and the toroidal 
Alfv\'en Mach number $M_{\rm A \phi}$ for some selected parts of the jets.

Figure \ref{fig:A2_KS} (1) shows a space-time $(\lambda, t)$ diagram of 
the power spectrum of $|\tilde{{\bolJ}^{\rm jc}}(1, \lambda)|$ for model A -- 2.
The shortest wavelength physically growing $m=1$ mode is 
$\lambda_{\rm min} \sim 1.0$; perturbations with wavelength longer than this 
already have started growing by $t=15.0$.
An enlarged snapshot of $J_{z}$ for A -- 2 at $t=19.5$ (from Fig. \ref{fig:Jz_before}) 
also is shown in Fig. \ref{fig:A2_KS} (2).
Note the wiggled structure behind (upstream of) the {\sf R--S} point with a destabilization 
length of $\lambda \sim 2.0$.  
The {\em 3rd} panel of Fig. \ref{fig:A2_KS} (3) shows the transverse profiles from panel (2) 
of $\Phi (r)$ (for $\lambda = 2.0$) and of 
$M_{\rm A \phi}$ at two different axial positions in the jet ($z$=6.1 and 8.3). 

The value of $\Phi (r)$ at the point $z=6.1$ increases 
gradually towards the inner jet.
It exceeds the K-S critical value $\Phi_{\rm crit} = 2 \pi$ 
for radial position $|x| \lesssim 0.4$ and has a maximum ($\sim 10$) near the jet axis. 
At $z=8.3$, on the other hand, $\Phi (r)$ shows a sharp rise inward 
and has already exceeded $\Phi_{\rm crit}$ by $|x| \sim 1.0$.
Between $0.4 \lesssim |x| \lesssim 1.0$, the magnetic twist is very large 
($\gtrsim 20$), followed by a sharp fall to around $\sim 10$. 

The values of $M_{\rm A \phi}$ at both axial points decrease gradually towards 
the inner jet, approaching close to 0 within $|x| \lesssim 0.4$.  Because there 
is little rotation in the region $|x| \lesssim 0.4$, the classical K--S 
criterion (\ref{eq:K-S_limit}) can be used to analyze the stability of this shallow-atmosphere, 
mildly PFD case jet.  Noting that the twist $\Phi (r)$ exceeds the K--S limit
at $z=8.3$ throughout most of the body of the jet, we conclude that the jet at that 
point is already unstable to the CD kink mode.  On the other hand, because at $z=6.1$ 
the K--S criterion is moderately violated in the region $|x| \lesssim 0.4$, we conclude that 
the flow in this model is beginning to go unstable to the CD kink mode at that point.

Figure \ref{fig:A1_KS} performs a similar analysis for 
another model from Fig. \ref{fig:Jz_before} --- the shallow-atmosphere, 
highly PFD case A -- 1.  In this case,
the shortest wavelength physically growing $m=1$ mode changes with time, 
gradually shifting from $\lambda_{\rm min} \sim 2.5$ to $\sim 1.0$. 
However, at the chosen model time ($t=21.5$), perturbations with $\lambda \lesssim
2.0$ have not yet started growing. [Compare Fig. \ref{fig:Jz_before} and 
Fig. \ref{fig:A1_KS} (1) and (2)].  Yet, in Fig. \ref{fig:A1_KS} (3), the values 
of the magnetic twist $\Phi (r)$ at {\em both} axial positions 
($z=6.1$ and 8.3) appear to exceed the critical $\Phi_{\rm crit}$ for 
$|x| \lesssim 0.4$ and have a maximum ($\sim 10$) near the center. 

The reason for this seeming violation of the K--S stability criterion, and yet 
apparent jet stability, can be seen in the $M_{\rm A \phi}$ plot in Fig. \ref{fig:A1_KS} (3). 
The value of the azimuthal Mach number at both axial points is {\em large 
for all radii in the jet}, never dipping below $M_{\rm A \phi} \sim 0.5$. 
The large rotational velocity of the jet in this case provides a stabilizing 
influence over the classical K--S criterion throughout this jet region. 

Our nonlinear numerical results are consistent with the linear theory.
A linear stability analysis of the CD $m=1$ mode was performed for a
non-rotating cold (strongly magnetized) jet (in force-free equilibrium) by
\citet[]{APP96} and \citet[]{APP00}.  They showed that the most unstable
wavenumber $k_{\rm max}$, as well as the maximum growth rate ${\rm
Im}\,(\omega)_{\rm max}$, are only slightly affected by the jet (axial)
velocity $V_{\rm j}$ and the radial profile of the magnetic pitch
$r B_{z}/B_{\phi}$.  (Note that the decrease of ${\rm Im}\,(\omega)$ with
increasing $M_{\rm A}=V_{\rm j}/V_{\rm A}$ in their papers is associated
with a decrease in $V_{\rm A}$, not an increase in $V_{\rm j}$.)  In our
results, unstable axial modes are excited over a large wavelength range.
However, the wavelength $\lambda_{\rm max}$ corresponding to $k_{\rm max}$
seems to occur in the range $2 < \lambda_{\rm max} < 3$ for both the
super-Alfv\'enic (A -- 2) and sub-Alfv\'enic (A -- 1) cases [see Figs.
\ref{fig:A2_KS} (1) and \ref{fig:A1_KS} (1)].  Therefore, we do not see a
strong correlation between $k_{\rm max}$ and $M_{\rm A}$ in our results,
and this is qualitatively consistent with the linear analysis.

These results lead us to conclude that the classical K--S criterion holds in 
approximately non-rotating jets (even the nonlinear regime), but that 
rotation in a jet creates a stabilizing effect against the CD kink mode. 
{\em Rotating jets can remain stable even when the magnetic twist exceeds the classical 
K--S criterion.}

\subsection{Sudden Destabilization of a Jet by Angular Momentum Loss}

Having shown that the $V_{\phi}$ velocity component 
plays an important role in stabilizing the jet, we now investigate how 
the jet behaves if that rotation is suddenly removed.  When $\beta \ll 1$, 
if $V_{\phi}$ were to decrease gradually
as the jet propagates, {\it e.g.}, $V_{\phi} \propto z^{-1}$, then the 
magnetic field in the jet would shift quasi-statically toward
a force-free (${\bolJ} \times {\bolB} \simeq 0$) configuration. 
However, if part of the jet suddenly loses its toroidal velocity
though some dynamical means, then that part will subject to a 
non force-free magnetic field (${\bolJ} \times {\bolB} \neq 0$).
Here we investigate the sudden loss of rotation as a possible trigger 
process for the CD instability.  

\subsubsection{The principal driving force of the CD kink}

First, we shall determine the principal driving force of the CD kink ($m=1$) mode.
Transverse profiles of the physical quantities in the intermediate stage  
at axial position $z=8.3$ are plotted in Fig. \ref{fig:SF_radial_before}.
Panel (1) shows the specific centrifugal ($V_{\phi}^{2}/r$) 
and hoop-stress ($-B_{\phi}^{2}/r/\rho$) forces for all four models.  
Panel (2) shows the distribution 
of magnetic pressure contributed by both the azimuthal and axial components 
$p_{\rm m}^{\ast} = (B_{\phi}^{2}+B_{z}^{2})/2$. 

In both models B -- 1 and B -- 2, the centrifugal force is balanced by the
hoop-stress, so the flows are {\em azimuthally} trans-Alfv\'enic
($M_{\rm A \phi} \sim 1$).
In addition, the magnetic pressure in these two models has quite a small
gradient and so contributes little to the radial force balance. 
The flow is apparently CD stable at this axial position 
($z=8.3$), so we conclude that the azimuthal rotation has stabilized the 
flow. 

In model A -- 1, the centrifugal force is smaller than the hoop-stress,
so the flow is azimuthally sub-Alfv\'enic ($M_{\rm A \phi} < 1$) and has 
a significant magnetic pressure gradient working with the centrifugal 
force to balance the hoop-stress.  It is noteworthy, however, that 
the position of the magnetic pressure peak lies at the center of the jet 
($x,\,y$)=(0,\,0):  the force balance on either side of the jet is therefore 
symmetric.  This situation is also apparently CD stable. 

In model A -- 2, however, the relative contribution of the centrifugal force is 
essentially zero ($M_{\rm A \phi} \simeq 0$).  The only forces trying to remain 
in balance are the magnetic hoop-stress and the radial magnetic pressure gradient. 
Figure \ref{fig:SF_radial_before}, however, shows that these forces are not
able to maintain a symmetric distribution as they try to balance each other.  
The hoop-stress at this axial position is asymmetric, and the (otherwise symmetric) 
magnetic pressure profile peak has been pushed to a position offset from the axis.  
{\em The 3-D response of a force-free magnetic configuration to asymmetric hoop-stress, 
then, is not to create an asymmetric magnetic pressure profile but rather to push that 
symmetric profile off to one side.} 
This asymmetry indicates that the CD modes have already begun to grow to a measurable 
level, and that this part of the jet is being accelerated in a negative radial 
direction ($-x$).

We conclude that the CD instability is driven by asymmetric hoop-stress, created in a region 
of large magnetic twist, which cannot 
be balanced stably by a corresponding asymmetric magnetic pressure gradient.

\subsubsection{Origin of the very large magnetic twist}

We now examine the origin of this high magnetic twist in the jet. The twist 
is characterized by a relatively large $|B_{\phi}|$ and a small $B_{z}$. 

Figure \ref{fig:B_radial_before} shows the intermediate-stage radial profiles (parallel to
the $r$-axis along $\phi=0$) of each component of the magnetic field ($|B_{\phi}|$ 
and $B_{z}$) and the magnetic pressure $p_{\rm m}^{\ast}$, as well as the initial
distribution of the axial field component $B_{z}^{\rm init}$ (same for all models).  
As we saw in Fig. \ref{fig:SF_radial_before} (2), the distributions of the magnetic
pressure for B -- 1 and B -- 2 are flat.  This creates separate $|B_{\phi}|$ and $B_{z}$ 
distributions that are low and high at different radii. 

For models A -- 1 and A -- 2 we see that, in the region $0.3 \lesssim r \lesssim 1.0$, 
the magnetic pressure follows $\kappa \sim -1$ and $\kappa \sim -2$ distributions, respectively. 
Of particular interest in these plots is the behavior of $B_{z}$ at these intermediate
radii.  Compared with the initial distribution, the axial field becomes stronger at small 
radii ($r \lesssim 0.3$) and remains about the same strength at large radii ($r \lesssim 1.5$).  
However, in the region $0.3 \lesssim r \lesssim 1.5$ it dips {\em substantially}, particularly 
for model A -- 2 in Fig. \ref{fig:B_radial_before}. 

We conclude that the very large magnetic twist (which leads to asymmetric hoop stress 
and a CD instability) is caused not so much by a large increase in $B_{\phi}$ as by a 
large {\em decrease} in $B_{z}$ at intermediate radii [see, especially, Fig. \ref{fig:A2_KS}(3)].

\subsubsection{Origin of the weak axial field causing the large twist:
triggering of the CD kink instability by the reverse slow-mode shock}

Finally, we shall identify the physical mechanism that causes $B_{z}$
to decrease and create the strong magnetic twist.  In particular, this will 
occur when rotation is suddenly removed from the jet flow, triggering the 
pinch.  We find that this occurs at a number of places in the jet flow, 
but is particularly strong right behind the {\sf R--S} shock wave.  

Figure \ref{fig:B_axial_before} shows in the intermediate
evolutionary stage offset axial profiles (parallel to the $z$-axis, at
($r,\,\phi$)=($0.5,\,0$)) of each component of the magnetic field and
velocity.  All models begin with the same power-law distribution of $B_{z}$
with $z$ [see panel (1)].  However, only model A -- 2, and to a lesser extent
A -- 1, develop a large region of substantially low $B_{z}$, especially
in the region $8.0 \lesssim z \lesssim 10$.

As pointed out in Sec. \ref{sect:energy_conv}, there is a sudden rise in $|B_{\phi}|$,
$V_{z}$, and $V_{\phi}$ around $z=10.0$ for all models [see Fig. 
\ref{fig:B_axial_before} (2), (3), and (4)], which corresponds to the
passage of the {\sf F--F} compressive/shock wave front.  The models split into
two classes [see panels (2) and (4)]:  those with a large enhancement in
$|B_{\phi}|$ and a small enhancement of $V_{\phi}$ 
(A -- 1 and A -- 2) and those with a small enhancement in $|B_{\phi}|$
and a large increase in $V_{\phi}$ (B -- 1 and B -- 2).

A large enhancement in $|B_{\phi}|$, with a small enhancement in
$V_{\phi}$, indicates that a considerable amount of energy has been
converted from the toroidal (rotational) kinetic energy in the TAW to
toroidal magnetic energy behind the {\sf F--F} shock wave front.

The sudden drop of both $|B_{\phi}|$ and $V_{\phi}$ at $z=9.2$ in
model A -- 2 (dashed lines) and at $z=5.5$ for B -- 2 (dotted lines)
occurs with the passage of the {\sf F--S} shock wave front.
We also can identify the {\sf R--S} shock wave front near
axial position $z=8.5$ for model A -- 2.  At this point $|B_{\phi}|$
is enhanced once again (panel (2)), but $V_{\phi}$ has a sharp dip before
rising to its injected value for $z < 8.0$.  The contact discontinuity lies in between
the two slow-mode shock waves --- at $z \sim 8.7$.

The behavior of these quantities at the {\sf R--S} shock wave
is especially interesting.  While $|B_{\phi}|$ simply increases, the dip in 
$V_{\phi}$ has important consequences.  
In this transition, part of the jet near $z=8.3$ loses
toroidal velocity (angular momentum) but still gains enhanced toroidal
magnetic field.  The centrifugal support against the hoop-stress,
therefore, is lost, causing a sudden squeezing of of the poloidal
magnetic flux $|{\bolB}_{\rm p}| \equiv (B_{r}^{2}+B_{z}^{2})^{1/2}$
towards the central ($z$) axis. As a result, a concentration of $B_{z}$
flux occurs at the jet center ($r \lesssim 0.3$), {\em at the expense
of $B_{z}$ flux at intermediate radii} ($0.3 \lesssim r \lesssim 1.5$) 
(see also A -- 2 in Fig. \ref{fig:B_radial_before}).

To summarize, we believe that the CD kink mode is triggered as follows.
The process begins when rotating, magnetized jet material experiences a sudden 
loss of kinetic angular momentum to the magnetic field.  This destroys the (stable) 
balance between centrifugal force and hoop-stress, producing a strong pinch 
of $B_{z}$ flux toward the $z$ axis, depleting $B_{z}$ flux in the 
intermediate radial region of the jet, and further enhancing the magnetic twist.
With no rotation to stabilize this large twist, the K--S criterion
is suddenly strongly violated and the jet becomes unstable to the CD kink mode.  
Any slight radial asymmetry in the hoop-stress produces a helical kink in the current,
with no stabilizing restoring force.  
This process can occur at various places in the jet flow, but appears 
strongest right behind the {\sf R--S} shock wave, where there can be a pronounced 
dip in the rotational velocity.  
We believe that this is one of the most promising physical processes
for producing a CD kink instability in rotating jets.

\subsection{Suppression of MHD KH Instabilities due to the External 
Magnetized Winds}

We now show how the helically magnetized winds that surround
the core of the jet can suppress MHD KH instabilities.

The modified stability criterion for MHD KH instabilities is as follows \citep[]{HAR02}:
\beqn
\label{eq:MHDKHI}
\Delta V^{2} - V_{\rm A s}^{2} < 0,
\eeqn
where $\Delta V \equiv V_{\rm j}-V_{\rm e}$ is the velocity shear
between the jet and the external medium, and $V_{\rm A s} \equiv
[(\rho_{\rm j}+\rho_{\rm e})(B_{\rm j}^{2}+B_{\rm e}^{2})/ (
\rho_{\rm j}\,\rho_{\rm e})]^{1/2}$ is the surface Alfv\'en speed.
Inequality (\ref{eq:MHDKHI}) shows that a magnetic field and an outgoing
wind in the external medium act to stabilize the flow against
MHD KH surface modes.  Figure \ref{fig:MHDKHI} shows, for all four of
our simulations, the transverse profiles in the $x$ direction of the
difference between $\Delta V^{2}$ and $V_{\rm A s}^{2}$, as well as the
total Alfv\'en Mach Number $M_{\rm A}~(\equiv |{\bolV}|/V_{\rm A})$.
Two different evolutionary stages are compared:  the intermediate
stage at axial position $z=8.0$ and the final stage at $z=14.0$.

In the intermediate stage, the MHD flow becomes trans- to
super-Alfv\'enic for model A -- 2, but remains sub-Alfv\'enic for all
other models, as shown in Fig. \ref{fig:MHDKHI} (1-2).  Moreover, for
model A -- 2 $M_{\rm A}$ varies smoothly at a radius ($r \sim |x| \sim
0.4$), roughly the boundary between the current-carrying jet and the
outer magnetized wind (see also Fig. \ref{fig:Jz_transverse_before}).
As is seen in Fig. \ref{fig:MHDKHI} (1-1), inequality (\ref{eq:MHDKHI})
is satisfied for all models, independent of the absolute value of
$M_{\rm A}$.  (Note that the distribution for A -- 2 is already slightly
asymmetric, even though the flow is stable to MHD KH instabilities.
This indicates that the helical kink that develops must be caused by
not the MHD KH instability but another possible mechanism --- the MHD 
CD instability in this case.)

In the final stages, the flow of jet still remains sub-Alfv\'enic
for model A -- 1, but trans- to super-Alfv\'enic for all other models,
as shown Fig. \ref{fig:MHDKHI} (2-2).  By this time all of these
current-carrying jet systems, except model B -- 2, have been distorted 
by the CD instability.  As a result, the distributions of $M_{\rm A}$ for the three
distorted models are decidedly asymmetric.  Note that, whether the instability 
has grown (A -- 1, A -- 2, and B -- 1) or not (B -- 2), inequality
(\ref{eq:MHDKHI}) still holds everywhere for these models.

We conclude that the MHD KH surface modes 
are effectively suppressed, because the external magnetized winds 
reduce the velocity shear in the transverse direction.  Our MHD
jets, therefore, are distorted not by KH instabilities but pure CD
instabilities, even when the flow itself becomes super-Alfv\'enic.

\subsection{Entrainment of Ambient Material in PFD Jets}

In Fig. \ref{fig:final_zaxis} we plot the axial profiles of the
density and the ratio of Alfv\'en and sound speeds in the final stage
(similar to Fig. \ref{fig:init_zaxis}).  Note that the distributions of
mass entrainment and $V_{\rm A}/C_{\rm s} (\approx \beta^{-1/2}$) depend
not on the initial conditions (initial gradients in the ambient medium), but
on the nature of the flow itself (on  whether it is strongly
Poynting flux dominated or not).  As the jets propagate through the
ambient medium, they adjust themselves to the decreasing external 
pressure.  If the ambient medium has no significant
field strength ($\beta \gg 1$), the 
jets expand, the density ratio $\rho_{\rm j}/\rho_{\rm e}$
becomes smaller, and the jets are rarefied.
In this case, the gradient of the external medium may play a crucial
role in the entrainment of matter.  

However, in the opposite case 
($\beta \ll 1$), the jets have considerable field strength and
are prevented from expanding by the self-confining hoop-stress.
This means the external density gradient might not have as much influence
on mass entrainment.

The mass entrained in the mildly PFD flows is much greater than in the highly
PFD flows.  In particular, $\rho$ decreases slower than $\sim z^{-1}$ for both 
models A -- 2 and B -- 2, but roughly as $\sim z^{-2}$ for models A -- 1 and 
B -- 1 as seen in the {\em Upper} panel of Fig. \ref{fig:final_zaxis}. 
This is partly due to the difference in injected mass flux into
the computational domain.  (Low mass flux implies high Poynting flux.)
However, some of the effect also appears to be related to the initial 
ambient density gradient, with a steep atmosphere causing 
a higher mass entrainment in jets than the slowly decreasing case.
The highest entrainment occurs for model B -- 2, where the flow is 
both mildly PFD {\em and} the atmosphere has an initially steep density gradient. 
An enhanced mass distribution also can be seen in model B -- 1, even 
though it is highly Poynting flux dominated.  
On the other hand, for model A -- 1, which neither is mildly PFD nor 
has an initially steep density gradient, essentially no enhancement 
(in fact, a slight decrease) can be seen. 
Denser jets in general are predicted to remain stable against MHD KH instabilities
beyond the Alfv\'en surface and have been found to be more robust than
their less dense counterparts \citep[]{HAR97a}.

The $V_{\rm A}/C_{\rm s}$ distributions also depend strongly on the
nature of the flow as seen in the {\em Lower} panel of Fig. \ref{fig:final_zaxis}. 
Compared with their initial distributions of $V_{\rm A}/C_{\rm s}$, 
all models except A -- 2 become significantly flattened in the $z$ direction. 
Only this shallow-atmosphere, mildly PFD model, develops an
even steeper distribution $\beta^{-1/2}$.

Even with this mass entrainment, however, our jets still keep their highly
magnetized state $V_{\rm A}/C_{\rm s} \gtrsim 3~ (\beta \lesssim 0.1)$
during their dynamical evolution, except in the region right behind 
the {\sf F--F} compressive/shock wave front where compression is high.

\subsection{Saturation and Advection of Kinked Jet Structures}

\begin{figure}
\begin{center}
\includegraphics[scale=0.89, angle=0]{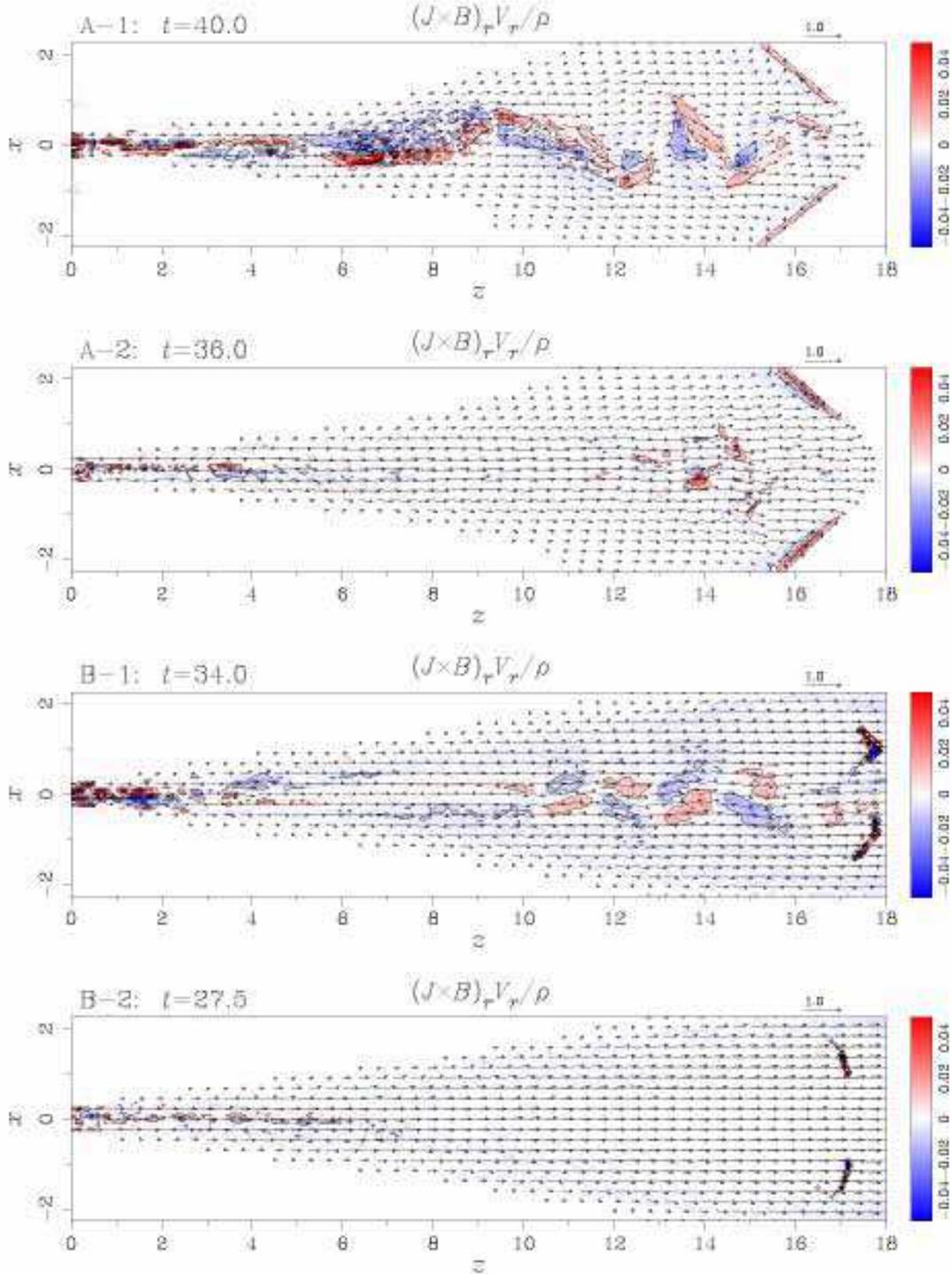}
\caption{ \label{fig:JxB_after}
Snapshots of the radial specific power generated by the Lorentz force
 (${\bolJ}\times{\bolB})_{r} V_{r}/\rho$ in the final stage for each of the models.
The $r$-component of the Lorentz force is composed of both (i) the magnetic
 pressure gradient of the azimuthal and axial components
 $-({\partial}/{\partial r}) [(B_{\phi}^{2}+B_{z}^{2})/2]$ and of (ii) the
 hoop-stress $-B_{\phi}^{2}/r$.
Positive power ({\em Red}) indicates the net acceleration (increase in
 radial velocity); negative power ({\em Blue}) indicates net deceleration.
There is a strong correlation between the power distribution and the
 wiggled structures seen in models A -- 1, A -- 2, and B -- 1.
}
\end{center}
\end{figure}

\begin{figure}
\begin{center}
\includegraphics[scale=0.85, angle=0]{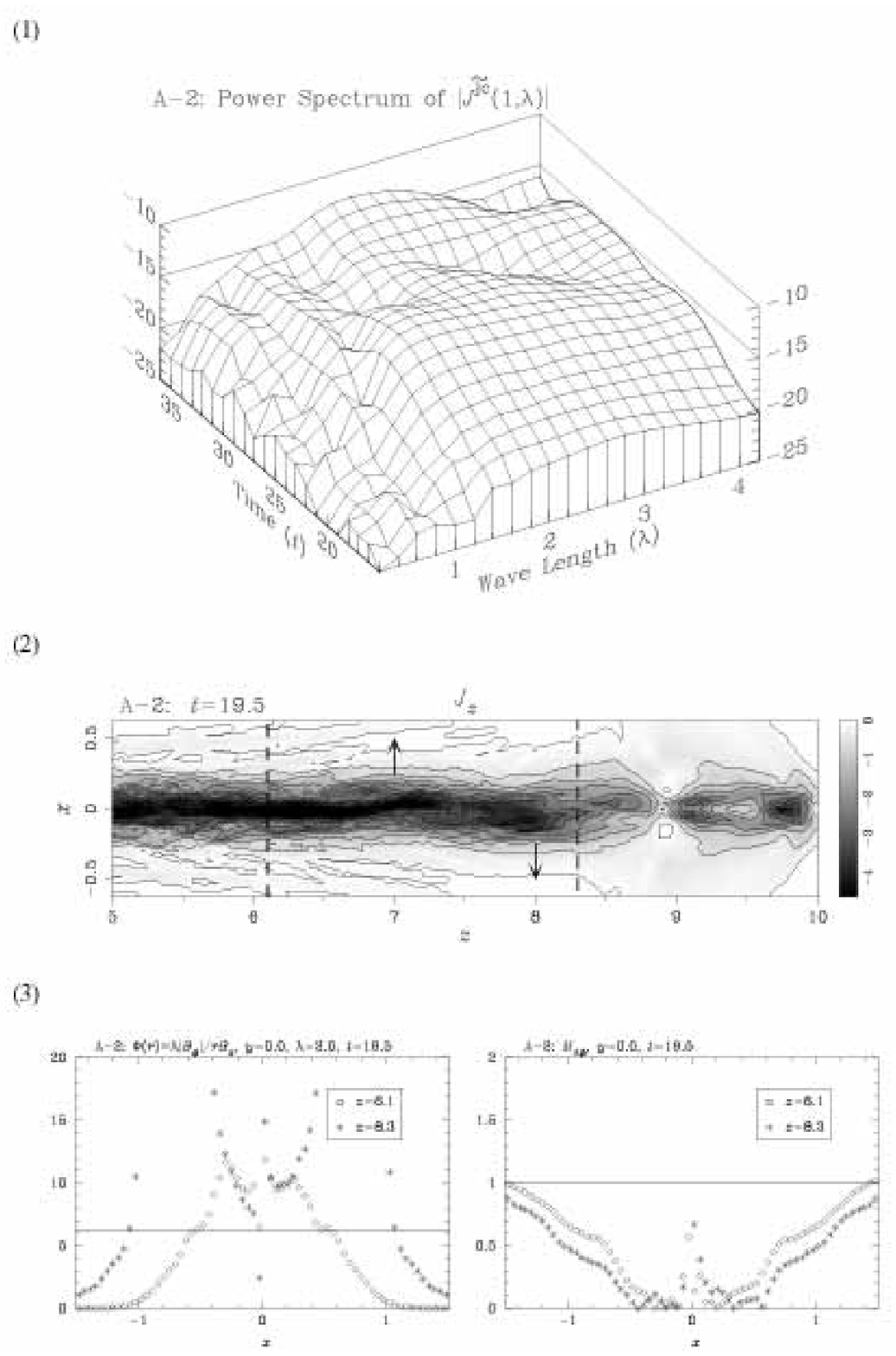}
\caption{ \label{fig:A2_KS}
Analysis of the K--S criterion for Model A -- 2 in the intermediate stage.  
(1) Space-time ($\lambda,\,t$) diagram of the Growing CD kink ($m=1$) mode showing 
the variation in growth with axial mode number ($k=\lambda/2\pi$).
(2) Enlarged snapshot of $J_{z}$ ($|x| \leq 0.635,\,5.0 \leq z \leq 10.0$), similar to 
Figs. \ref{fig:Jz_before} and \ref{fig:Jz_after}. The vertical
arrows indicate the direction of transverse motion currently underway.  
(3) Transverse profiles (parallel to the $x$-axis) of the magnetic twist 
$\Phi(r) \equiv \lambda |B_{\phi}|/r B_{z}$ ({\em Left}) and the toroidal Alfv\'en Mach number 
$M_{\rm A \phi}$ ({\em Right}) at the axial positions $z=6.1$ and 8.3, indicated with the vertical 
dashed lines in panel (2).  The horizontal lines in panel (3) indicate $\Phi(r) = \Phi_{\rm crit}
 = 2 \pi$ ({\em Left}) and $M_{\rm A \phi} = 1$ ({\em Right}). 
}
\end{center}
\end{figure}

\begin{figure}
\begin{center}
\includegraphics[scale=0.9, angle=0]{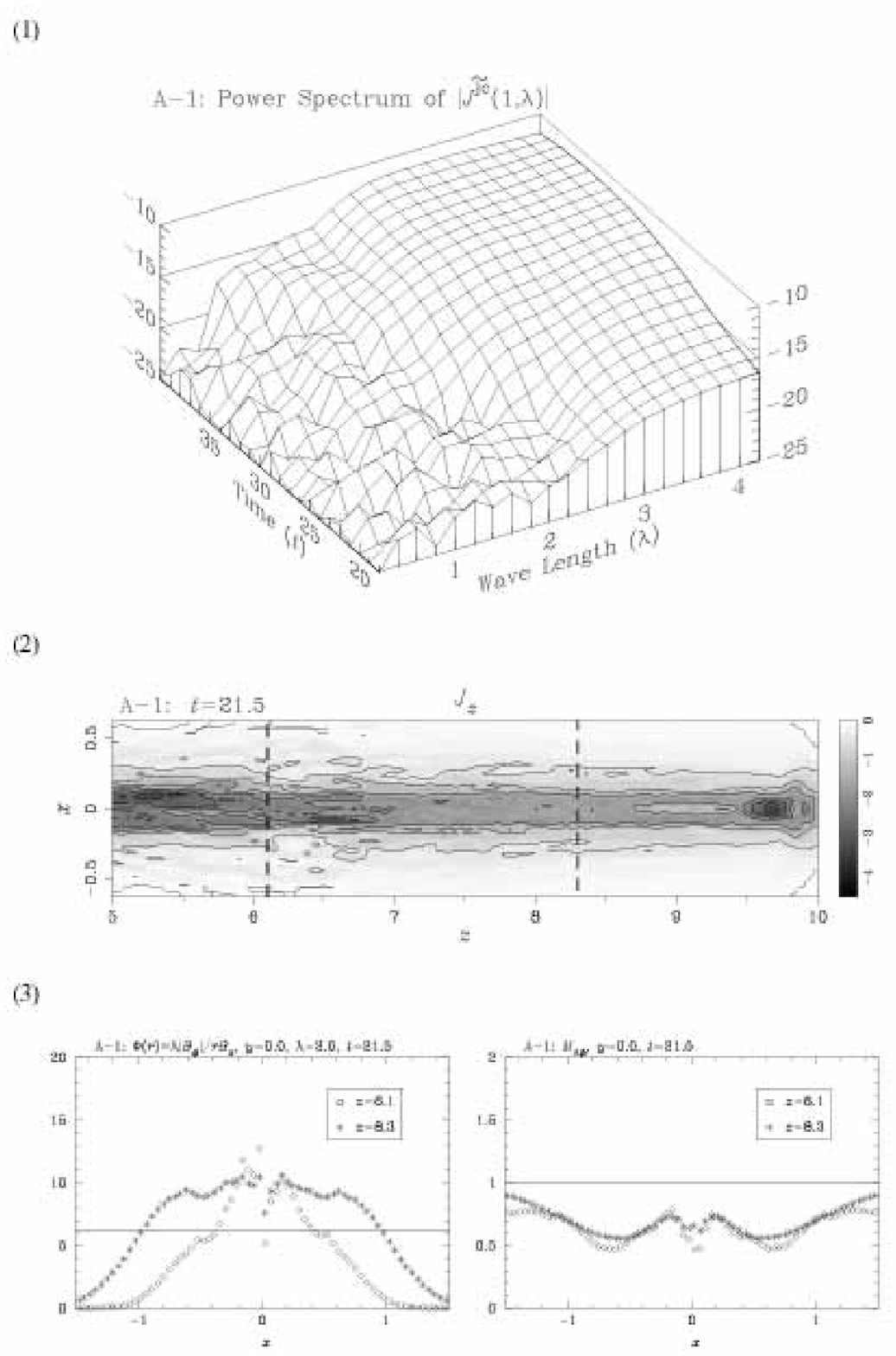}
\caption{ \label{fig:A1_KS}
Similar to Fig. \ref{fig:A2_KS}, but for Model A -- 1.
}
\end{center}
\end{figure}

\begin{figure}
\begin{center}
\includegraphics[scale=1.0, angle=-90]{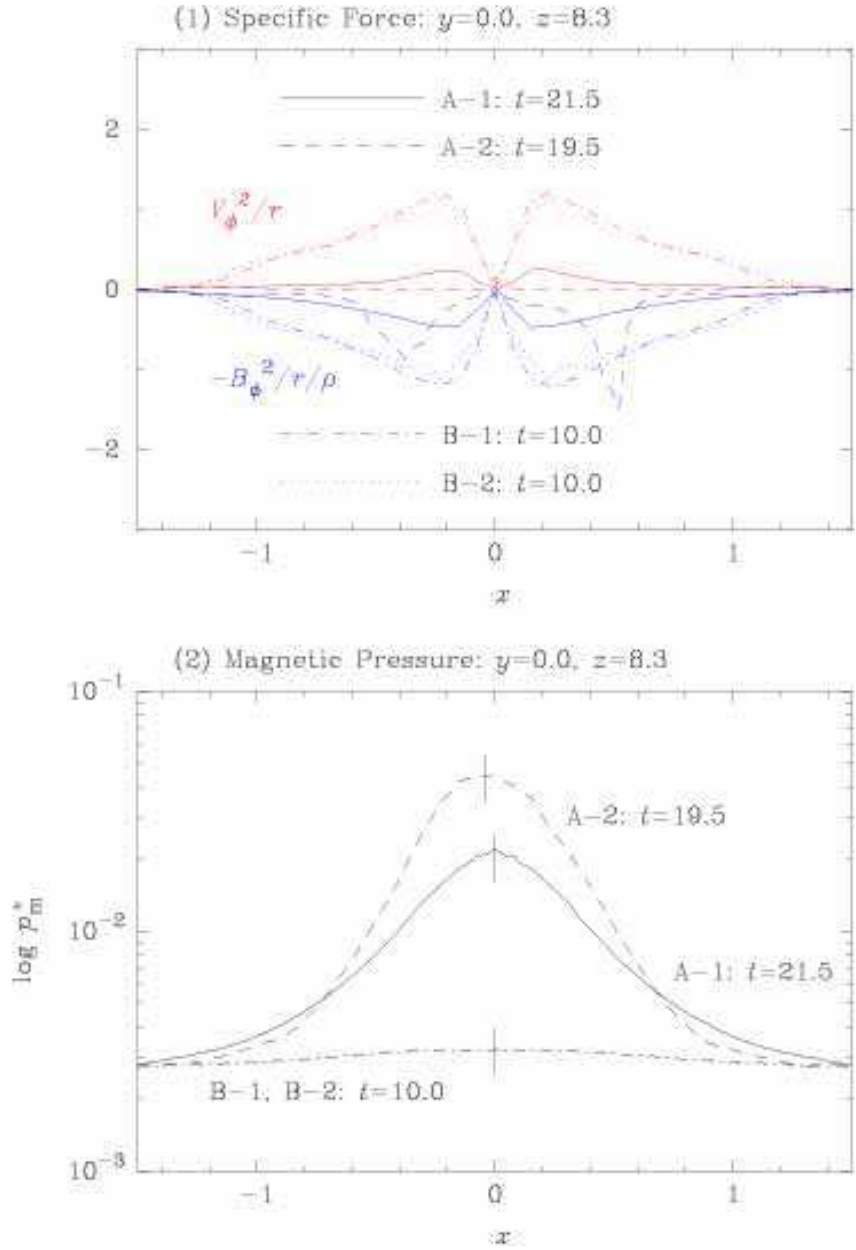}
\caption{ \label{fig:SF_radial_before}
Transverse profiles (parallel to the $x$-axis) in the intermediate stage at axial position $z=8.3$. 
(1) Profiles of the specific centrifugal force $V_{\phi}^{2}/r$ 
 ({\em Red}) and the specific hoop-stress $-B_{\phi}^{2}/r/\rho$ ({\em Blue}). 
(2) Profiles of the magnetic pressure $p_{\rm m}^{\ast} = (B_{\phi}^{2}+B_{z}^{2})/2$ (logarithmic scale) 
 for each of the models.  (Note that the curves for models B -- 1 and B -- 2 coincide in the 
 {\em dotted} / {\em dot-dashed} line in panel(2).) 
The short vertical solid lines in the {\em Lower} panel indicate the position of the 
peak of each transverse magnetic pressure profile. 
}
\end{center}
\end{figure}

\begin{figure}
\begin{center}
\includegraphics[scale=0.7, angle=-90]{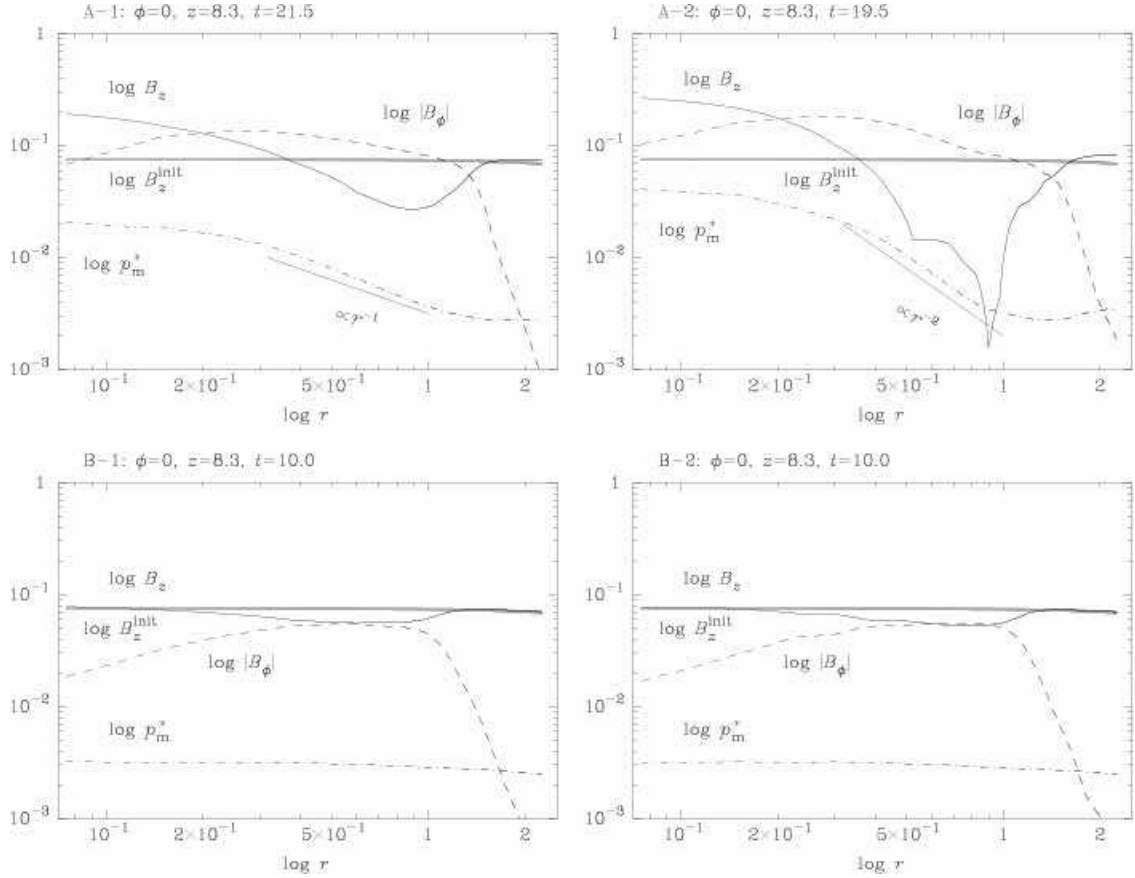}
\caption{ \label{fig:B_radial_before}
Radial profiles (parallel to the ``$r$''-axis along $\phi=0$) 
of the absolute value of magnetic field components $B_{z}$, $|B_{\phi}|$, and 
the magnetic pressure $p_{\rm m}^{\ast}$ (logarithmic scale) 
in the intermediate stage at axial position $z=8.3$.
$B_{z}^{\rm init}$ ({\em Light gray thick solid line}) represents the initial ($t=0.0$) distribution of the axial field.  
}
\end{center}
\end{figure}

\begin{figure}
\begin{center}
\includegraphics[scale=0.7, angle=-90]{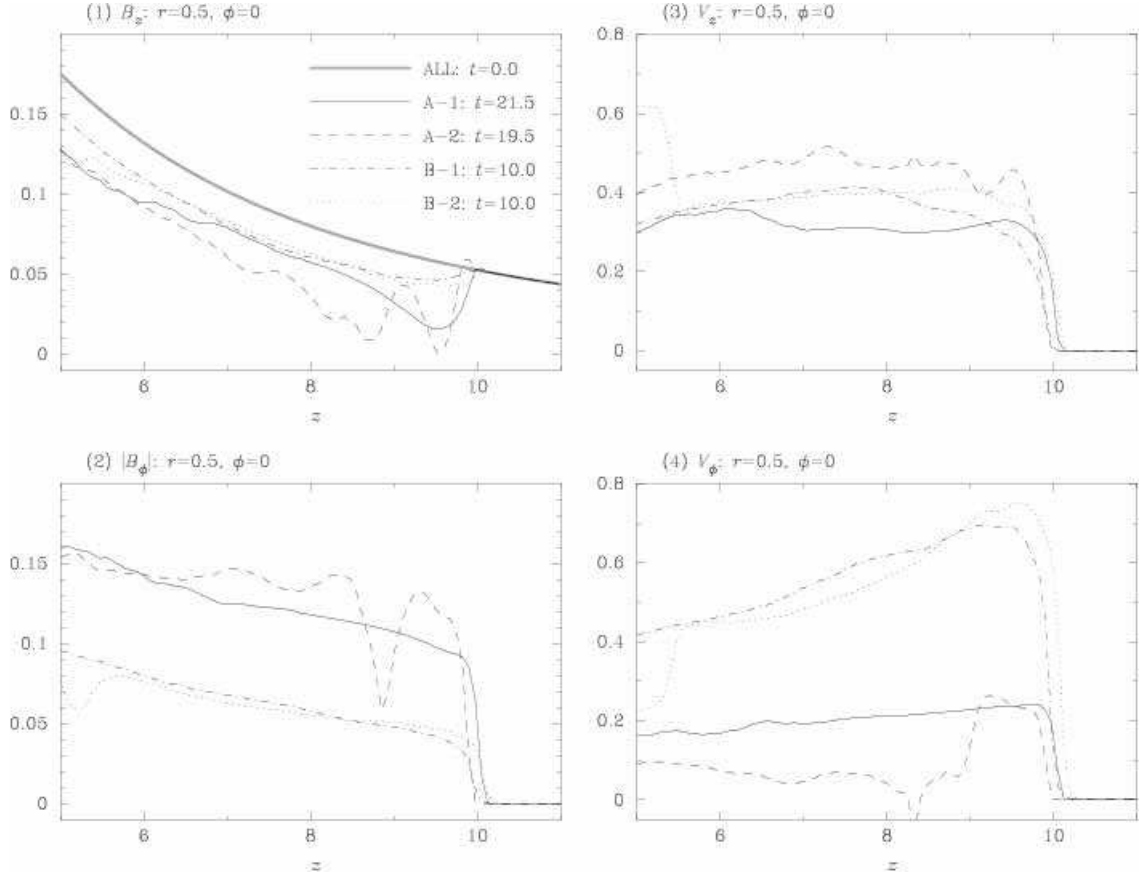}
\caption{ \label{fig:B_axial_before}
Offset axial profiles, parallel to the $z$-axis, at $(r,\,\phi)=(0.5,\,0)$ 
in the intermediate stage for each of the models.
(1) Axial magnetic field component $B_{z}$ (the initial distribution is also plotted 
by {\em light gray thick solid line}). 
(2) Absolute value of azimuthal magnetic field component $|B_{\phi}|$. 
(3) Axial velocity $V_{z}$. 
(4) Azimuthal velocity $V_{\phi}$. 
}
\end{center}
\end{figure}

\begin{figure}
\begin{center}
\includegraphics[scale=0.7, angle=-90]{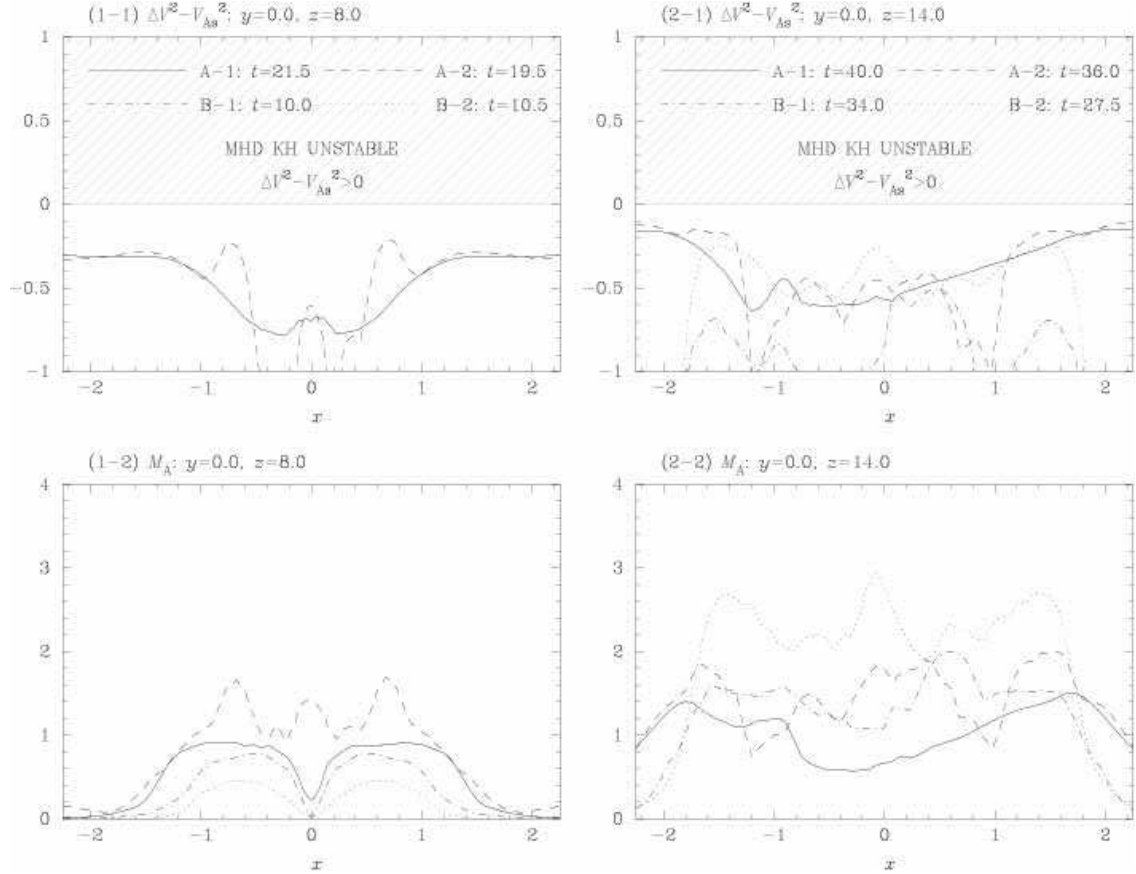}
\caption{ \label{fig:MHDKHI}
Transverse profiles in the $x$ direction of velocity shear and Alfv\'en Mach number 
in the intermediate stage ({\em Left} panels at axial position $z=8.0$) and 
final stage ({\em Right} panels at $z=14.0$) of evolution.  
{\em Upper} panels:  the difference between the square of the velocity shear $\Delta V^{2}$ 
and the square of the surface Alfv\'en velocity $V_{\rm A s}^2$.   
{\em Lower} panels:  the total Alfv\'en Mach number $M_{\rm A}$.  
The shaded area in the {\em Upper} figures shows the MHD KH unstable
region ($\Delta V^{2} \geq V_{\rm A s}^{2}$).  In the first panel, models B -- 1 
and B -- 2 do not appear, because their velocity difference lies below $-1$. 
}
\end{center}
\end{figure}

\begin{figure}
\begin{center}
\includegraphics[scale=1.0, angle=-90]{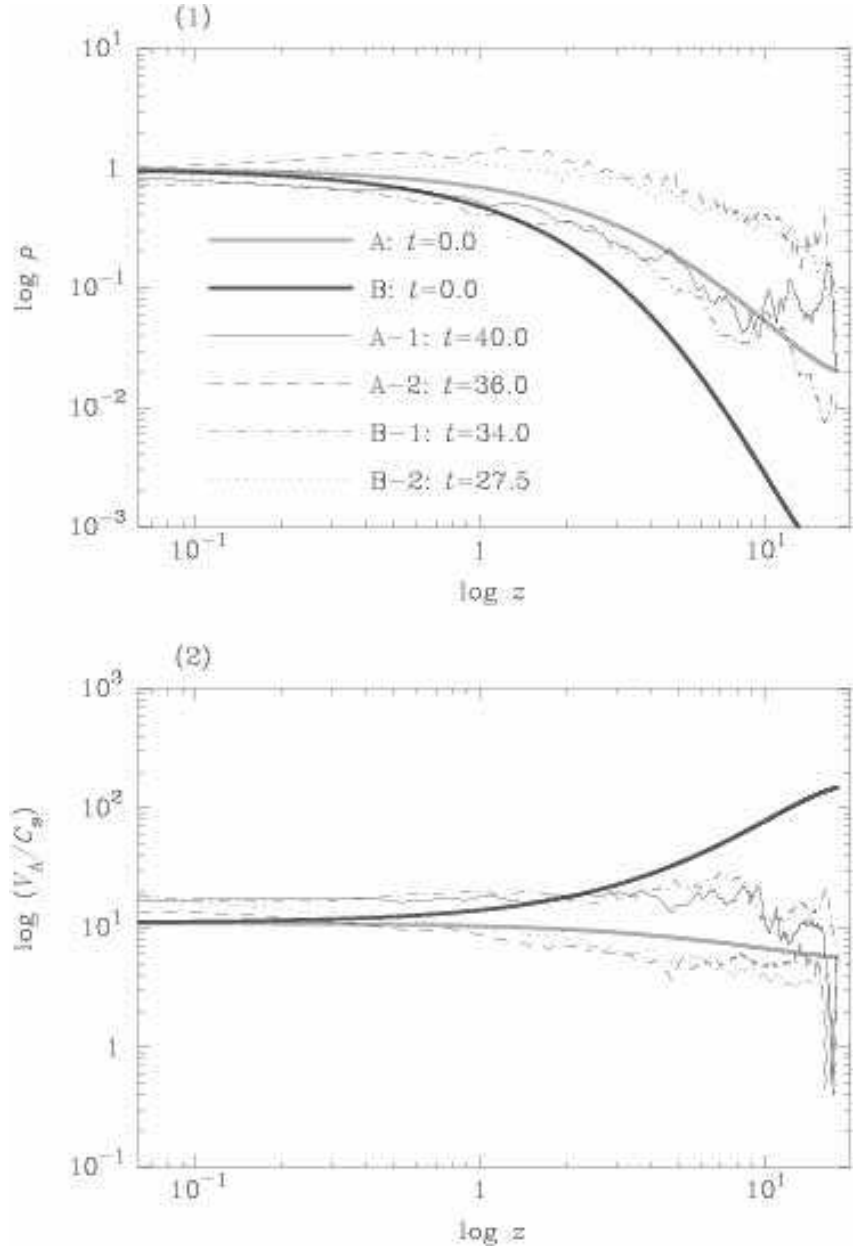}
\caption{\label{fig:final_zaxis}
Axial profiles of (1) the density $\rho$ ({\em Upper}) and 
(2) the ratio of Alfv\'en velocity to sound speed $V_{\rm A}/C_{\rm s}$ 
({\em Lower}) in the final stage of evolution.
Each thick solid line represents the initial ($t=0.0$) distribution for
Model A ({\em light gray}) and Model B ({\em dark gray}).
}
\end{center}
\end{figure}

As we have seen in Fig. \ref{fig:space_time_twist} and \ref{fig:JxB_after},
the patterns created by the CD instability can be advected by the
propagating jet even after saturation of that instability has occurred.
This means that {\em the CD instability does not destroy the interior of the jet.}
Rather, it only distorts the interior into a semi-permanent wiggled structure.
In the final stage of the dynamical evolution, the jets still
remain strongly magnetically dominated ($\beta \lesssim 0.1$), and the 
flow never attains a state of MHD turbulence (which requires $\beta \gg 1$). 
If the magnetic diffusion time scale ({\it i.e.}, dissipation time scale of the current) 
is much longer than the jet propagation time scale for trans-Alfv\'enic flow 
(the Alfv\'en transit time scale) \citep[]{KOI96}, 
these patterns will persist for some time as the flow advances.  
And the plasma flow itself will appear to travel in a true helical 
pattern as it follows the magnetic backbone of the helix (see the 
velocity vectors in Figs. \ref{fig:Jz_after} and \ref{fig:JxB_after}). 

\section{CONCLUSIONS AND APPLICATION}

\subsection{Summary and Conclusions}

By performing 3-D non-relativistic MHD simulations we have investigated in detail 
the nonlinear dynamics of PFD jets and the development of asymmetric instabilities. 
The PFD jets, powered by TAWs, propagate into extended stratified atmospheres.
The presence of various compressive MHD waves in jets, which can steepen into MHD shock waves, 
plays an important role in the nonlinear evolution of the system. 
Effects such as shock waves or re-distribution of the axial current profile in the 
post-shocked region strongly affect the excitation of the CD instabilities.
The growth of the CD $m=1$ mode causes a kinked, 3-D spatial helical structure, which may 
be responsible for the wiggled structures seen in images of some observed jets.

We have performed four simulations with different combinations of the following initial parameters:  
highly ($F_{E \times B}/F_{\rm tot} \sim 0.9$) or mildly ($F_{E \times B}/F_{\rm tot} \sim 0.6$) 
PFD jets with decreasing or constant $V_{\rm A}$ external ambient medium along the central axis ($z$).  
From these simulations we conclude the following:  
\begin{enumerate}
\item 
Due to a centrifugal effect, rotating jets can be stabilized against the CD kink 
mode beyond the point predicted by the classical K-S criterion.  Radial force 
balance in strongly magnetized ($C_{\rm s}^{2} \ll V_{\rm A}^{2}$) rotating jets, 
therefore, can involve a centrifugal force component; it need not involve only a 
pure electromagnetic force-free field.  
\item 
Non-rotating jets will be subject to the CD kink mode when the classical K-S 
criterion is violated. The kink ($m=1$) mode grows faster than the higher order modes 
($m > 1$). 
\item 
The driving force of the CD kink mode is an asymmetrically distribution of 
hoop-stress (the magnetic tension force).  This is caused by a sudden decrease 
of jet rotation and a concentration the poloidal magnetic flux toward the central
axis via nonlinear processes that occur at a reverse slow-mode MHD shock wave. 
As a result, the magnetic twist become larger than the critical value specified by 
the K-S criterion.    
\item 
The linear growth rate of the CD kink mode does not depend on the flow speed.  
Instead, it grows on time scales of order of the local Alfv\'en crossing time.
Kinetic-flux dominated jets that have a large Alfv\'en Mach number $M_{\rm A}$ 
($\sim M_{\rm FM}$ in $C_{\rm s}^{2} \ll V_{\rm A}^{2}$) are not subject to the 
CD kink mode, because the jet propagation time is shorter than the CD instability
growth time.  
\item
We do not see growing surface modes of the MHD KH instability even when our flows become 
super-Alfv\'enic.  
The naturally-occurring, external helically magnetized wind, which is (quasi-) axially
current-free ($J_{z} \simeq 0$), surrounds the well-collimated
current-carrying jet.
This wind reduces velocity shear between the jet and the external
medium, suppressing the growth of KH surface modes. 
Therefore, we have been able to investigate exclusively the nonlinear behavior of
pure CD instabilities.
\item
Under strongly magnetized ambient conditions, the amount of mass entrained and the 
ratio $V_{\rm A}/C_{\rm s}$ in the jet depend critically on the nature itself of 
the flow ({\it i.e.}, on the level of Poynting flux domination) and much less 
on the external density or gas pressure gradients.  
The properties of jets observed thousands of Schwarzschild radii from the galactic 
center, therefore, may nevertheless carry important clues about the nature of the 
central engine itself. 
\end{enumerate}

The results presented in this paper are valid in the non-relativistic regime.
For relativistic motion ($V \approx c$), the electric field ${\bolE}$ may be
comparable to the magnetic field ${\bolB}$, but it will never dominate
(${\bolB}^{2}-{\bolE}^{2}>0$ in the force-free limit).
As in these non-relativistic simulations, therefore, 
the CD instabilities will not propagate in the rest frame of the jet (jet
co-moving frame).  If we choose the inertial frame to be the rest
frame of the jet fluid, ${\bolE}$ will vanish for a perfectly conducting
fluid (the MHD condition).
With ${\bolE}$ (and the displacement current ${\bolD}$) negligible, the 
rotation of a relativistic PFD jet (in which the energy density of the 
electromagnetic field is much greater than that of the particles)
still will play an important role in stabilizing the flow against the growing 
CD instability, even in the relativistic limit.
Full treatment of nonlinear CD instabilities in relativistic PFD jets will be
given in a forthcoming paper.

\subsection{The CD Instability as the Origin of Kinked Structure in AGN and Pulsar Jets}

The behavior of our simulations is 
consistent with helical flows seen in such sources as 3C 345 \citep[]{ZEN95} and 
3C 120 \citep[]{GOM01}.  Helical 
structures appear to persist for longer than component propagation times, and 
the plasma components themselves appear to propagate on helical trajectories. 

Of course, a three-dimensional helical shape in a real jet could be 
produced by precessional motion at the jet origin instead of by 
asymmetric CD ($m > 0$) modes. 
In both cases the center of the actual jet flow would deviate from the jet 
central axis during dynamical evolution.  
However, precessional jets do not show true helical motion as their components 
move outward from the galactic center.  They simply move ballistically, 
with the turning radius becoming larger as the jet moves further from the origin.

In the case of asymmetric KH surface modes, such a deviation of the flow from 
the jet axis is not even possible \citep[see, {\it e.g.}, Fig. 5 in][]{HAR97b}. 
While it might be so for the asymmetric KH ``body'' modes, 
the growth range for these modes is limited (see, 2.4.1).

Therefore, if most parsec-scale helical jets in AGN are confirmed to be true helical flows, 
then the CD instability in a Poynting flux dominated jet may be a viable model 
for producing the kinked, helical structures seen in these objects. 
An excited wave disturbance, such as occurs in a KH mode, may not always be 
needed in order to explain the kinked morphological features in these jets. 

High angular resolution, time-dependent X-ray observations by the {\it Chandra} 
X-ray observatory displayed year-scale variations in jet morphology in the 
Vela \citep[]{PAV03} and Crab \citep[]{MOR04} pulsar wind nebulae.  
These sources showed a growth of their overall kinked structure as the jet 
moves outward.  The Crab jet is 10 times larger and varies 10 times more slowly 
than the Vela jet. (Time scales and jet widths are 150-500 days and $2.9 \times 10^{17}$ cm 
for the Crab, and 10-30 days and $3.0 \times 10^{16}$ cm for Vela, respectively.) 
These kinked patterns seem to be co-moving with the main flow, and 
the time scales reported are proportional to the Alfv\'en crossing time (with $V_A$
being approximately the speed of light for these ultra-relativistic plasmas in 
equipartition with the magnetic field).  
All of these features are consistent with our results on CD instabilities in 
PFD collimated flows.  
Pulsar jets, therefore, are one of the strongest cases for a Poynting flux dominated 
astrophysical flow that has been distorted by current-driven instabilities.  

\acknowledgments
We thank P. E. Hardee for many useful comments and suggestions.
M.N. thanks S. Hirose for useful discussions.
He is supported, in part, by a National Research Council Resident Research 
Associateship, sponsored by the National Aeronautics and Space Administration.  
D.M. is supported, in part, by a NASA Astrophysics Theory Program grant.  
This research was performed at the Jet Propulsion Laboratory, 
California Institute of Technology, under contract to NASA.
Numerical computations were carried out on Fujitsu VPP5000/32R at the Astronomical Data 
Analysis Center of the National Astronomical Observatory, Japan.  Visualizations were
performed on SGI Origin 2000 at the Supercomputing and Visualization Facility of 
Jet Propulsion Laboratory and on other departmental computers.

\appendix

\section{THE INITIAL FIELD CONFIGURATION}

The initial current-free field is explicitly given by the following formula:
\beqn
&&{\bolB}=\left(B_{x},\,B_{y},\,B_z \right)=
(B_{r}\cos \phi-B_{\phi} \sin \phi,\, B_{r} \sin \phi + B_{\phi} \cos \phi,\, B_{z}),\\
&&\quad B_{r} \equiv \sum_{n=1}^2\left\{
\frac{2(z-z_{n})}{r \sqrt{(r_{n}+r)^2+(z-z_{n})^2}}
\left(-K(k)+\frac{r_{n}^2+r^2+(z-z_{n})^2}
{(r_{n}-r)^2+(z-z_{n})^2}E(k)\right)\right\} \nonumber \\
&&\quad \quad \left/\sum_{n=1}^2\left\{\frac{2\pi r_{n}^2}
{(r_{n}^2+z_{n}^2)^{3/2}}\right\} \right., \\
&&\quad B_{\phi} \equiv 0, \\ 
&&\quad B_z \equiv \sum_{n=1}^2\left\{
\frac{2}{\sqrt{(r_{n}+r)^2+(z-z_{n})^2}}
\left(K(k)+\frac{r_{n}^2-r^2-(z-z_{n})^2}
{(r_{n}-r)^2+(z-z_{n})^2}E(k)\right)\right\} \nonumber \\
&&\quad \quad \left/\sum_{n=1}^2\left\{\frac{2\pi r_{n}^2}
{(r_{n}^2+z_{n}^2)^{3/2}}\right\} \right.,\\
&&\quad K(k)=\int_{0}^{\pi/2}\frac{d\theta}{\sqrt{1-k^2\sin^2\theta}}, \ \
E(k)= \int_{\scriptstyle 0}^{\scriptstyle \pi/2}\sqrt{1-k^2\sin^2\theta} \ d\theta, \\
&&\quad k^2=\frac{4 r_{n} r}{(r_{n}+r)^2+z^2}, \ \ r \equiv \sqrt{x^2+y^2}.
\eeqn
On the $z$-axis ($r=0$) we have 
\beqn
&&B_{r} \equiv 0, \\
&&B_{\phi} \equiv 0, \\
&&B_{z} \equiv \sum_{n=1}^2\left\{
\frac{2 \pi r_{n}^2}{\left\{r_{n}^2+(z-z_{n})^2 \right\}^{3/2}}\right\}\left/\sum_{n=1}^2\left\{\frac{2\pi r_{n}^2}
{(r_{n}^2+z_{n}^2)^{3/2}}\right\} \right..
\eeqn
Parameters $r_{n}$ and $z_{n}$ are the radius and position of the $n$th loop (ring) current,
respectively, which are set as follows: $r_{n}=1.0$ for all $n$, and the positions of 
pair of current loops are at $z_{n}=-3.0$ and $2 z_{\rm max}+3.0$.



\onecolumn

\clearpage

\clearpage

\clearpage

\clearpage

\clearpage

\clearpage

\clearpage

\clearpage

\clearpage

\clearpage

\clearpage

\clearpage

\clearpage

\clearpage

\clearpage

\clearpage

\clearpage

\clearpage

\clearpage

\clearpage

\clearpage

\clearpage

\clearpage






\clearpage

\begin{deluxetable}{rlc}
\tabletypesize{\scriptsize}
\tablecaption{Units of Physical Quantities for Normalization. \label{tbl:unit}}
\tablewidth{0pt}
\tablehead{
\colhead{Physical Quantity} & \colhead{Description}   & \colhead{Normalization Unit}}
\startdata
$t$\dotfill                   & Time            & $\tau_{\rm A0}\,(\equiv L_{\rm 0}/V_{\rm A0})$ \\
$L\,(\equiv \sqrt{x^2+y^2+z^2})$\dotfill        & Length          & $L_{0}                           $ \\
$\rho$\dotfill                & Density         & $\rho_{0}                        $ \\
$p$\dotfill                   & Pressure        & $\rho_{0} V_{\rm A0}^{2}         $ \\
${\bolV}$\dotfill & Velocity        & $V_{\rm A0}                      $ \\
${\bolB}$\dotfill & Magnetic field  & $\sqrt{4 \pi \rho_{0} V^{2}_{\rm A0}}  $ \\
${\bolJ}$\dotfill & Current Density & $\sqrt{4 \pi \rho_{\rm 0} V^{2}_{\rm A0}}/L_{0} $ \\
\enddata
\tablecomments{
The initial value of the density $\rho_{0}$ and Alfv\'en velocity $V_{\rm A0}$ 
at the origin $(x,\,y,\,z)=(0,\,0,\,0)$ are chosen to be the typical density and velocity in the system.
That is, the initial dimensionless density $\rho^{\prime}$ and Alfv\'en velocity $V_{\rm A}^{\prime}$ 
at the origin are set to unity. 
The unit length scale $L_{0}$ is approximately the same as the initial jet diameter at $z=0$.
So, we can define a characteristic time scale, the initial Alfv\'en 
crossing time $\tau_{\rm A 0}$ at the origin, which is associated with 
$L_{0}$ and $V_{\rm A 0}$.
The time is normalized with $\tau_{\rm A 0}$, so the dimensionless Alfv\'en crossing time
also is set to unity: $\tau_{\rm A}^{\prime}=1$.
}
\end{deluxetable}

\begin{deluxetable}{rlccccc}
\tabletypesize{\scriptsize}
\tablecaption{Model Parameters and Flow Properties. \label{tbl:model}}
\tablewidth{0pt}
\tablehead{
\colhead{Model} & \colhead{${\alpha}$\tablenotemark{a}} & \colhead{${\Gamma}$\tablenotemark{b}} & \colhead{${\nabla C_{\rm s}}$\tablenotemark{b}} 
& \colhead{${\nabla V_{\rm A}}$\tablenotemark{d}} & \colhead{${\nabla \beta}$\tablenotemark{e}} 
& \colhead{${F_{E \times B}/F_{\rm tot}}$\tablenotemark{f}}
}
\startdata
A-1\dotfill & 1.0 & 5/3 & $\searrow$ & $\searrow$    & $\nearrow$ & $\sim 0.6$\\
A-2\dotfill & 1.0 & 5/3 & $\searrow$ & $\searrow$    & $\nearrow$ & $\sim 0.9$\\
B-1\dotfill & 2.0 & 5/3 & $\searrow$ & $\rightarrow$ & $\searrow$ & $\sim 0.6$\\
B-2\dotfill & 2.0 & 5/3 & $\searrow$ & $\rightarrow$ & $\searrow$ & $\sim 0.9$\\
\enddata

\tablenotetext{a} {Power index of $\rho \propto |{\bolB}|^{\alpha}$.}
\tablenotetext{b} {Polytropic index of $p \propto \rho^{\Gamma}$.}
\tablenotetext{c} {Gradient of the sound velocity along the $z$-axis.}
\tablenotetext{d} {Gradient of the Alfv\'en velocity along the $z$-axis.}
\tablenotetext{e} {Gradient of the plasma-$\beta$ along the $z$-axis.}
\tablenotetext{f} {Time averaged Poynting flux $F_{E \times B}$ injected into the system though the $z=0$ plane, 
normalized by the total energy flux $F_{\rm tot}$.}

\tablecomments{
Models A represent slowly decreasing atmospheres with decreasing $V_{\rm A}$. 
Models B represent steeply decreasing atmospheres with constant $V_{\rm A}$.
The signs ``$\searrow$'', ``$\nearrow$'', and ``$\rightarrow$'' indicate negative, positive, 
and no gradients along the $z$-axis, respectively.
}

\end{deluxetable}

\begin{deluxetable}{rlcc}
\tabletypesize{\scriptsize}
\tablecaption{Comparison of Growth Rates for All Models. \label{tbl:growth}}
\tablewidth{0pt}
\tablehead{
\colhead{Model} & \colhead{${\rm Im}\,(\omega)$} & \colhead{Growing azimuthal modes ($m$)} & \colhead{Saturation}}
\startdata
A-1\dotfill & 0.31 & 1,\,2 & No\\
A-2\dotfill & 0.54 & 1     & Yes\\
B-1\dotfill & 0.51 & 1     & No\\ 
B-2\dotfill & 0.42 & 1     & Yes\\
\enddata
\end{deluxetable}



\end{document}